\documentclass[fleqn, usenatbib]{mnras}

\usepackage{newtxtext, newtxmath}

\usepackage[T1]{fontenc}
\usepackage[utf8]{inputenc}

\DeclareRobustCommand{\VAN}[3]{#2}
\let\VANthebibliography\thebibliography
\def\thebibliography{\DeclareRobustCommand{\VAN}[3]{##3}\VANthebibliography}

\usepackage{ae,aecompl}
\usepackage{graphicx}
\usepackage[dvipsnames,hyperref]{xcolor}
\usepackage{xspace}
\usepackage{soul}
\usepackage{booktabs}
\usepackage{adjustbox}

\usepackage{amsmath}
\usepackage{commath}
\usepackage{siunitx}
\usepackage{nth}

\usepackage[capitalise, noabbrev]{cleveref}
\usepackage{enumitem}
\usepackage{csvsimple}
\usepackage{multirow}
\usepackage[acronym]{glossaries}
\hypersetup{unicode}

\crefname{equation}{equation}{equations}

\bibpunct[]{(}{)}{;}{a}{}{,}

\defcitealias{1981PASP...93....5B}{BPT}
\defcitealias{1987ApJS...63..295V}{VO87}

\newcommand{\orcidsymb}[2]{#1\href{http://orcid.org/#2}{\adjustbox{trim={-.15\width} {0\height} {-.15\width} {0\height},clip}{\includegraphics[height=10pt]{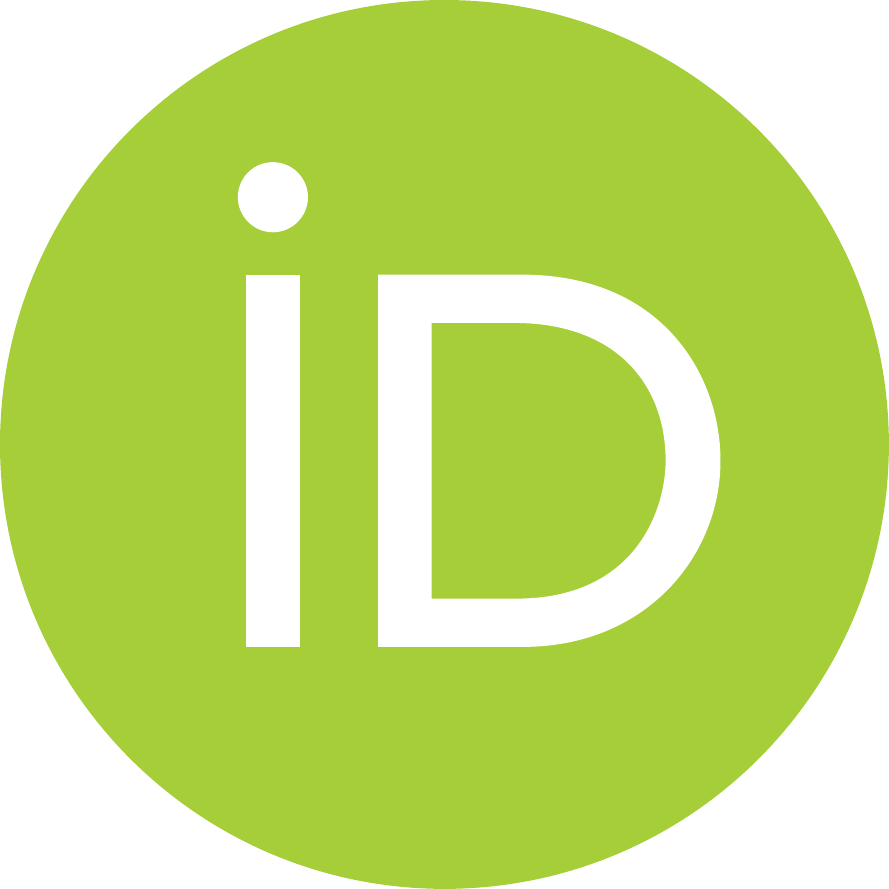}}}}

\newcommand{\program}{\textsc}
\newcommand{\ssim}{\sim \!}

\newcommand{\Lymana}{{Lyman-\ensuremath{\upalpha}}\xspace}
\newcommand{\Lymanatext}{Lyman-α}
\newcommand{\Lya}{{Ly\ensuremath{\upalpha}}\xspace}
\newcommand{\Lyatext}{Lyα}
\newcommand{\HI}{\hbox{H\,{\sc i}}\xspace}
\newcommand{\HII}{\hbox{H\,{\sc ii}}\xspace}
\newcommand{\CIV}{\hbox{C\,{\sc iv}}\xspace}
\newcommand{\HeI}{\hbox{He\,{\sc i}}\xspace}
\newcommand{\HeII}{\hbox{He\,{\sc ii}}\xspace}
\newcommand{\CIII}{\hbox{C\,{\sc iii}]}\xspace}
\newcommand{\CIIIs}{\hbox{C\,{\sc iii}]}\xspace}
\newcommand{\CIIIf}{\hbox{[C\,{\sc iii}]}\xspace}

\newcommand{\OII}{\hbox{[O\,{\sc ii}]}\xspace}
\newcommand{\OIII}{\hbox{[O\,{\sc iii}]}\xspace}
\newcommand{\OIIIs}{\hbox{O\,{\sc iii}]}\xspace}
\newcommand{\NeIII}{\hbox{[Ne\,{\sc iii}]}\xspace}
\newcommand{\MgII}{\hbox{Mg\,{\sc ii}}\xspace}
\newcommand{\NII}{\hbox{[N\,{\sc ii}]}\xspace}

\newcommand{\SII}{\hbox{[S\,{\sc ii}]}\xspace}
\newcommand{\CII}{\hbox{[C\,{\sc ii}]}\xspace}

\newcommand{\Halpha}{\ensuremath{\mathrm{H}\upalpha}\xspace}
\newcommand{\Hbeta}{\ensuremath{\mathrm{H}\upbeta}\xspace}
\newcommand{\Hgamma}{\ensuremath{\mathrm{H}\upgamma}\xspace}
\newcommand{\Hdelta}{\ensuremath{\mathrm{H}\updelta}\xspace}
\newcommand{\Hepsilon}{\ensuremath{\mathrm{H}\upepsilon}\xspace}

\newcommand{\hst}{\textit{HST}\xspace}
\newcommand{\jwst}{\textit{JWST}\xspace}

\newcommand{\JGNzeightzeroLA}{JADES-GN-z8-0-LA\xspace}
\newcommand{\JGSzeightzeroLA}{JADES-GS-z8-0-LA\xspace}
\newcommand{\JGSzeightoneLA}{JADES-GS-z8-1-LA\xspace}

\hypersetup{
    unicode=true,
    final=true,
    plainpages=false,
    pdfstartview=FitV,
    pdftoolbar=false,
    pdfmenubar=true,
    bookmarksopen=true,
    bookmarksnumbered=true,
    breaklinks=true,
    linktocpage,
    colorlinks=true,
    linkcolor=Blue,
    urlcolor=Blue,
    citecolor=Blue,
    anchorcolor=green
}
\AtBeginDocument{
    \hypersetup{
        pdftitle={JADES: Primaeval \Lymanatext emitting galaxies reveal early sites of reionisation},
        pdfauthor={J. Witstok et al.},
        pdfsubject={Primaeval \Lymanatext emitting galaxies reveal early sites of reionisation},
        pdfkeywords={{galaxies: high-redshift} -- {dark ages, reionization, first stars} -- {methods: observational} -- {techniques: spectroscopic}}
    }
}

\title[Primaeval LAEs reveal early reionisation sites]{JADES: Primaeval \texorpdfstring{\Lymana}{\Lymanatext} emitting galaxies reveal early sites of reionisation out to redshift $z \sim 9$}

\author[J. Witstok et al.]{{\orcidsymb{Joris Witstok}{0000-0002-7595-121X}$^{\hyperlink{inst:Kavli}{1}, \hyperlink{inst:Cav}{2}, \hyperlink{inst:DAWN}{3}, \hyperlink{inst:NBI}{4}}$\thanks{E-mail: \href{mailto:joris.witstok@nbi.ku.dk}{joris.witstok@nbi.ku.dk}}, \orcidsymb{Roberto Maiolino}{0000-0002-4985-3819}$^{\hyperlink{inst:Kavli}{1}, \hyperlink{inst:Cav}{2}, \hyperlink{inst:UCL}{5}}$\thanks{E-mail: \href{mailto:rm665@cam.ac.uk}{rm665@cam.ac.uk}}, \orcidsymb{Renske Smit}{0000-0001-8034-7802}$^{\hyperlink{inst:LJMU}{6}}$, \orcidsymb{Gareth C. Jones}{0000-0002-0267-9024}$^{\hyperlink{inst:Oxford}{7}}$,
    }
    \newauthor{\orcidsymb{Andrew J.\ Bunker}{0000-0002-8651-9879}$^{\hyperlink{inst:Oxford}{7}}$, \orcidsymb{Jakob M. Helton}{0000-0003-4337-6211}$^{\hyperlink{inst:Steward}{8}}$, \orcidsymb{Benjamin D.\ Johnson}{0000-0002-9280-7594}$^{\hyperlink{inst:CfA}{9}}$, \orcidsymb{Sandro Tacchella}{0000-0002-8224-4505}$^{\hyperlink{inst:Kavli}{1}, \hyperlink{inst:Cav}{2}}$,
    }
    \newauthor{\orcidsymb{Aayush Saxena}{0000-0001-5333-9970}$^{\hyperlink{inst:Oxford}{7}, \hyperlink{inst:UCL}{5}}$, \orcidsymb{Santiago Arribas}{0000-0001-7997-1640}$^{\hyperlink{inst:CAB}{10}}$, \orcidsymb{Rachana Bhatawdekar}{0000-0003-0883-2226}$^{\hyperlink{inst:ESAC}{11}}$, \orcidsymb{Kristan Boyett}{0000-0003-4109-304X}$^{\hyperlink{inst:Melbourne}{12}, \hyperlink{inst:ARC3D}{13}}$,
    }
    \newauthor{\orcidsymb{Alex J. Cameron}{0000-0002-0450-7306}$^{\hyperlink{inst:Oxford}{7}}$, \orcidsymb{Phillip A. Cargile}{0000-0002-1617-8917}$^{\hyperlink{inst:CfA}{9}}$, \orcidsymb{Stefano Carniani}{0000-0002-6719-380X}$^{\hyperlink{inst:SNS}{14}}$, \orcidsymb{Stéphane Charlot}{0000-0003-3458-2275}$^{\hyperlink{inst:IAP}{15}}$,
    }
    \newauthor{\orcidsymb{Jacopo Chevallard}{0000-0002-7636-0534}$^{\hyperlink{inst:Oxford}{7}}$, \orcidsymb{Mirko Curti}{0000-0002-2678-2560}$^{\hyperlink{inst:ESO}{16}}$, \orcidsymb{Emma Curtis-Lake}{0000-0002-9551-0534}$^{\hyperlink{inst:Herts}{17}}$, \orcidsymb{Francesco D'Eugenio}{0000-0003-2388-8172}$^{\hyperlink{inst:Kavli}{1}, \hyperlink{inst:Cav}{2}}$,
    }
    \newauthor{\orcidsymb{Daniel J.\ Eisenstein}{0000-0002-2929-3121}$^{\hyperlink{inst:CfA}{9}}$, \orcidsymb{Kevin Hainline}{0000-0003-4565-8239}$^{\hyperlink{inst:Steward}{8}}$, \orcidsymb{Ryan Hausen}{0000-0002-8543-761X}$^{\hyperlink{inst:JHU}{18}}$, \orcidsymb{Nimisha Kumari}{0000-0002-5320-2568}$^{\hyperlink{inst:AURA}{19}}$, \orcidsymb{Isaac Laseter}{0000-0003-4323-0597}$^{\hyperlink{inst:Wisconsin}{20}}$,
    }
    \newauthor{\orcidsymb{Michael V. Maseda}{0000-0003-0695-4414}$^{\hyperlink{inst:Wisconsin}{20}}$, \orcidsymb{Marcia Rieke}{0000-0002-7893-6170}$^{\hyperlink{inst:Steward}{8}}$, \orcidsymb{Brant Robertson}{0000-0002-4271-0364}$^{\hyperlink{inst:UCSC}{21}}$, \orcidsymb{Jan Scholtz}{0000-0001-6010-6809}$^{\hyperlink{inst:Kavli}{1}, \hyperlink{inst:Cav}{2}}$, \orcidsymb{Irene Shivaei}{0000-0003-4702-7561}$^{\hyperlink{inst:CAB}{10}}$,
    }
    \newauthor{\orcidsymb{Christina C. Williams}{0000-0003-2919-7495}$^{\hyperlink{inst:NSF}{22}}$, \orcidsymb{Christopher N. A. Willmer}{0000-0001-9262-9997}$^{\hyperlink{inst:Steward}{8}}$, and \orcidsymb{Chris Willott}{0000-0002-4201-7367}$^{\hyperlink{inst:NRC}{23}}$
    }
    \\
    \\
    {\normalsize Affiliations are listed at the end of the manuscript.}
}

\pubyear{2024}

\begin{document}
\label{firstpage}
\pagerange{\pageref{firstpage}--\pageref{lastpage}}
\maketitle

\begin{abstract}
    Given the sensitivity of the resonant \Lymana (\Lya) transition to absorption by neutral hydrogen, observations of \Lya emitting galaxies (LAEs) have been widely used to probe the ionising capabilities of reionisation-era galaxies and their impact on the intergalactic medium (IGM). However, prior to \jwst our understanding of the contribution of fainter sources and of ionised `bubbles' at earlier stages of reionisation remained uncertain. Here, we present the characterisation of three exceptionally distant LAEs at $z > 8$, newly discovered by \jwst/NIRSpec in the JADES survey. These three similarly bright ($M_\text{UV} \approx -20 \, \mathrm{mag}$) LAEs exhibit small \Lya velocity offsets from the systemic redshift, $\Delta v_\text{\Lya} \lesssim 200 \, \mathrm{km \, s^{-1}}$, yet span a range of \Lya equivalent widths ($15 \, \Angstrom$, $31 \, \Angstrom$, and $132 \, \Angstrom$). The former two show moderate \Lya escape fractions ($f_\text{esc, \Lya} \approx 10\%$), whereas \Lya escapes remarkably efficiently from the third ($f_\text{esc, \Lya} \approx 72\%$), which moreover is very compact (half-light radius of $90 \pm 10 \, \mathrm{pc}$). We find these LAEs are low-mass galaxies dominated by very recent, vigorous bursts of star formation accompanied by strong nebular emission from metal-poor gas. We infer the two LAEs with modest $f_\text{esc, \Lya}$, one of which reveals evidence for ionisation by an active galactic nucleus, may have reasonably produced small ionised bubbles preventing complete IGM absorption of \Lya. The third, however, requires a $\ssim 3 \, \text{physical Mpc}$ bubble, indicating faint galaxies have contributed significantly. The most distant LAEs thus continue to be powerful observational probes into the earlier stages of reionisation.
\end{abstract}

\begin{keywords}
    {galaxies: high-redshift} -- {dark ages, reionization, first stars} -- {methods: observational} -- {techniques: spectroscopic}
\end{keywords}



\section{Introduction}

Cosmic reionisation is the crucial phase transition initiated by the formation of the first astrophysical objects at Cosmic Dawn, a few hundred million years after the Big Bang \citep[$z \gtrsim 15$;][]{2018PhR...780....1D}. In recent years, a consensus has been established that the ionising radiation of star-forming galaxies at $6 \lesssim z \lesssim 10$ was likely responsible for reionising the majority of hydrogen in the Universe \citep{2015ApJ...802L..19R, 2019ApJ...879...36F, 2020ApJ...892..109N}. The foundational observations favouring this timeline, on the one hand, are the optical-depth measurements inferred from the cosmic microwave background \citep[e.g.][]{2020A&A...641A...6P}. These measurements suggest the fraction of neutral hydrogen (\HI) within the intergalactic medium (IGM) approached $\bar{x}_\text{\HI} = 1/2$ -- in other words, reionisation reached the halfway mark -- when the age of the Universe was approximately $0.6 \, \mathrm{Gyr}$ ($z \sim 8$). On the other hand, a highly effective observational probe focussed on galaxies in the later stages of reionisation ($z \lesssim 7$) comes in the form of the principal $2p \rightarrow 1s$ electronic transition of hydrogen, \HI~\Lymana\ (\Lya) with resonance wavelength $\lambda_\text{\Lya} = 1215.67 \, \Angstrom$ \citep{1967ApJ...147..868P}.

The sensitivity of \Lya as a tracer of the ionisation state of gas in and around galaxies can be attributed to the vast number of hydrogen atoms in the Universe, combined with the Lorentzian wings that characterise the non-Gaussian absorption cross-section of neutral hydrogen \citep{2014PASA...31...40D}. This leads to severe resonant scattering of \Lya photons travelling through the interstellar medium (ISM) and IGM, unless these media are very highly ionised \citep{1965ApJ...142.1633G}. Barring strong evolution in the mechanisms regulating internal \Lya production within and \Lya escape from galaxies, reionisation should, therefore, cause heavy attenuation of \Lya emission in the IGM at sufficiently high redshifts \citep{2004MNRAS.354..695F, 2014PASA...31...40D, 2020ARA&A..58..617O}. A swift decline in the fraction of \Lya emitting galaxies (LAEs) has indeed been confirmed beyond $z \gtrsim 6$ \citep{2010MNRAS.408.1628S, 2017MNRAS.464..469S, 2011ApJ...743..132P, 2014ApJ...795...20S, 2014ApJ...794....5T}, strongly supporting a significant evolution in the IGM neutral hydrogen fraction within the first billion years of cosmic time \citep{2018ApJ...856....2M, 2019MNRAS.485.3947M, 2020ApJ...891L..10T, 2020MNRAS.495.3602W}.

Similarly, \Lya absorption due to residual neutral hydrogen seen in quasar spectra points towards an increasingly neutral state of the IGM at high redshifts \citep*[see][ for a review]{2023ARA&A..61..373F}. However, a small number of sightlines reveal the persistence of long `dark gaps', representing large, neutral islands of gas, towards late times ($z \sim 5$), which is interpreted as evidence for inhomogeneous reionisation \citep{2015MNRAS.447.3402B, 2019MNRAS.485L..24K, 2020MNRAS.491.1736K, 2022MNRAS.514...55B}. Conversely, the most densely populated environments in the Universe are therefore predicted to readily host large-scale ionised regions early on \citep[e.g.][]{2018ApJ...857L..11M, 2022MNRAS.510.3858Q, 2022MNRAS.515.5790L}. Crucially, these first ionised `bubbles' become transparent to \Lya emission \citep{2018MNRAS.479.2564W, 2021MNRAS.502.6044E, 2022MNRAS.517.5642E, 2022arXiv221209850J, 2022ApJ...933...87J}, which in turn renders the most distant LAEs powerful observational probes of the earliest sites of reionisation \citep{2021NatAs...5..485H, 2022MNRAS.511.6042E, 2023MNRAS.524.5891T, 2024MNRAS.528.4872L}.

Recently, the successful deployment of the long-awaited \jwst \citep{2023PASP..135e8001M, 2023PASP..135d8001R}, whose near-infrared imaging and spectroscopy capabilities are specifically aimed at finding and characterising the earliest galaxies \citep{2022ARA&A..60..121R, 2023PASP..135f8001G}, has reinvigorated the search for such systems. In particular, the \jwst/Near-Infrared Spectrograph \citep[NIRSpec;][]{2022A&A...661A..80J, 2023PASP..135c8001B} has proven efficient at uncovering new populations of LAEs in the early Universe \citep{2024A&A...683A.238J, 2024A&A...688A.106N, 2024ApJ...967...73J, 2024MNRAS.529..855W}, which in combination with the \jwst/Near-Infrared Camera \citep[NIRCam;][]{2023PASP..135b8001R} provides an invaluable view into the physical conditions in and around LAEs in the epoch of reionisation \citep[EoR; e.g.][]{2023A&A...678A..68S, 2024NatAs...8..384W, 2024A&A...682A..40W}.

While observational efforts prior to \jwst had already been reasonably successful in identifying several of the brightest $z > 7$ sources as LAEs \citep[e.g.][]{2012ApJ...744...83O, 2013Natur.502..524F, 2015ApJ...804L..30O, 2016ApJ...823..143R}, \Lya emission observed beyond $z \sim 8$ remained elusive \citep[e.g.][]{2019A&A...627A..84L}. The discovery of \Lya emission in two $z \approx 8.6$ galaxies in close vicinity, EGSY8p7 and EGS-z910-44164, were the only two significant detections reported by \citet{2015ApJ...810L..12Z} and \citet{2022ApJ...930..104L} respectively. Exploiting \jwst/NIRSpec measurements from the Cosmic Evolution Early Release Science (CEERS) programme \citep[ID 1345, PI: Finkelstein;][]{2023ApJ...946L..13F}, \citet{2023MNRAS.526.1657T} found systemic redshifts based on rest-frame optical lines of $z = 8.678$ for EGSY8p7 (CEERS-1019) and $z = 8.610$ for EGS-z910-44164 (CEERS-1029) and confirmed the presence of \Lya emission in EGSY8p7, though only yielding a tentative \Lya detection for EGS-z910-44164 (not recovered by a later analysis; \citealt{2024ApJ...967...28N}).

Remarkably, however, the enigmatic galaxy GN-z11 \citep{2016ApJ...819..129O} soon after proved that \Lya emission can readily be observed out to $z = 10.6$ \citep{2023A&A...677A..88B, 2024A&A...687A.283S}. GN-z11 was recently eclipsed by the discovery of a galaxy with an exceptionally strong \Lya line at $z = 13.0$, JADES-GS-z13-1-LA \citep{2024arXiv240816608W}, at a time when the Universe was a mere $300 \, \mathrm{Myr}$ old and, most likely, the IGM was still highly neutral \citep{2024arXiv240801507T}. Although the latter does not strictly preclude us from observing \Lya emission from a theoretical point of view \citep{2023ApJ...954L..14H}, it does starkly contrast with recent indications of extremely strong \Lya absorption among the population of very early galaxies ($z \gtrsim 9$) uncovered by \jwst \citep{2024arXiv240402211H, 2024Sci...384..890H, 2024A&A...689A.152D, 2024arXiv240404325H}. Empirically, the surprisingly early presence of LAEs therefore raises many questions on the mechanisms regulating its production and escape in the early Universe. Is the possible influence of an active galactic nucleus \citep[AGN;][]{2024Natur.627...59M} a necessary condition to observe \Lya, by producing an excess of ionising photons and high-velocity outflowing gas that facilitates the escape of \Lya? Or are these sources unique by virtue of being located in overdense environments, as is the case for GN-z11 \citep{2023ApJ...952...74T, 2024A&A...687A.283S}, which may accommodate a small ionised bubble enhancing \Lya transmission?

\begin{figure}
	\centering
	\includegraphics[width=\linewidth]{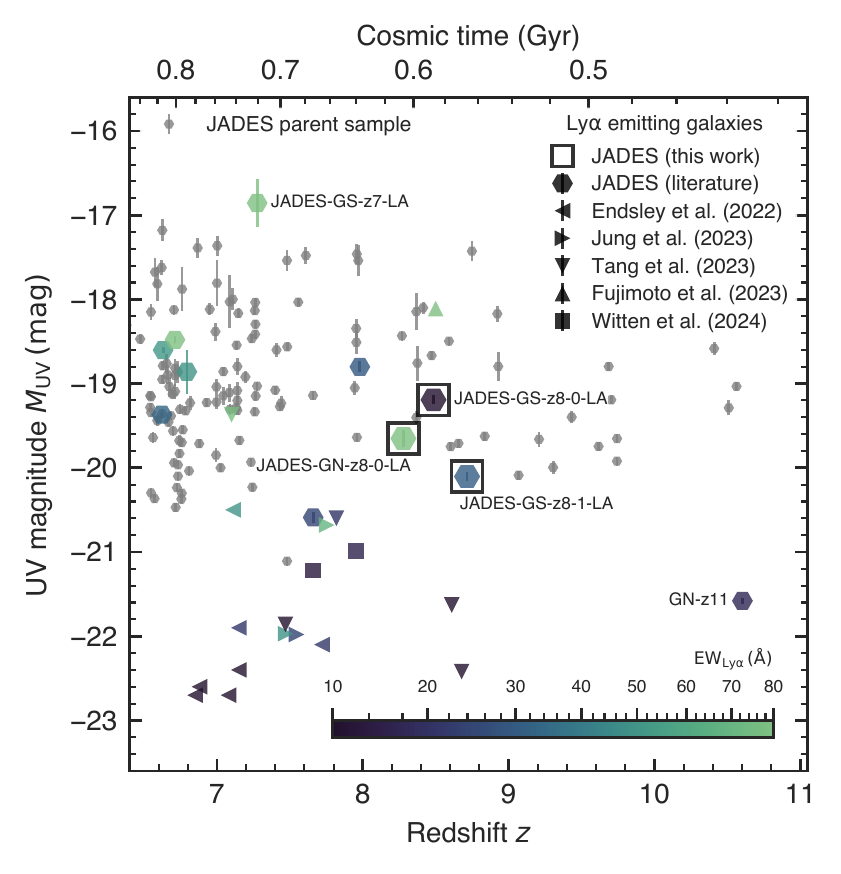}
	\caption{UV magnitudes, $M_\text{UV}$, of spectroscopically confirmed galaxies in the JADES parent sample (small grey points) as a function of cosmic time \citep{2024arXiv240406531D}. Distant LAEs ($z \gtrsim 6.5$) from the literature include JADES-GS-z7-LA \citep{2023A&A...678A..68S}, GN-z11 \citep{2023A&A...677A..88B}, as well as those samples compiled in \citet{2022MNRAS.517.5642E}, \citet{2024ApJ...967...73J}, \citet{2023MNRAS.526.1657T}, \citet{2023arXiv230811609F}, \citet{2024NatAs...8..384W}, and previous JADES works including \citet{2023A&A...677A..88B}, \citet{2024A&A...683A.238J}, \citet{2024A&A...684A..84S}, and \citet{2024A&A...682A..40W}. Points are coloured according to the \Lya equivalent width (colourbar on the right). The three sources studied in this work are highlighted with black squares.
	}
	\label{fig:LAE_M_UV_redshift}
\end{figure}

To address such questions, in this work we present a detailed investigation of three of the most distant ($z > 8$) reionisation-era LAEs newly discovered by \jwst, including a $z \approx 8.5$ LAE reported recently by \citet{2024MNRAS.531.2701T}. The outline of this work is as follows: in \cref{sec:Observations}, we discuss the observations underlying this work, and \cref{sec:Methods_and_analysis} discusses our methods and analysis. The main findings are placed in the context of results in the literature in \Cref{sec:Discussion}. Finally, we summarise our findings in \cref{sec:Summary}. We adopt a flat $\Lambda$CDM cosmology based on the latest results of the Planck collaboration (i.e. $H_0 = 67.4 \, \mathrm{km \, s^{-1} \, Mpc^{-1}}$, $\Omega_\text{m} = 0.315$, $\Omega_\text{b} = 0.0492$; \citealt{2020A&A...641A...6P}) and a cosmic hydrogen fraction of $f_\text{H} = 0.76$ throughout. On-sky separations of $1\arcsec$ and $1\arcmin$ at $z = 8.5$ hence correspond to $4.73 \, \text{physical kpc}$ (pkpc) and $0.284 \, \text{physical Mpc}$ (pMpc), respectively. Magnitudes are quoted in the AB system \citep{1983ApJ...266..713O}.
\begin{figure*}
	\centering
	\includegraphics[width=\linewidth]{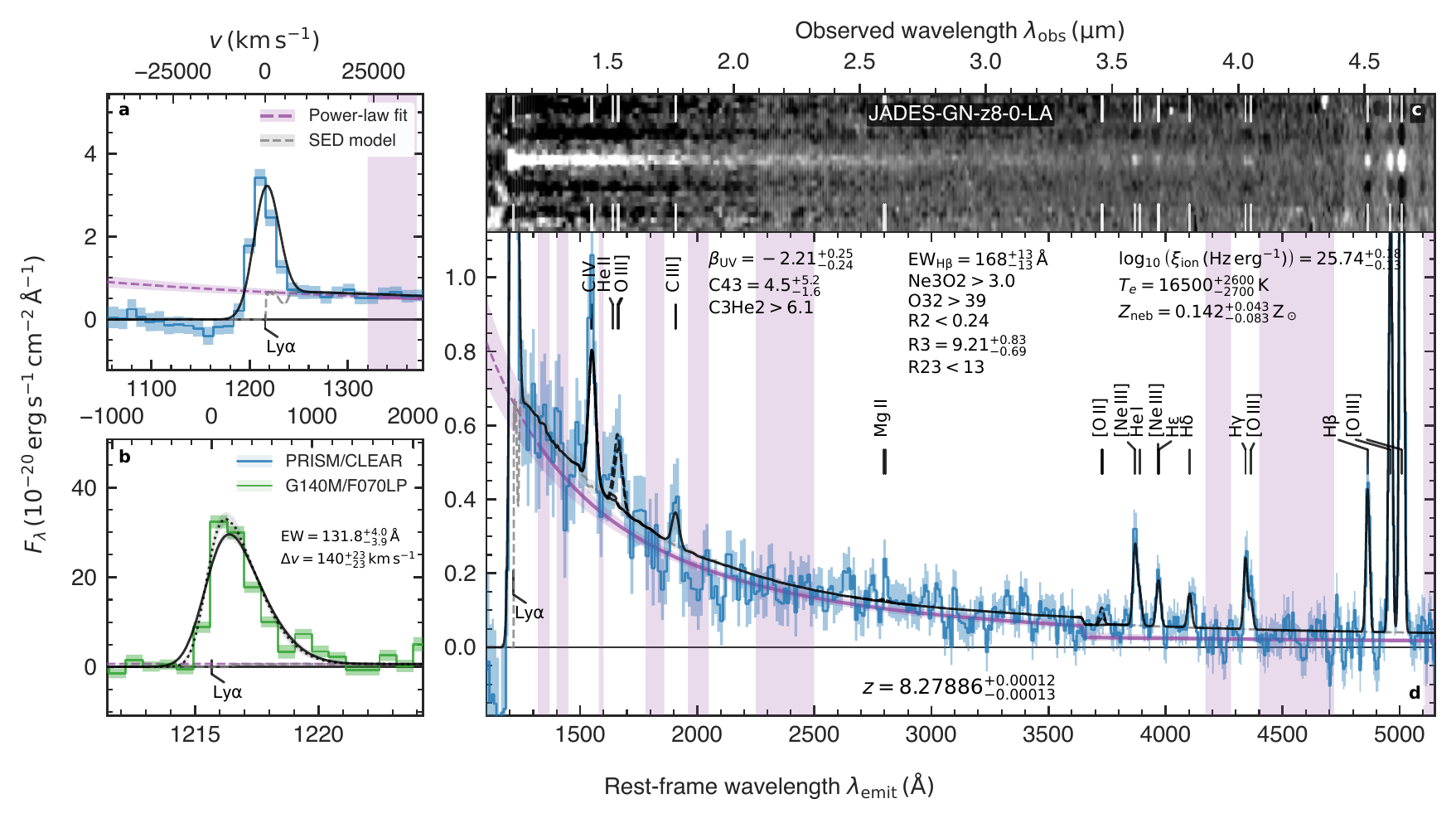}
	\caption{NIRSpec measurements of \JGNzeightzeroLA. The \Lya emission line is seen both in the low-resolution PRISM (panel~a) and medium-resolution G140M/F070LP (panel~b) spectra. The two-dimensional PRISM spectrum is shown in panel~c, while panel~d shows the one-dimensional spectrum extracted from a 3-pixel aperture, both highlighting the locations of notable emission lines other than \Lya. A broken power-law, fitted to line-free spectral regions in the rest-frame UV and optical (shaded light purple regions; see \cref{tab:Continuum_windows}), is shown as a purple line in all panels (solid within the fitting range, dashed in the extrapolation). A grey dashed line shows the best-fit SED model continuum (\cref{ssec:SED_modelling}). An empirical emission-line fit, consisting of the latter with superimposed (asymmetric) Gaussian emission lines at the appropriate spectral resolution (\cref{ssec:Continuum_and_line_emission}), is shown by the solid black line. A dotted black line shows the unconvolved \Lya profile in panel~b; dashed black lines in panel~d indicate UV lines for which an upper limit is obtained (\cref{ssec:Continuum_and_line_emission}).
	}
	\label{fig:Line_overview_1899}
\end{figure*}

\section{Observations}
\label{sec:Observations}

\subsection{Data sets}
\label{ssec:Data_sets}

The observations considered here were primarily obtained as part of the \jwst Advanced Deep Extragalactic Survey \citep[JADES;][]{2023arXiv230602465E}, supplemented with other relevant \jwst programmes. JADES is designed and carried out by a collaboration between the NIRCam and NIRSpec instrument science teams and encompasses \jwst guaranteed time observations (GTO) programme IDs (PIDs) 1180 and 1181 (PI: Eisenstein) as well as 1210, 1286 and 1287 (PI: Luetzgendorf), all located within the Great Observatories Origins Deep Survey (GOODS) extragalactic legacy fields \citep{2004ApJ...600L..93G} in the equatorial North (GOODS-N) and South (GOODS-S), the latter of which contains the Hubble Ultra Deep Field \citep[HUDF;][]{2006AJ....132.1729B}.

In addition to extensive multi-wavelength ancillary data in both fields, GOODS-S in particular benefits from publicly available NIRCam observations from JEMS, the \jwst Extragalactic Medium-band Survey \citep[PID 1963, PIs: Williams, Maseda \& Tacchella;][]{2023ApJS..268...64W}. Moreover, the First Reionization Epoch Spectroscopic COmplete Survey (FRESCO; PID 1895, PI: Oesch) provides NIRCam imaging and wide-field slitless spectroscopy (WFSS) over both GOODS-N and GOODS-S \citep{2023MNRAS.525.2864O}. As in \citet{2024A&A...682A..40W}, we consider the WFSS observations taken as part of FRESCO to obtain a large sample of high-redshift galaxies within the JADES footprint that are spectroscopically confirmed via their $\OIII \, \lambda \, 5008 \, \Angstrom$ emission at $6.8 < z < 8.9$. The WFSS data reduction routine is described in \citet{2023ApJ...953...53S}, while \citet{2024ApJ...962..124H} outlines the emission-line identification algorithms \citep[see also][]{2024A&A...682A..40W}. Finally, we complement our data set with a dedicated follow-up programme targeting the JADES Origins Field (contained within GOODS-S) in observations completed in Cycle 2 \citep[PID 3215, PIs: Eisenstein \& Maiolino;][]{2023arXiv231012340E}.

We refer to \citet{2023ApJS..269...16R}, \citet{2024A&A...690A.288B}, and \citet{2024arXiv240406531D} for details on the JADES survey strategy, data reduction, and first public release of high-level science products derived from the NIRCam and NIRSpec measurements, respectively. Additional data sets discussed above are folded into the current analysis after having been processed in a consistent manner. In this work, we mainly focus on the NIRSpec spectroscopy, for which we briefly outline the relevant details in the following.
\begin{figure*}
	\centering
	\includegraphics[width=\linewidth]{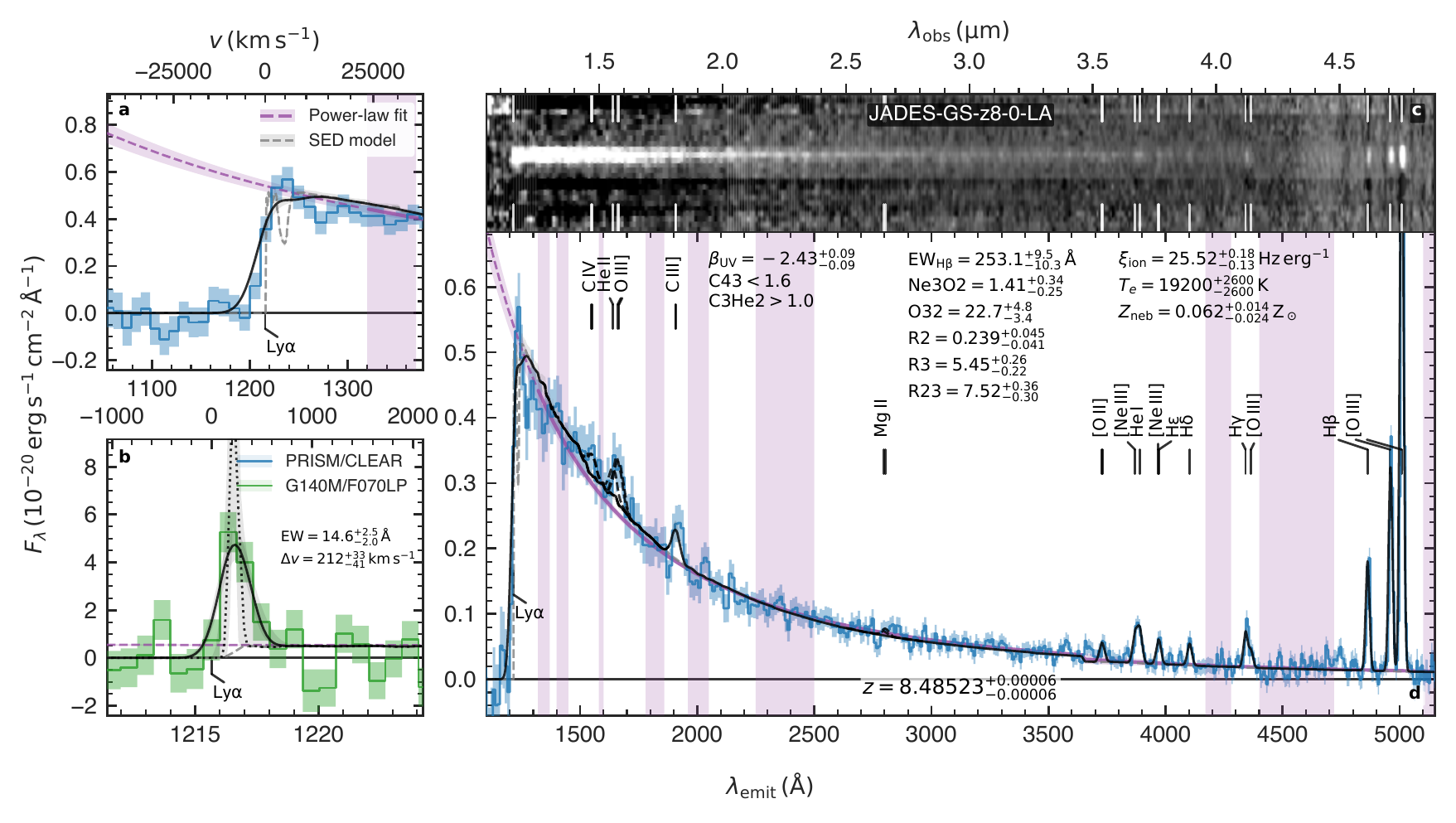}
	\caption{As \cref{fig:Line_overview_1899}, but for \JGSzeightzeroLA. The narrow \Lya line seen in the G140M/F070LP spectrum gets largely `washed out' by the spectral break in the low-resolution PRISM spectrum \citep[see e.g.][]{2024A&A...683A.238J}.
	}
	\label{fig:Line_overview_20213084}
\end{figure*}

\subsection{NIRSpec data reduction}
\label{ssec:NIRSpec_data_reduction}

The primary NIRSpec targets in JADES were drawn from samples of high-redshift galaxy candidates selected in a combination of publicly available \textit{Hubble Space Telescope} (\hst) and NIRCam imaging \citep{2024A&A...690A.288B}, for which multi-object spectroscopy was obtained in the micro-shutter array (MSA) mode \citep{2022A&A...661A..81F}. The spectral configurations considered here are the low-resolution PRISM/CLEAR (`PRISM' hereafter; resolving power $30 \lesssim R \lesssim 300$ over a wavelength range between $0.6 \, \mathrm{\upmu m}$ and $5.3 \, \mathrm{\upmu m}$) and the medium-resolution grating-filter combinations G140M/F070LP, G235M/F170LP, and G395M/F290LP (collectively `R1000 gratings', with resolving power $R \approx 1000$).

Within the JADES NIRSpec observing strategy, there are the DEEP tiers, in which a limited number of sources were observed with long exposure times (between $9.3$ and $28$ hours in PRISM mode, depending on slit placement across three visits; R1000 gratings each collect a third of the PRISM integration time), whereas the MEDIUM tiers targeted larger numbers of sources with exposure times of a few hours in both the PRISM and R1000 configurations. Additionally, tier identifiers distinguish between the two fields (GN for GOODS-N and GS for GOODS-S) as well as between two types of target selection, specifically those tiers based solely on \hst imaging (e.g. MEDIUM-HST-GN) and those where \jwst imaging was available (e.g. DEEP-JWST-GS). The NIRSpec observations from PID 3215 consist of exposures totalling $53.2 \, \mathrm{h}$ in the PRISM mode, $11.2 \, \mathrm{h}$ in G140M/F070LP, and $33.6 \, \mathrm{h}$ in G395M/F290LP, and are classified as the ULTRA-DEEP-GS-3215 tier.

The current analysis is based on version 3 of the data reduction pipelines that were developed between the ESA NIRSpec Science Operations Team and the NIRSpec GTO team, which has been discussed in several previous JADES works \citep[e.g.][]{2024arXiv240406531D} and will be described in more detail in a forthcoming paper \citetext{Carniani \& NIRSpec GTO collaboration in prep.}. We applied path-loss corrections assuming a point-source light profiles, taking into account the intra-shutter location of the targets. Throughout this work, one-dimensional spectra are extracted from the central $3$ pixels along the central micro-shutter to maximise signal-to-noise ratio (SNR) on relative spectroscopic measurements (based on flux ratios) for these compact sources, while integrated galaxy properties such as stellar mass are derived in combination with the photometry (\cref{ssec:SED_modelling}) to account for any additional path losses in the $3$-pixel extracted spectroscopy.
\begin{figure*}
    \centering
    \includegraphics[width=\linewidth]{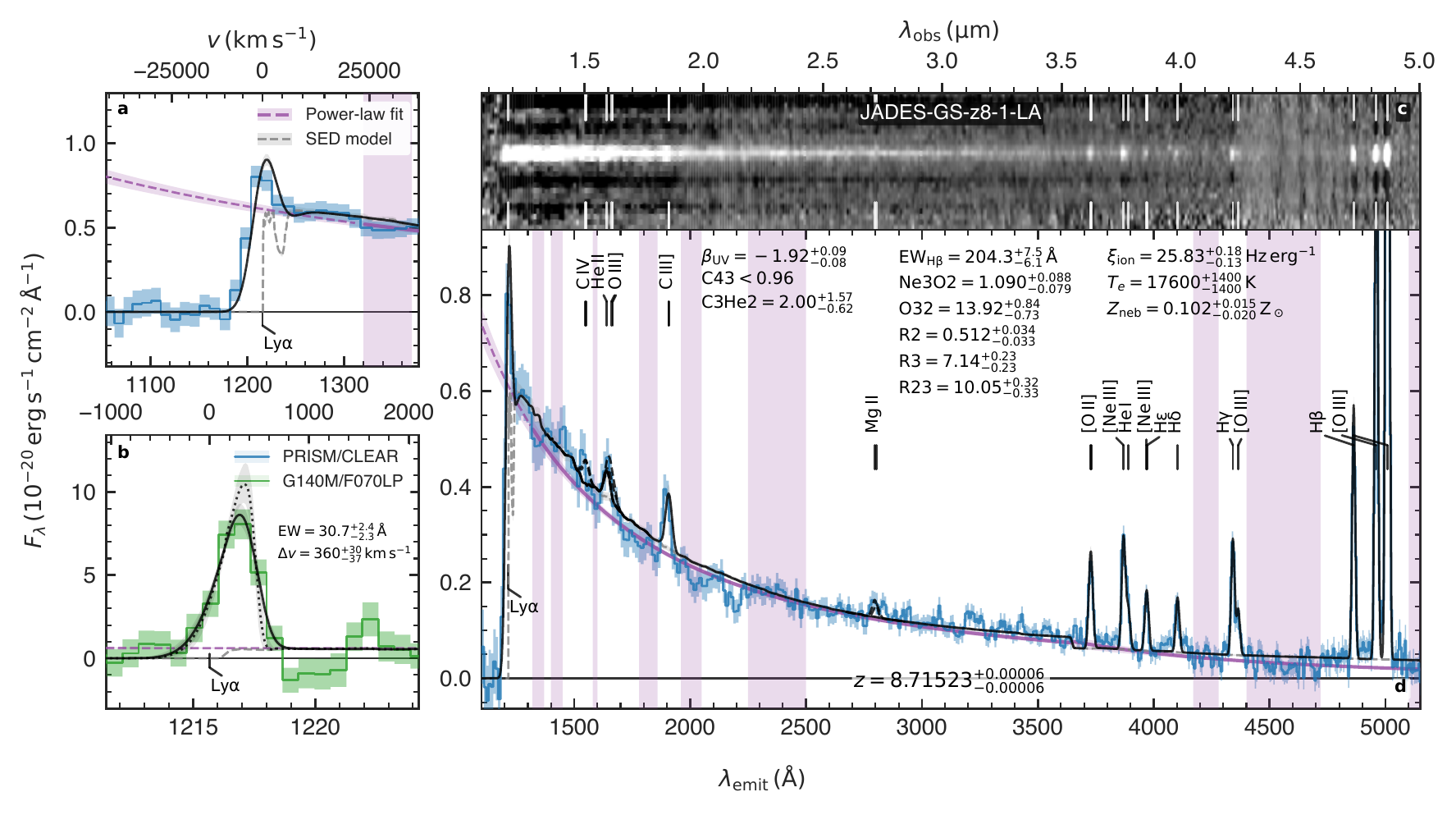}
    \caption{As \cref{fig:Line_overview_1899}, but for \JGSzeightoneLA.
    }
    \label{fig:Line_overview_20006347}
\end{figure*}

While these losses are well corrected for in terms of the UV continuum, as seen from the excellent agreement in UV magnitudes measured by NIRCam and NIRSpec (with the exception of \JGSzeightoneLA where the slit only captures approximately half of the galaxy; \cref{ssec:Continuum_and_line_emission} and \cref{fig:LAE_SED_models}), we note \Lya emission is often found to be extended \citep{2017A&A...608A...8L, 2018Natur.562..229W, 2022A&A...660A..44K, 2024A&A...688A..37G} and occasionally exhibits spatial offsets from the UV continuum \citep[e.g.][]{2022A&A...666A..78C, 2023MNRAS.523.5468S}. In these cases, the applied path loss corrections are not optimised to retrieve all \Lya flux, such that the EWs may, in reality, be slightly higher. However, we have verified that an alternative background subtraction scheme where only the two off-centre nodding positions are considered yields very similar results. Although there is some hint of \Lya emission being more extended than the UV continuum from the two-dimensional spectra (e.g. \cref{fig:Line_overview_1899}), this similarity suggests the extension is not significantly larger than the extent of the micro-shutters \citep[$0.46 \arcsec$ along the nodding direction;][]{2022A&A...661A..80J} given that it would be expected to suffer from self-subtraction in the default three-nod scheme, which is further corroborated by the compact morphology of the LAEs in the F115W filter containing \Lya (discussed further in \cref{ssec:SED_modelling}).

\section{Methods and analysis}
\label{sec:Methods_and_analysis}

\subsection{LAE sample}
\label{ssec:LAE_sample}

The various tiers of NIRSpec observations within JADES (including the ultra-deep spectra from PID 3215; \cref{ssec:Data_sets}) comprised a total of $4100$ targets, all of which were visually inspected to yield $174$ galaxies robustly confirmed to be at $z > 6$, shown as the parent sample in \cref{fig:LAE_M_UV_redshift}, out of which $36$ have redshifts $z > 8$ \citep{2024arXiv240406531D}. Within the full JADES sample, approximately $30$ LAEs at $z > 6$ were identified from PRISM data, with three $z > 8$ galaxies showing well-detected ($\text{SNR} > 3$) \Lya emission in the R1000 spectra. More detailed statistics on the fraction of LAEs within the wider galaxy population across cosmic time are presented in \citet{2024arXiv240906405J}. An extraordinary LAE at $z = 13.0$, whose \Lya emission however is too faint to be seen in the R1000 spectrum, is discussed separately in \citet{2024arXiv240816608W}.

Firstly, we consider JADES-GS+53.15891-27.76508 (\JGSzeightzeroLA hereafter), the $z \simeq 8.49$ LAE discussed in \citet{2024MNRAS.531.2701T}. Originally, this source was identified as a $Y_{105}$-band dropout in \hst imaging over the HUDF, placing its photometric redshift at $z_\text{phot} \approx 8.6$ \citep{2010ApJ...709L.133B, 2010MNRAS.409..855B, 2010MNRAS.403..960M}. Based on deep spectroscopy with the Spectrograph for INtegral Field Observations in the Near Infrared \citep[SINFONI;][]{2003SPIE.4841.1548E} on the Very Large Telescope (VLT), \citet{2010Natur.467..940L} reported a $6\sigma$ detection of strong \Lya emission in this source at $1.16156 \pm 0.00024 \, \mathrm{\upmu m}$, implying $z_\text{\Lya} = 8.5549 \pm 0.0002$, with a line flux of $6.1 \times 10^{-18} \, \mathrm{erg \, s^{-1} \, cm^{-2}}$ and an observed equivalent width (EW) of $1900 \, \Angstrom$ ($200 \, \Angstrom$ in the rest frame).

In PID 3215, \JGSzeightzeroLA was one of six galaxies selected to be a sufficiently bright target for NIRSpec ($m \lesssim 30 \, \mathrm{mag}$) with robust photometric redshift $8 < z_\text{phot} < 10$. Our new NIRSpec observations confirm the source to be at a different redshift of $z \approx 8.48523 \pm 0.00006$, with \Lya emission lying more than $\ssim 2000 \, \mathrm{km \, s^{-1}}$ bluewards of the claimed \citet{2010Natur.467..940L} line (many times the spectral resolution of the NIRSpec grating and the SINFONI spectrograph). As will be discussed in \cref{ssec:Continuum_and_line_emission}, we measure a line flux of \Lya from NIRSpec to be $0.67 \times 10^{-18} \, \mathrm{erg \, s^{-1} \, cm^{-2}}$, which is considerably lower ($\ssim 10 \times$) than for the line reported in \citet{2010Natur.467..940L}, and a rest-frame EW of $14.7 \Angstrom$ (compared to $200 \Angstrom$ in \citealt{2010Natur.467..940L}). Hence we conclude that the previously-reported $z\approx 8.6$ line emission \citep{2010Natur.467..940L} is unlikely to be real (as we would have seen this at a $\text{SNR} > 50$ in our NIRSpec spectrum), and indeed \citet{2013MNRAS.430.3314B} argued this early detection was most likely to be spurious, as independent spectroscopic observations with VLT/FORS2 and Subaru/MOIRCS were not able to reproduce the feature. We note that the $2\sigma$ upper limit on the non-detection of the line flux of $F_\text{\Lya} \approx 2 \times 10^{-18} \, \mathrm{erg \, s^{-1} \, cm^{-2}}$ placed by \citet{2013MNRAS.430.3314B} is fully consistent with our NIRSpec detection of faint \Lya flux $0.67 \times 10^{-18} \, \mathrm{erg \, s^{-1} \, cm^{-2}}$.

\begin{figure*}
    \centering
    \includegraphics[width=\linewidth]{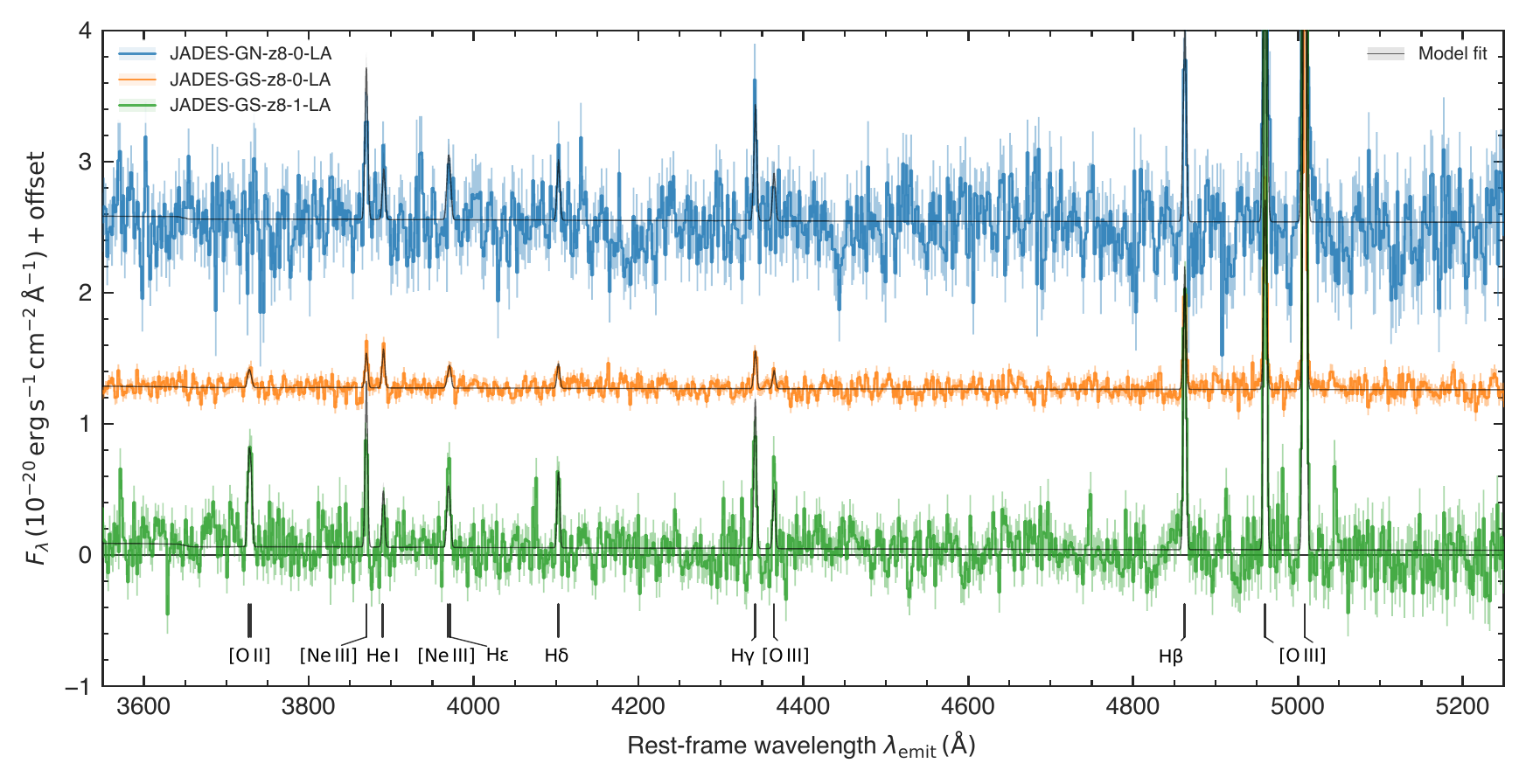}
    \caption{NIRSpec G395M/F290LP measurements of the three LAEs considered in this work, offset by $1.25$ ($2.5$) $\times 10^{-20} \, \mathrm{erg \, s^{-1} \, cm^{-2} \, \Angstrom^{-1}}$ in the case of \JGSzeightzeroLA (\JGSzeightoneLA) for visualisation purposes. The solid black line shows an empirical emission-line fit (see \cref{ssec:Continuum_and_line_emission} for details).
    }
    \label{fig:Line_R1000_overview}
\end{figure*}

Finally, our sample includes JADES-GN+189.19774+62.25696 (\JGNzeightzeroLA), which was placed on the MSA mask in the MEDIUM-HST-GN tier as one of $16$ $z > 5.7$ galaxy candidates with $H_{160} < 27.5 \, \mathrm{mag}$ (photometric redshift $z_\text{phot} = 7.3$, priority class 2; cf. \citealt{2024arXiv240406531D} for details on the prioritisation system in the MEDIUM tiers). The NIRSpec measurements revealed this to be an LAE with remarkably high EW at $z \simeq 8.28$, independently spectroscopically confirmed via $\OIII \, \lambda \, 5008 \, \Angstrom$ in the FRESCO data \citep{2024ApJ...962..124H}.\footnote{The FRESCO spectroscopy shows a marginal \OIII detection in \JGSzeightzeroLA, while \JGSzeightoneLA falls outside the FRESCO footprint.} Finally, JADES-GS+53.10900-27.90084 (\JGSzeightoneLA), selected as a third priority class target among $13$ other galaxies in the DEEP-JWST-GS tier, was found to be a third LAE at $z \simeq 8.72$. We summarise specific observational details for the three main LAEs in \cref{tab:Source_properties}. Their observed PRISM and G140M/F070LP spectra are shown in \cref{fig:Line_overview_1899,fig:Line_overview_20213084,fig:Line_overview_20006347}, while \cref{fig:Line_R1000_overview} shows their G395M/F290LP spectra.

In the following, we discuss stellar population synthesis (SPS) modelling (\cref{ssec:SED_modelling}) and further spectral fitting (\cref{ssec:Continuum_and_line_emission}), the results of which are summarised in \cref{tab:Source_properties}. Specifically, we adopted a three-stage approach to model the available \jwst spectroscopy and photometry. Firstly, we determined the systemic redshift of each source based on several strong, well-detected, rest-frame optical emission lines (\Hdelta, \Hgamma, \Hbeta, $\OIII \, \lambda \, 4960, 5008 \, \Angstrom$) in the G395M/F290LP grating (\cref{fig:Line_R1000_overview}) using the emission-line fitting routines described in \cref{ssec:Continuum_and_line_emission}. Secondly, we employed SPS models to infer the stellar and nebular properties from a combination of NIRSpec and NIRCam measurements, discussed in more detail in \cref{ssec:SED_modelling}. Finally, we performed a fitting routine of all observed emission lines in PRISM and R1000 gratings, taking into account the underlying continuum observed in the PRISM spectra (\cref{ssec:Continuum_and_line_emission}).

\subsection{Stellar population synthesis modelling}
\label{ssec:SED_modelling}

NIRCam photometry was extracted from circular $0.3\arcsec$-diameter apertures, except for \JGSzeightoneLA where we performed full-scene modelling of the crowded field of sources with the \program{ForcePho} code \citetext{Johnson et al. in prep.} to obtain accurate photometry (see \cref{app:Neighbour_SED_modelling} for details). \program{ForcePho} simultaneously fits to the individual exposures in multiple photometric bands, using a single \citet{1963BAAA....6...41S} light profile per source. To explore potential wavelength dependencies in the morphology of all three LAEs, we additionally fit \citeauthor{1963BAAA....6...41S} profiles separately to the F115W, F200W, and F444W filters employing the \program{pysersic} code \citep{2023JOSS....8.5703P}. We do not find evidence for such wavelength dependency in any of the LAEs, however, with the three different filters yielding sizes fully consistent within the uncertainties. The resulting half-light radii found by \program{pysersic} in F200W, probing rest-frame wavelengths around $\lambda_\text{emit} = 2000 \, \Angstrom$ at $z \sim 8.5$, are reported as $R_\text{UV}$ in \cref{tab:Source_properties}.

We then modelled the full spectral energy distribution (SED) of the three LAEs combining the NIRSpec PRISM measurements (masking the spectral region around \Lya, $1150 \, \Angstrom < \lambda_\text{emit} < 1280 \, \Angstrom$) and NIRCam photometry, largely following the methodology described in \citet{2023Natur.621..267W} and \citet{2024Natur.629...53L}. Briefly, we employed the Bayesian Analysis of Galaxies for Physical Inference and Parameter EStimation (\program{bagpipes}) code \citep{2018MNRAS.480.4379C} with Binary Population and Spectral Synthesis (\program{bpass}) v2.2.1 models \citep{2017PASA...34...58E}. We chose \program{bpass} models that include binary stars and use the default \program{bpass} initial mass function (IMF; stellar mass from $1 \, \mathrm{M_\odot}$ to $300 \, \mathrm{M_\odot}$ and slope of $-2.35$ for $M > 0.5 \, \mathrm{M_\odot}$).

We assumed a non-parametric star formation history (SFH) under the prescription of \citet{2019ApJ...876....3L}. Following \citet{2023MNRAS.522.6236T}, we considered $6$ bins in lookback time $t$, the first of which are located between $0 \, \mathrm{Myr} < t < 5 \, \mathrm{Myr}$ and $5 \, \mathrm{Myr} < t < 10 \, \mathrm{Myr}$, and the remaining $4$ bins spaced logarithmically up until $t(z = 20)$. Logarithmic decrements in star formation rate (SFR) between adjacent bins, $x_i = \log_{10} \left( \text{SFR}_i / \text{SFR}_{i+1} \right)$, are modelled using a Student's-$t$ distribution with $\nu = 2$ degrees of freedom as a prior \citep{2019ApJ...876....3L}. We adopted a `bursty-continuity' prior \citep{2022ApJ...927..170T} where the scale is $\sigma = 1.0$, having verified a more smoothly varying SFH ($\sigma = 0.3$; `standard continuity') yields consistent results. This flexible SFH accommodates stochastic star formation and underlying older stellar populations, both of which likely play an important role in the EoR \citep[e.g.][]{2023MNRAS.519.5859W, 2024Natur.629...53L}, in a minimally biased framework weighing against very abrupt changes across SFH bins \citep[e.g.][]{2019ApJ...873...44C, 2022ApJ...927..170T}.

\begin{figure*}
	\centering
	\includegraphics[width=\linewidth]{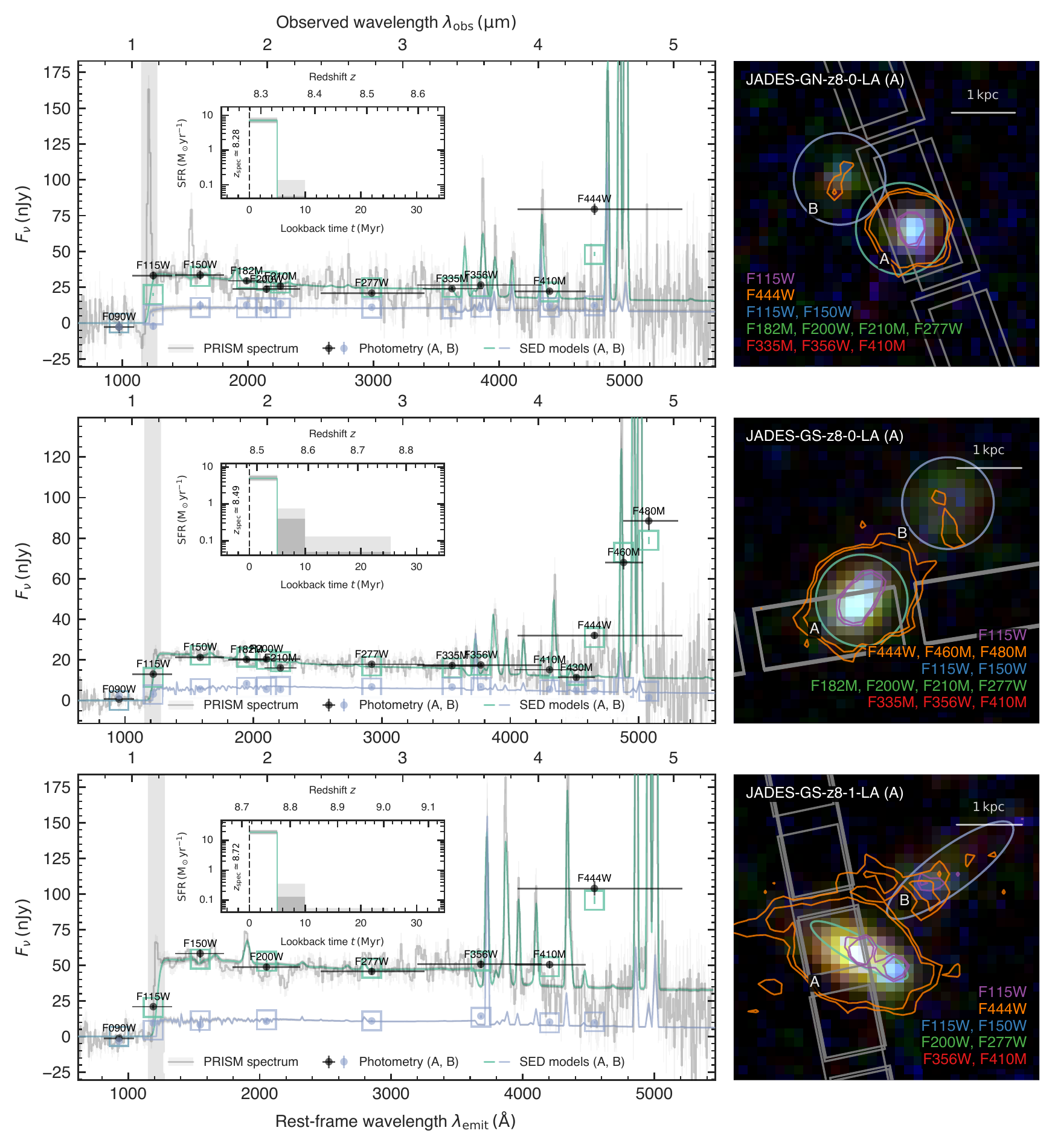}
	\caption{SED models of the three LAEs considered in this work, \JGNzeightzeroLA (top row), \JGSzeightzeroLA (middle row), and \JGSzeightoneLA (bottom row). \textit{Left column}: the low-resolution NIRSpec/PRISM spectra (light-grey line, with correction applied; see \cref{ssec:SED_modelling}) and NIRCam photometry of the main sources (black points). Throughout, best-fit \program{bagpipes} models are shown by cyan lines (dark and lighter shading representing $1\sigma$ and $2\sigma$ uncertainty), with open cyan squares indicating the predicted photometric data points. Photometry and accompanying SED models of neighbouring sources (B) are shown in blue. Inset panels show the modelled SFHs. \textit{Right column}: false-colour images (North up, East left) created from stacks of available NIRCam filters as annotated (all convolved by $0.04\arcsec$). The F115W image, into which \Lya is redshifted for all three LAEs, is overlaid in purple contours representing $3\sigma$ and $4\sigma$. Similarly, orange contours show filters boosted by strong rest-frame optical lines (F444W, F460M, F480M). The placement of the NIRSpec micro-shutters is indicated by thin grey rectangles. Ellipses indicate the apertures from which the photometric data points are extracted (circular apertures $0.3\arcsec$ in diameter, except for \JGSzeightoneLA). A scale bar in the top right shows the projected size of $1 \, \mathrm{kpc}$.
	}
	\label{fig:LAE_SED_models}
\end{figure*}

The total stellar mass formed and stellar metallicity were varied respectively across $0 < M_* < 10^{15} \, \mathrm{M_\odot}$, and $0.001 \, \mathrm{Z_\odot} < Z_* < 1.5 \, \mathrm{Z_\odot}$ (both log-uniform priors). We included nebular emission with freely varying ionisation parameter ($-3 < \log_{10} U < -0.5$), which \program{bagpipes} derives from a grid of \program{Cloudy} \citep{2017RMxAA..53..385F} models computed self-consistently, with the relevant SPS models as incident radiation field and the nebular metallicity tied to that of the stars. A flexible \citet{2000ApJ...539..718C} prescription was included to model dust attenuation, with a Gaussian prior on the V-band attenuation ($\mu_{A_V} = 0.15$, $\sigma_{A_V} = 0.15$, limited to $0 \, \mathrm{mag} \leq A_V \leq 7 \, \mathrm{mag}$) and the power-law slope over $0.4 < n < 1.5$, fixing the attenuation fraction arising from the diffuse ISM and stellar birth clouds to 40\% and 60\%, respectively \citep{2019MNRAS.483.2621C}. Finally, we assumed the spectroscopic data to follow the PRISM resolution curve of a uniformly illuminated micro-shutter\footnote{Provided by the Space Telescope Science Institute (STScI) as part of the \jwst documentation: \url{https://jwst-docs.stsci.edu/jwst-near-infrared-spectrograph/nirspec-instrumentation/nirspec-dispersers-and-filters}.}, and we allowed them to be corrected by a first-order polynomial to take potential remaining aperture and flux calibration effects into account \citep{2019MNRAS.490..417C}. The resulting best-fit models are shown in \cref{fig:LAE_SED_models}.

From the NIRCam imaging, we identify several faint nearby sources (within $1.5\arcsec$ or $\ssim 7 \, \mathrm{kpc}$) from which we extract NIRCam photometry in the same way as for the three main LAEs (see \cref{fig:LAE_SED_models}). While some are likely foreground galaxies (see \cref{app:Neighbour_SED_modelling} in particular), each LAE has one neighbouring source with photometry that appears consistent with being co-located in redshift (i.e. dropping out of the F090W filter). We fit the photometry of these neighbours fixing the redshift to that of the relevant LAE using the same methodology as described above (also shown in \cref{fig:LAE_SED_models}), the results of which are discussed in \cref{app:Neighbour_SED_modelling}.
\begingroup
    \setlength{\tabcolsep}{6pt} 
    \renewcommand{\arraystretch}{1.25} 
    \begin{table}
        \centering
        \caption
        {Rest-frame wavelength windows for continuum measurements.}
        \label{tab:Continuum_windows}
        \begin{tabular}{cc}
            \hline
            \multicolumn{2}{c}{$\lambda_\text{emit} \, (\Angstrom)$}
            \\
            \hline
            UV & Optical
            \\
            \hline
            1320$-$1370 & 4170$-$4280
            \\
            1400$-$1450 & 4400$-$4720
            \\
            1580$-$1600 & 5100$-$5300
            \\
            1780$-$1860 & 
            \\
            1960$-$2050 & 
            \\
            2250$-$2500 & 
            \\
            \hline
        \end{tabular}
    \end{table}
\endgroup

\subsection{Empirical fitting of continuum and line emission}
\label{ssec:Continuum_and_line_emission}

Since continuum emission is not significantly detected in the R1000 measurements, we focussed on the PRISM spectra to empirically determine the strength and shape of the continuum emission. To account for the potential presence of a discontinuity at the Balmer edge, $\lambda_\text{emit} = 3645.1 \, \Angstrom$, we separately fitted power-law continua ($F_\lambda \propto \lambda^\beta$) to line-free spectral regions in the rest-frame UV and optical. The rest-frame UV windows loosely follow those by \citet{1994ApJ...429..582C}, but given the limited spectral resolution of the PRISM, we merged several and avoided contamination by strong emission lines such as the $\CIV \, \lambda \, 1548, 1551 \, \Angstrom$ and $\CIIIf \, \lambda \, 1907 \, \Angstrom, \CIIIs \, \lambda \, 1909 \, \Angstrom$ (simply \CIII hereafter) doublets, as summarised in \cref{tab:Continuum_windows}. We find good agreement between the broken power law and the best-fit \program{bagpipes} continua (consisting of a combination of stellar and nebular continuum; \cref{ssec:SED_modelling}). Both indeed show evidence of Balmer `jumps', suggestive of significant contribution of the nebular continuum \citep[e.g.][]{2024MNRAS.534..523C}. We note there are some minor residuals between the model and observed continua (e.g. around \CIII in the spectrum of \JGNzeightzeroLA; \cref{fig:Line_overview_1899}). However, these are largely within the estimated uncertainties and therefore can likely be attributed to noise fluctuations, corroborated by the good agreement seen in the other two sources (\cref{fig:Line_overview_20213084,fig:Line_overview_20006347}), which received longer integration times (\cref{tab:Source_properties}).

To measure the emission-lines properties, we simultaneously fitted the PRISM and R1000 spectra with the best-fit stellar continua as the underlying continuum in a Bayesian inference framework based on a \program{python} implementation of the \program{multinest} multimodal nested-sampling algorithm \citep{2009MNRAS.398.1601F}, \program{pymultinest} \citep{2014A&A...564A.125B}.\footnote{Code available at \url{https://github.com/joriswitstok/emission_line_fitting}.} We adopted a log-normal prior on the velocity dispersion such that $x = \log_{10} \sigma_v \, (\mathrm{km \, s^{-1}})$ is normally distributed with $\mu_x = \log_{10} (75)$ and $\sigma_x = 0.2$. Since the NIRSpec instrument possesses a point spread function (PSF) that is typically smaller than the width of a micro-shutter \citep[$0.2\arcsec$;][]{2022A&A...661A..80J}, its spectral resolution for compact sources can be significantly enhanced relative to the `nominal' STScI resolution curves based on a uniformly illuminated micro-shutter \citep{2024A&A...684A..87D}. To account for this effect, we allowed the nominal STScI resolution curves based on a uniformly illuminated micro-shutter (\cref{ssec:SED_modelling}) to be multiplied by a common scaling factor linear in wavelength (independent of the grating-filter combination). Motivated by the findings of \citet{2024A&A...684A..87D}, we let this scaling factor vary uniformly from unity up to a factor of $2$ and $1.6$ at the blue and red ends of the NIRSpec wavelength coverage, $0.6 \, \mathrm{\upmu m}$ and $5.3 \, \mathrm{\upmu m}$ respectively (\cref{ssec:NIRSpec_data_reduction}).

We fitted Gaussian profiles to the rest-frame UV $\CIV \, \lambda \, 1548, 1551 \, \Angstrom$, $\HeII \, \lambda \, 1640 \, \Angstrom$ (\HeII), $\OIIIs \, \lambda \, 1660, 1666 \, \Angstrom$, $\CIIIf \, \lambda \, 1907 \, \Angstrom, \CIIIs \, \lambda \, 1909 \, \Angstrom$, $\MgII \, \lambda \, 2796, 2804 \, \Angstrom$ lines, the rest-frame optical $\OII \, \lambda \, 3727, 3730 \, \Angstrom$, $\NeIII \, \lambda \, 3870 \, \Angstrom$, $\HeI \, \lambda \, 3890 \, \Angstrom$, $\NeIII \, \lambda \, 3969 \, \Angstrom$, $\OIII \, \lambda \, 4364 \, \Angstrom$, $\OIII \, \lambda \, 4960, 5008 \, \Angstrom$ lines, and the \HI Balmer lines from \Hbeta down to \Hepsilon. We fixed line ratios set by atomic physics between the \NeIII lines at $3870 \, \Angstrom$ to $3969 \, \Angstrom$ ($F_{3870}/F_{3969} = 3.319$), and between the \OIII lines at $4960 \, \Angstrom$ and $5008 \, \Angstrom$ ($F_{5008}/F_{4960} = 2.98$ as calculated in \program{PyNeb}; \citealt{2015A&A...573A..42L}). If well detected, we varied the \CIII doublet ratio between $0 < F_{1907}/F_{1909} < 1.7$ \citep{2019ApJ...880...16K} and the \OII doublet ratio between $0.685 < F_{3727}/F_{3730} < 3.0$ \citep{2016ApJ...816...23S}. Otherwise, a log-uniform distribution over $10^{-22}$ to $10^{-16} \, \mathrm{erg \, s^{-1} \, cm^{-2}}$ was adopted for the integrated emission-line fluxes.

Meanwhile, \Lya was fit simultaneously in the PRISM and R1000 spectra as a skewed Gaussian profile\footnote{We opted to model the line profile as a skew-normal distribution \citep{2009arXiv0911.2093A} to ensure its normalisation is well defined.} with full convolution of the continuum and line emission with the line spread function (LSF), to properly account for the smoothed continuum break in the low-resolution PRISM spectra \citep[see][ for details]{2024A&A...683A.238J}. We take the peak of the profile to be the \Lya velocity offset from the systemic redshift, $\Delta v_\text{\Lya}$. We adopted uniform priors on the velocity offset from the systemic redshift (\cref{ssec:LAE_sample}), $-500 \, \mathrm{km \, s^{-1}} < \Delta v_\text{\Lya} < 1000 \, \mathrm{km \, s^{-1}}$, and intrinsic line width, $5 \, \mathrm{km \, s^{-1}} < \sigma_{v, \, \text{\Lya}} < 2000 \, \mathrm{km \, s^{-1}}$.

Where appropriate, upper limits were calculated over a spectral window three times the size of velocity dispersion by integrating the continuum-subtracted observed spectrum, if higher than zero, to which we added three times the standard deviation on the total flux (i.e. $3 \sigma$) across the same window (found by summing individual channel uncertainties in quadrature). Line EWs were computed over a spectral region spanning five times the velocity dispersion around the central wavelength, except for \Lya where we simply divided the integrated line flux by the median value of the continuum at $1216.5 \, \Angstrom < \lambda_\text{emit} < 1225 \, \Angstrom$. We show the best-fit models at PRISM resolution in \cref{fig:Line_overview_1899,fig:Line_overview_20213084,fig:Line_overview_20006347}. Integrated fluxes, EWs, and other inferred properties of individual lines, as well as a detailed view of the observed rest-frame UV lines, are presented in \cref{app:Emission-line_measurements}.

We note that the observed emission line profiles occasionally appear to show minimal discrepancies between the PRISM and R1000 spectra, which is likely due to minor remaining wavelength and flux calibration issues \citep{2024arXiv240406531D}. Our measurement of the systemic redshifts of \JGNzeightzeroLA and \JGSzeightzeroLA, $z = 8.27886 \pm 0.00012$ and $z = 8.48523 \pm 0.00006$, are consistent (within $<1.5\sigma$) with the values reported by \citet{2024arXiv240801507T} and \citet{2024MNRAS.531.2701T} respectively, $z = 8.279$ and $z = 8.4858 \pm 0.0004$, which were obtained based on an independent data reduction. While the use of multiple high-SNR emission lines, in our case the well-detected \Hdelta, \Hgamma, \Hbeta, and $\OIII \, \lambda \, 4960, 5008 \, \Angstrom$ emission lines (\cref{ssec:LAE_sample}) or \Hbeta and $\OIII \, \lambda \, 4960, 5008 \, \Angstrom$ lines in the case of \citet{2024MNRAS.531.2701T}, allow a common line-centre measurement precision below the spectral pixel size of the G395M/F290LP grating ($\Delta \lambda_\text{obs} \approx 18 \, \Angstrom$ or $\Delta z_\text{\OIII} \approx 0.0036$), the minor calibration issues not captured by the statistical uncertainty likely explain residual marginal differences.
\begingroup
    \setlength{\tabcolsep}{6pt} 
    \renewcommand{\arraystretch}{1.25} 
    \begin{table*}
        \centering
        \caption
        {Sources studied in this work: observational details and measured and derived properties from a combination of NIRCam and NIRSpec.}
        \label{tab:Source_properties}
        \begin{tabular}{llp{3.5cm}p{3.5cm}p{3.5cm}}
\toprule
\multicolumn{2}{l}{\textit{General}} & JADES-GN-z8-0-LA & JADES-GS-z8-0-LA & JADES-GS-z8-1-LA
\\
\midrule \vspace{-2.5ex}
\\ & NIRSpec ID & 1899 & 20213084 & 20006347
\\ & NIRCam ID & 1010260 & 213084 & 6347
\\ & $\alpha_{2000} \, (\mathrm{deg})$ & $189.197740$ & $53.158906$ & $53.109000$
\\ & $\delta_{2000} \, (\mathrm{deg})$ & $62.256964$ & $-27.765076$ & $-27.900842$
\\ & NIRSpec tier & MEDIUM-HST-GN & ULTRA-DEEP-GS-3215 & DEEP-JWST-GS
\\ & Date of obs. & 7 February 2023 & 16 October 2023 & 10 January 2024
\\ & $t_\text{exp, PRISM} \, (\mathrm{h})$ & $2.9$ & $53.2$ & $18.7$
\\ & $t_\text{exp, R1000} \, (\mathrm{h})$ & $3 \times 1.7$ & $11.2$ (G140M), $33.6$ (G395M) & $3 \times 4.7$
\\ & $z_\text{spec}$ & $8.27886_{-0.00013}^{+0.00012}$ & $8.48523_{-0.00006}^{+0.00006}$ & $8.71523_{-0.00006}^{+0.00006}$
\vspace{0.5ex}
\\
\multicolumn{5}{l}{\textit{Stellar populations}}
\\
\midrule \vspace{-2.5ex}
\\ NIRCam and NIRSpec & $M_* \, (10^{7} \, \mathrm{M_\odot})$ & $3.48_{-0.40}^{+0.56}$ & $2.58_{-0.30}^{+0.49}$ & $9.22_{-1.04}^{+1.40}$
\\ & $Z_* \, (10^{7} \, \mathrm{Z_\odot})$ & $0.14_{-0.02}^{+0.01}$ & $0.12_{-0.02}^{+0.03}$ & $0.15_{-0.01}^{+0.01}$
\\ & $\mathrm{{SFR}}_{{10}} \, (\mathrm{{M_\odot \, yr^{{-1}}}})$ & $3.52_{-0.42}^{+0.56}$ & $2.57_{-0.29}^{+0.36}$ & $9.36_{-1.03}^{+1.36}$
\\ & $\Sigma_\mathrm{{SFR, \, 10}} \, (\mathrm{{M_\odot \, yr^{{-1}} \, kpc^{{-2}}}})$ & $69.2 \pm 26.6$ & $1.88 \pm 0.34$ & $4.45 \pm 1.05$
\\ & $\mathrm{{sSFR}}_{{10}} \, (\mathrm{{Gyr^{{-1}}}})$ & $101.84_{-1.72}^{+0.10}$ & $101.43_{-7.70}^{+0.54}$ & $101.88_{-1.52}^{+0.08}$
\\ & $t_* \, (\mathrm{Myr})$ & $2.75_{-0.15}^{+2.07}$ & $3.43_{-0.81}^{+8.22}$ & $2.70_{-0.10}^{+2.19}$
\\ & $A_V \, (\mathrm{mag})$ & $0.13_{-0.06}^{+0.08}$ & $0.12_{-0.06}^{+0.06}$ & $0.28_{-0.08}^{+0.08}$
\\ & $\log_{10} U$ & $-1.09_{-0.20}^{+0.13}$ & $-0.81_{-0.21}^{+0.18}$ & $-1.40_{-0.09}^{+0.12}$
\vspace{0.5ex}
\\
\multicolumn{5}{l}{\textit{Rest-frame UV}}
\\
\midrule \vspace{-2.5ex}
\\ NIRCam (F150W) & $M_\text{UV} \, (\mathrm{mag})$ & $-19.65 \pm 0.10$ & $-19.19 \pm 0.05$ & $-20.33 \pm 0.05$
\\ NIRCam (F200W) & $R_\text{UV} \, (\mathrm{pc})$ & $90 \pm 10$ & $466 \pm 30$ & $579 \pm 50$
\\ NIRSpec & $M_\text{UV} \, (\mathrm{mag})$ & $-19.41_{-0.08}^{+0.09}$ & $-19.23_{-0.03}^{+0.03}$ & $-19.57_{-0.03}^{+0.03}$
\\ & $\beta_\text{UV}$ & $-2.21_{-0.24}^{+0.25}$ & $-2.43_{-0.09}^{+0.09}$ & $-1.92_{-0.08}^{+0.09}$
\\ & $\text{EW}_\text{\Lya} \, (\Angstrom)$ & $131.8_{-3.9}^{+4.0}$ & $14.6_{-2.0}^{+2.5}$ & $30.7_{-2.3}^{+2.4}$
\\ & $\Delta v_\text{\Lya} \, (\mathrm{km \, s^{-1}})$ & $140_{-23}^{+23}$ & $212_{-41}^{+33}$ & $360_{-37}^{+30}$
\\ & $f_\text{esc, \Lya}$ (case A) & $0.494_{-0.041}^{+0.044}$ & $0.089_{-0.013}^{+0.016}$ & $0.0841_{-0.0069}^{+0.0071}$
\\ & $f_\text{esc, \Lya}$ (case B) & $0.715_{-0.060}^{+0.063}$ & $0.129_{-0.018}^{+0.022}$ & $0.122_{-0.010}^{+0.010}$
\\ & $\text{C43}$ & $4.5_{-1.6}^{+5.2}$ & $< 1.6$ & $< 0.96$
\\ & $\text{C3He2}$ & $> 6.1$ & $> 1.0$ & $2.00_{-0.62}^{+1.57}$
\vspace{0.5ex}
\\
\multicolumn{5}{l}{\textit{Rest-frame optical}}
\\
\midrule \vspace{-2.5ex}
\\ NIRSpec & $\text{EW}_\text{\Hbeta} \, (\Angstrom)$ & $168_{-13}^{+13}$ & $253.1_{-10.3}^{+9.5}$ & $204.3_{-6.1}^{+7.5}$
\\ & $\text{EW}_\mathrm{\OIII \, \lambda 4960 \, \Angstrom} \, (\Angstrom)$ & $525.5_{-8.1}^{+7.7}$ & $479.8_{-4.2}^{+4.9}$ & $505.5_{-3.5}^{+3.5}$
\\ & $\text{EW}_\mathrm{\OIII \, \lambda 5008 \, \Angstrom} \, (\Angstrom)$ & $1600_{-25}^{+24}$ & $1479_{-13}^{+15}$ & $1534_{-10}^{+10}$
\\ & $\text{Ne3O2}$ & $> 3.0$ & $1.41_{-0.25}^{+0.34}$ & $1.090_{-0.079}^{+0.088}$
\\ & $\text{O32}$ & $> 39$ & $22.7_{-3.4}^{+4.8}$ & $13.92_{-0.73}^{+0.84}$
\\ & $\text{R2}$ & $< 0.24$ & $0.239_{-0.041}^{+0.045}$ & $0.512_{-0.033}^{+0.034}$
\\ & $\text{R3}$ & $9.21_{-0.69}^{+0.83}$ & $5.45_{-0.22}^{+0.26}$ & $7.14_{-0.23}^{+0.23}$
\\ & $\text{R23}$ & $< 13$ & $7.52_{-0.30}^{+0.36}$ & $10.05_{-0.33}^{+0.32}$
\\ & $\log_{10} \left( \xi_\mathrm{ion, \, 0} \, (\mathrm{Hz \, erg^{-1}}) \right)$ & $25.58_{-0.13}^{+0.18}$ & $25.37_{-0.13}^{+0.18}$ & $25.67_{-0.13}^{+0.18}$
\\ & $T_e \, (\mathrm{K})$ & $16500_{-2700}^{+2600}$ & $19200_{-2600}^{+2600}$ & $17600_{-1400}^{+1400}$
\\ & $12+\log_{10}(\text{O/H})$ & $7.84_{-0.38}^{+0.11}$ & $7.481_{-0.209}^{+0.089}$ & $7.699_{-0.097}^{+0.059}$
\\ & $Z_\text{neb} \, (\mathrm{Z_\odot})$ & $0.142_{-0.083}^{+0.043}$ & $0.062_{-0.024}^{+0.014}$ & $0.102_{-0.020}^{+0.015}$
\\
\bottomrule
\end{tabular}
        \flushleft
        \textbf{Notes.} Listed general properties are the NIRSpec and NIRCam IDs, right ascension ($\alpha_{2000}$) and declination ($\delta_{2000}$), NIRSpec tier, date the NIRSpec observations were taken, exposure times ($t_\text{exp}$) in the PRISM and R1000 modes, and photometric and spectroscopic redshifts ($z_\text{phot}$ and $z_\text{spec}$). Stellar population properties (\cref{ssec:SED_modelling}) include the stellar mass ($M_*$), stellar metallicity ($Z_*$), (specific) star formation rate and its surface density averaged over the last $10 \, \mathrm{Myr}$, $\text{(s)SFR}_{10}$ and $\Sigma_\text{SFR, 10}$, mass-weighted stellar age ($t_*$), V-band attenuation ($A_V$), and ionisation parameter ($U$). Rest-frame UV properties are the absolute magnitude ($M_\text{UV}$), deconvolved half-light radius ($R_\text{UV}$), UV slope ($\beta_\text{UV}$), \Lya equivalent width ($\text{EW}_\text{\Lya}$), \Lya velocity offset ($\Delta v_\text{\Lya}$), \Lya escape fraction ($f_\text{esc, \Lya}$) assuming case A and case B, and the C43 and C3He2 line ratios (see \cref{sssec:ISM_ionisation_excitation_and_metal_enrichment}). Rest-frame optical properties are the equivalent widths of the \Hbeta, $\OIII \, \lambda \, 4960, 5008 \, \Angstrom$ lines ($\text{EW}_\text{\Hbeta}$, $\text{EW}_{\OIII \, \lambda \, 4960 \, \Angstrom}$, $\text{EW}_{\OIII \, \lambda \, 5008 \, \Angstrom}$), the Ne3O2, O32, R2, R3, and R23 line ratios, the ionising photon production efficiency ($\xi_\mathrm{ion, \, 0}$), electron temperature ($T_e$), and the nebular oxygen abundance and metallicity ($12 + \log_{10} \left( \text{O/H} \right)$ and $Z_\text{neb}$; \cref{sssec:ISM_ionisation_excitation_and_metal_enrichment,sssec:Production_and_escape_of_ionising_photons}).
    \end{table*}
\endgroup
\begin{figure}
	\centering
	\includegraphics[width=\linewidth]{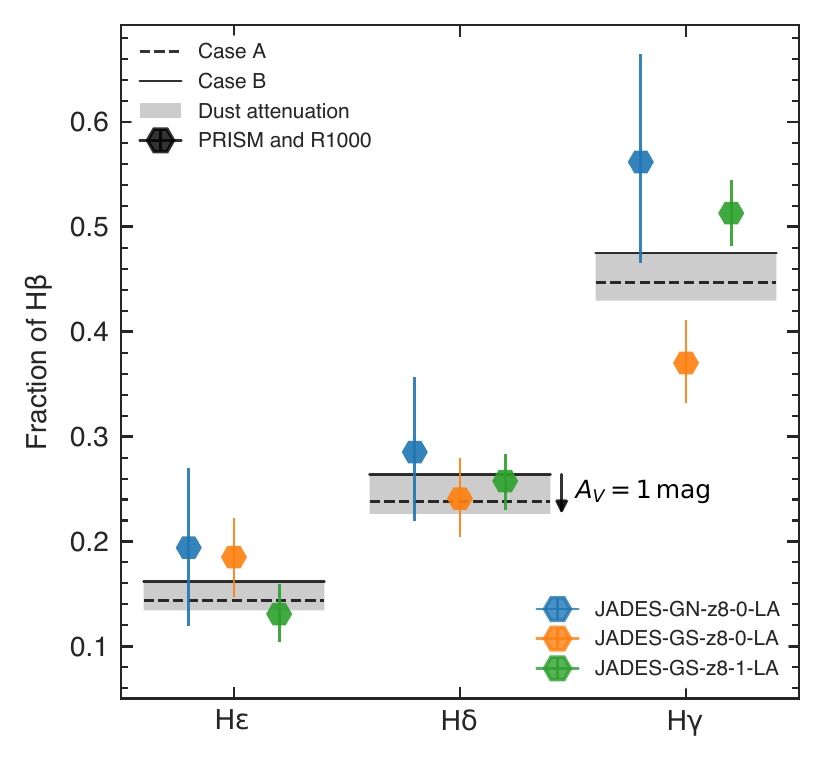}
	\caption{Observed Balmer line ratios, as measured in the PRISM and R1000 spectra, with respect to \Hbeta. Solid and dashed black lines represent the theoretical case-A and case-B values respectively (\cref{sssec:Dust_attenuation}), with grey shading indicating a shift from the case-B value caused by a corresponding V-band attenuation of $A_V = 1 \, \mathrm{mag}$ (assuming the SMC attenuation curve presented by \citealt{2003ApJ...594..279G}). While an accurate measure of the attenuation is hindered by the uncertainties and small wavelength leverage, the observed line ratios (particularly those of \Hdelta and \Hepsilon) are broadly consistent with the intrinsic case-B ratios, indicating little to no dust attenuation.
	}
	\label{fig:Balmer_decrements}
\end{figure}

\section{Discussion}
\label{sec:Discussion}

\subsection{Nebular properties}
\label{ssec:Nebular_properties}

\subsubsection{Dust attenuation}
\label{sssec:Dust_attenuation}

The observed strengths of the Balmer lines \Hgamma, \Hdelta, and \Hepsilon relative to \Hbeta are shown in \cref{fig:Balmer_decrements}. We compare these values with the theoretically predicted case-A and case-B line ratios, as computed in \program{Cloudy} v23.01 \citep{2017RMxAA..53..385F, 2023RMxAA..59..327C}, under physical conditions typical of \HII regions (electron density $n_e = 300 \, \mathrm{cm^{-3}}$ and temperature of $T_e = \num{15000} \, \mathrm{K}$, motivated by the directly measured value; \cref{sssec:ISM_ionisation_excitation_and_metal_enrichment}). Given the short spectral baselines between the \Hbeta and subsequent Balmer lines, we conclude that we cannot determine a very accurate measure of the attenuation, noting that even at higher SNR, a spread in Balmer decrements is observed with a non-negligible fraction of galaxies showing `negative' (i.e. strictly unphysical) decrements \citep[e.g.][]{2015ApJ...806..259R}. Still, the observed line ratios, particularly those of \Hdelta and \Hepsilon spanning a longer wavelength range, are consistent within $1\sigma$ uncertainties with the intrinsic case-B ratios (except for \Hgamma in \JGSzeightzeroLA at $\ssim 2\sigma$). This indicates little to no dust attenuation, in line with the observed blue UV slopes ($\beta_\text{UV} \lesssim -2$) and as would be expected for galaxies with considerable \Lya escape fractions \citep[e.g.][]{2011ApJ...730....8H}. In the following, we will therefore conservatively set $A_V = 0.1 \, \mathrm{mag}$ to apply the appropriate dust-attenuation corrections, which we based on the average Small Magellanic Cloud (SMC) attenuation curve presented by \citet{2003ApJ...594..279G}.

\subsubsection{Production and escape of ionising photons}
\label{sssec:Production_and_escape_of_ionising_photons}

The efficiency with which stellar populations or AGN produce ionising photons is commonly quantified as $\xi_\text{ion} \equiv \dot{N}_\text{ion} / L_\mathrm{\nu, \, UV}$, where $\dot{N}_\text{ion}$ is the rate at which hydrogen-ionising photons are emitted and $L_\mathrm{\nu, \, UV}$ is the UV luminosity density. We determine this quantity through the standard procedure \citep[e.g.][]{2022ApJ...935...94S, 2023A&A...678A..68S, 2024A&A...684A..84S}, measuring the UV luminosity density at a rest-frame frequency of $\nu_\text{emit} = 1.999 \, \mathrm{PHz}$ (i.e. $\lambda_\text{emit} = 1500 \, \Angstrom$), and taking the luminosity of one of the hydrogen recombination lines (in this case \Hbeta) as a proxy for the rate of ionising photons \citep{2006agna.book.....O}, having applied a dust correction to both. Again assuming physical conditions typical of \HII regions ($T_e = \num{15000} \, \mathrm{K}$ and $n_e = 300 \, \mathrm{cm^{-3}}$; \cref{sssec:Dust_attenuation}), the conversion between \Hbeta luminosity $L_\text{\Hbeta}$ and $\dot{N}_\text{ion}$ is given by
\begin{align}
    \label{eq:N_ion_H_beta_caseA}
    \dot{N}_\text{ion} & = \frac{5.72 \times 10^{12} \, \mathrm{erg^{-1}}}{1 - f_\text{esc, LyC}} \, L_\text{\Hbeta} \quad \text{(case A),}
    \\
    \label{eq:N_ion_H_beta_caseB}
    \dot{N}_\text{ion} & = \frac{2.11 \times 10^{12} \, \mathrm{erg^{-1}}}{1 - f_\text{esc, LyC}} \, L_\text{\Hbeta} \quad \text{(case B),}
\end{align}
\noindent where $f_\text{esc, LyC}$ is the escape fraction of Lyman-continuum (LyC) photons. In the following, we will adopt the conversion given by \cref{eq:N_ion_H_beta_caseB}, noting that case A instead would imply that $\dot{N}_\text{ion}$ (and hence $\xi_\text{ion}$) increases by $\ssim 0.4 \, \mathrm{dex}$.

Following convention in literature studies \citep[e.g.][]{2023A&A...672A.186P, 2023MNRAS.523.5468S, 2024arXiv240703399P}, we set $f_\text{esc, LyC} = 0$ for this calculation to derive $\xi_\mathrm{ion, \, 0}$, where the subscript underlines it refers to the rate of non-escaping ionising photons $\dot{N}_\text{ion} (1 - f_\text{esc, LyC})$, also considering that significant deviation from this scenario would likely violate pure case-B theory where the gas is assumed to be optically thick to LyC radiation \citep[cf.][]{2024arXiv240515859M}. In reality, we may however expect $f_\text{esc, LyC} \neq 0$ to explain the likely presence of ionised regions around the the current sample of $z > 8$ LAEs (as will be discussed in \cref{sssec:Required_ionised_bubble_sizes}), in which case our $\xi_\mathrm{ion, \, 0}$ estimates effectively represent a lower limit for their true intrinsic value $\xi_\text{ion}$. Nevertheless, we will argue in the following this calculation still yields a robust and useful estimate of the production efficiency of ionising photons.

As the IGM prevents direct measurements of LyC photons above $z \gtrsim 4$ \citep{2014MNRAS.442.1805I}, the LyC escape fraction is inherently unknown and notoriously hard to estimate for reionisation-era galaxies. Still, given the combined EW of the \OIII and \Hbeta lines is observed to be $\ssim 3 \times$ higher than the median value found in similarly bright ($M_\text{UV} \approx -20 \, \mathrm{mag}$) galaxies at $z \sim 8$ \citep{2024MNRAS.533.1111E}, we conclude that in any case $f_\text{esc, LyC}$ is unlikely to be extremely high, in which case the strengths of these lines would become suppressed. Indeed, based on the multivariate predictor developed by \citet{2024MNRAS.529.3751C}, we indirectly estimate this quantity may range from $1\%$ to $3\%$ among the current sample of $z > 8$ LAEs, while according to the relation between UV slope and $f_\text{esc, LyC}$ proposed by \citet{2022MNRAS.517.5104C} it ranges between $3\%$ and $10\%$. Even at the highest value among these ($f_\text{esc, LyC} = 10\%$), $\xi_\text{ion}$ would only increase by approximately $\ssim 0.05 \, \mathrm{dex}$ relative to $\xi_\mathrm{ion, \, 0}$.

The intrinsic \Lya and \Hbeta luminosities for case-B recombination under the same physical conditions as stated above are related as $L_\text{\Lya} / L_\text{\Hbeta} = 23.48$ ($34.00$ for case A). This can be used to estimate the \Lya escape fraction $f_\text{esc, \Lya}$, the observed fraction of \Lya flux which is intrinsically produced in \HII\ regions. We derived $f_\text{esc, \Lya}$ as the dust-corrected flux ratio of \Lya to \Hbeta divided by the theoretical intrinsic ratio in both case A and B \citep[e.g.][]{2023MNRAS.526.1657T, 2024MNRAS.531.2701T, 2023A&A...678A..68S, 2024A&A...684A..84S}.

All inferred quantities are summarised in \cref{tab:Source_properties}. The directly measured values of $\xi_\mathrm{ion, \, 0} \gtrsim 10^{25.4} \, \mathrm{Hz \, erg^{-1}}$ are $>1.5\times$ higher than reported values among statistical samples of EoR galaxies \citep{2024arXiv240703399P, 2024arXiv240901286S}, which match the canonical value, $10^{25.2} \, \mathrm{Hz \, erg^{-1}}$ \citep{2015ApJ...802L..19R}. Particularly \JGSzeightoneLA is a proficient producer of ionising photons, $\log_{10} ( \xi_\mathrm{ion, \, 0} \, (\mathrm{Hz \, erg^{-1}}) ) = 25.67_{-0.13}^{+0.18}$, reminiscent of values recently inferred for the faint $6 < z < 7$ galaxy population \citep[$M_\text{UV} \gtrsim -17 \, \mathrm{mag}$;][]{2024Natur.626..975A} even if \JGSzeightoneLA is substantially brighter ($>15\times$). This source, however, like \JGSzeightzeroLA does not exhibit overly efficient \Lya escape, $f_\text{esc, \Lya} \approx 10\%$, in stark contrast with \JGNzeightzeroLA, for which we infer a large escape fraction of $f_\text{esc, \Lya} = 72 \pm 6\%$ (case B) despite the IGM attenuation that may be expected to be rather large.

\subsubsection{Ionisation, excitation, and metal enrichment of the ISM}
\label{sssec:ISM_ionisation_excitation_and_metal_enrichment}

We define several commonly studied line-ratio diagnostics \citep[e.g.][]{2023A&A...677A.115C, 2023arXiv231118731S} across the rest-frame UV and optical as
\begin{align*}
    \text{C43} & = \CIV \, \lambda \, 1548, 1551 \, \Angstrom / \CIII \, ,
    \\
    \text{C3He2} & = \CIII / \HeII \, \lambda \, 1640 \, \Angstrom \, ,
    \\
    \text{Ne3O2} & = \NeIII \, \lambda \, 3870 \, \Angstrom / \OII \, ,
    \\
    \text{O32} & = \OIII \, \lambda \, 5008 \, \Angstrom / \OII \, ,
    \\
    \text{R2} & = \OII / \text{\Hbeta} \, ,
    \\
    \text{R3} & = \OIII \, \lambda \, 5008 \, \Angstrom / \text{\Hbeta} \, ,
    \\
    \text{R23} & = \OII + \OIII \, \lambda \, 4960, 5008 \, \Angstrom / \text{\Hbeta} \, ,
\end{align*}

\noindent where the line labels represent the corresponding integrated fluxes, and \CIII and \OII are shorthand respectively for the $\CIIIf \, \lambda \, 1907 \, \Angstrom, \CIIIs \, \lambda \, 1909 \, \Angstrom$ and $\OII \, \lambda \, 3727, 3730 \, \Angstrom$ doublets.

We detect the auroral $\OIII \, \lambda \, 4364 \, \Angstrom$ line in all three LAEs, which allows us to constrain the electron temperature $T_e$ and estimate the oxygen abundance via the direct method \citep{2020MNRAS.491..944C, 2023MNRAS.518..425C, 2023ApJ...943...75S, 2024ApJ...962...24S}. The strength of this transition, which corresponds to the $\mathrm{2s^2 \: 2p^{2} \: ^1S_0} \rightarrow \mathrm{2s^2 \: 2p^{2} \: ^1D_2}$ transition of $\mathrm{O^{2+}}$, is highly dependent on collisional excitation that populates the upper ($\mathrm{^1S_0}$) level, so that the ratio of $\OIII \, \lambda \, 4364 \, \Angstrom$ relative to the $\OIII \, \lambda \, 5008 \, \Angstrom$ line ($\mathrm{2s^2 \: 2p^{2} \: ^1D_2} \rightarrow \mathrm{2s^2 \: 2p^{2} \: ^3P_2}$) is strongly positively correlated with the electron temperature $T_e$.

We employed the \program{PyNeb} code \citep{2015A&A...573A..42L} to calculate temperatures and abundance ratios, adopting the electron impact excitation collision strengths from \citet{2006MNRAS.366L...6P} and \citet{2007ApJS..171..331T} for $\text{O}^+$, and those from \citet{1999ApJS..123..311A} and \citet{2012MNRAS.423L..35P} for $\text{O}^{2+}$. We derived the oxygen abundance via the common approximation $\mathrm{O/H} \approx \mathrm{O^+/H} + \mathrm{O^{2+}/H}$ \citep[e.g.][ and references therein]{2024A&A...681A..70L}, except for \JGNzeightzeroLA where we take $\mathrm{O/H} \approx \mathrm{O^{2+}/H}$ given the \OII doublet is not detected, noting that $\text{O}^{2+}$ typically dominates the oxygen budget in the low-metallicity regime \citep{2017MNRAS.465.1384C}. We assumed the relation from \citet{2009MNRAS.398..485P} to derive the temperature of \OII-emitting gas based on the $\text{O}^{2+}$ temperature, $T_{\text{O}^+} = 0.835 \, T_{\text{O}^{2+}} + 0.264$. We note there is likely some systematic bias in these estimates due to temperature fluctuations on the scale of individual \HII regions \citep{2024A&A...689A..78M}, which may cause the metallicity to be underestimated \citep{2023MNRAS.522L..89C}. We quote gas-phase metallicities $Z_\text{neb}$ adopting a solar oxygen abundance of $12 + \log_{10} \left( \text{O/H} \right)_\odot = 8.69$ \citep{2009ARA&A..47..481A}.

We infer via the direct-$T_e$ method that the ISM in these LAEs is relatively hot ($T_e > \num{15000} \, \mathrm{K}$), and is characterised by low (yet not extremely low) gas-phase metallicities of approximately $10\%$ solar, in agreement with the metallicities found in the SED modelling. As expected given these low metallicities and in line with previous work on LAEs \citep{2020MNRAS.493.5120M, 2023A&A...678A..68S, 2023MNRAS.523.5468S}, they are efficient at producing ionising photons, $\xi_\mathrm{ion, \, 0} \gtrsim 10^{25.4} \, \mathrm{Hz \, erg^{-1}}$ (\cref{sssec:Production_and_escape_of_ionising_photons}). In the following sections, we will investigate which intrinsic sources of ionisation are powering the strong nebular emission (\cref{sssec:Sources_of_ionisation}), and which escape mechanisms are in place that allow the \Lya emission to be observable well into the EoR, including the potential role of environment (\cref{ssec:Ionised_bubbles_and_large-scale_environments}).
\begin{figure*}
    \centering
    \includegraphics[width=\linewidth]{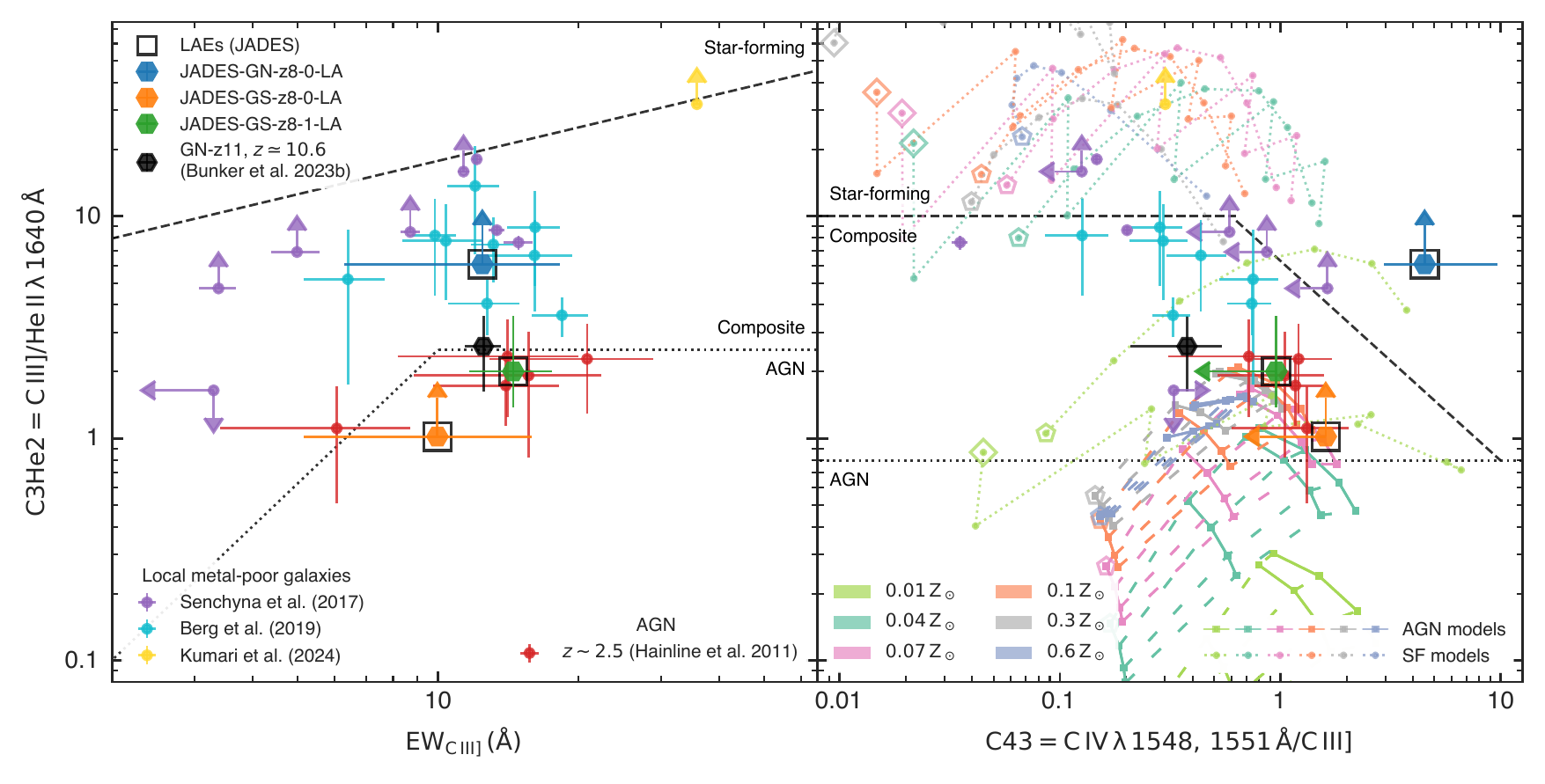}
    \caption{Rest-frame UV emission-line diagnostic diagrams from \citet{2019MNRAS.487..333H}. The LAEs considered in this work are highlighted by coloured hexagons outlined by black squares. GN-z11 \citep{2023A&A...677A..88B} is shown by a black hexagon. In both panels, the dashed and dotted black lines were designed by \citet{2019MNRAS.487..333H} to separate star-forming galaxies from those with composite origin of the nebular emission (AGN and star formation), and composite from AGN, respectively. We include measurements based on UV spectroscopy of local metal-poor galaxies, shown as small purple, blue, and yellow circles \citep{2017MNRAS.472.2608S, 2019ApJ...874...93B, 2024MNRAS.529..781K}, and of $z \sim 2.5$ AGN \citep{2011ApJ...733...31H}. \textit{Left panel}: C3He2 ratio versus the EW of \CIII. \textit{Right panel}: C3He2 ratio versus C43 ratio. Models for the predicted line ratios with star formation (SF; small circles) and AGN (small squares) as ionising sources are overplotted, with colours indicating varying metallicity according to the legend in the bottom left (see \cref{sssec:Sources_of_ionisation} for details). Starting from the larger open pentagons \citep{2016MNRAS.462.1757G, 2016MNRAS.456.3354F} or diamonds \citep{2022MNRAS.513.5134N}, the ionisation parameter increases along the dotted or dashed lines. For AGN models, solid lines additionally show variation in the spectral power-law slope $\alpha$. The emission-line properties of the $z > 8$ LAEs considered here can be explained as originating in star formation, except for \JGSzeightoneLA at $z \simeq 8.72$ which shows evidence of photoionisation by an AGN.
    }
    \label{fig:UV_line_ratios}
\end{figure*}

\subsection{On the nature of primaeval LAEs}
\label{sssec:On_the_nature_of_primaeval_LAEs}

\subsubsection{Stellar properties}
\label{sssec:Stellar_properties}

The main results of the SPS modelling (\cref{ssec:SED_modelling}), spectral fitting (\cref{ssec:Continuum_and_line_emission}), and derived physical quantities (\cref{ssec:Nebular_properties}) are shown in \cref{tab:Source_properties}. While there are some hints of slightly more evolved underlying stellar populations (\cref{fig:LAE_SED_models}), we find that the observed photometry and spectroscopy of the LAEs studied here, when interpreted purely as stellar in its origin, is largely dominated by the light of very young ($t_* < 5 \, \mathrm{Myr}$) and metal-poor ($Z_* \approx 0.15 \, \mathrm{Z_\odot}$) stars. Morphologically, each LAE is very compact: we measure deconvolved half-light radii (\cref{ssec:SED_modelling}) of $R_\text{UV} \approx 500 \, \mathrm{pc}$ for \JGSzeightzeroLA and \JGSzeightoneLA, and \JGNzeightzeroLA is smaller yet ($R_\text{UV} = 90 \pm 10 \, \mathrm{pc}$). We note the F115W filter containing \Lya shows a consistently compact morphology (\cref{ssec:SED_modelling}), in the case of \JGSzeightoneLA even revealing two sources (\cref{app:Neighbour_SED_modelling}). Dominating the rest-frame UV, they are reminiscent of the compact, young star clusters that have recently begun to be resolved in gravitationally lensed, high-redshift galaxies by \jwst and are speculated to form through gravitational instabilities in a more extended gaseous disc \citep{2024arXiv240218543F} and be precursors to globular clusters \citep{2024Natur.632..513A}. Indeed, given the young stellar ages and little dust obscuration ($A_V < 0.3 \, \mathrm{mag}$) inferred from the integrated SEDs, the comparatively UV-bright LAEs \citep[relative to e.g.][]{2024A&A...684A..84S, 2023A&A...678A..68S} are characterised by low mass-to-light ratios and thus relatively low stellar masses ($M_* < 10^{8} \, \mathrm{M_\odot}$).

These stellar properties are reflected in the extreme nebular emission-line properties observed with NIRSpec. We find the optical lines to exhibit very large EWs of around $\text{EW}_\text{\OIII + \Hbeta} \approx 2000 \, \Angstrom$, roughly $\ssim 3 \times$ higher than the median value found in similarly bright ($M_\text{UV} \approx -20 \, \mathrm{mag}$) galaxies at $z \sim 8$ \citep{2024MNRAS.533.1111E}. Such strong nebular line emission is indicative of high specific SFR \citep[sSFR;][]{2013ApJ...763..129S, 2014ApJ...784...58S, 2015ApJ...801..122S, 2024arXiv240116934B, 2024ApJ...972...56T}, as confirmed by the SED modelling ($\text{sSFR}_{10} \approx 100 \, \mathrm{Gyr^{-1}}$). Assuming half of the the inferred SFR is spread uniformly over $\pi R_\text{UV}^2$, the area enclosed by the half-light radius \citep[e.g.][]{2024ApJ...963....9M}, the compact sizes moreover imply high SFR surface densities of $\Sigma_\text{SFR} > 1 \, \mathrm{M_\odot \, yr^{-1} \, kpc^{-2}}$, in the case of \JGNzeightzeroLA even $\Sigma_\text{SFR} \approx 80 \, \mathrm{M_\odot \, yr^{-1} \, kpc^{-2}}$. Such intensely confined star formation is very rarely seen in local galaxies, but has been found to become the norm in reionisation-era galaxies \citep{2024ApJ...963....9M}, especially at the highest-redshift regime \citep{2023NatAs...7..611R, 2024ApJ...972..143C}, perhaps paving the way to the early formation of dense stellar cores \citep[e.g.][]{2024NatAs.tmp..246B}.

\subsubsection{Sources of ionisation}
\label{sssec:Sources_of_ionisation}

Traditionally, AGN activity has been diagnosed by determining whether specific optical line ratios fall within theoretical \citep{2001ApJ...556..121K} or empirical \citep{2003MNRAS.346.1055K} star-forming demarcations, specifically on the classification diagrams proposed by \citet*[; \citetalias{1981PASP...93....5B}]{1981PASP...93....5B} and \citet[; \citetalias{1987ApJS...63..295V}]{1987ApJS...63..295V}. While these diagnostic methods have been explored extensively and well calibrated for galaxies at $z \lesssim 3$ \citep{2014ApJ...795..165S}, they become more challenging to employ in the high-redshift regime, in the first place because key optical lines shift out of the spectral coverage: even with NIRSpec, the \Halpha, \NII, and \SII lines become inaccessible at $z \gtrsim 7$ (as is the case for the $z > 8$ LAEs considered here). Moreover, line ratios on the \citetalias{1981PASP...93....5B} and \citetalias{1987ApJS...63..295V} diagrams have been shown to significantly deviate from the sequences occupied by local star-forming galaxies and AGN, and hence lose their discriminating power, in the low-metallicity regime \citep{2023MNRAS.526.3610H, 2023A&A...677A.145U, 2023ApJ...959...39H, 2023arXiv231118731S, 2024arXiv240312683C}. We therefore instead consider two diagnostic diagrams designed for rest-frame UV lines, proposed by \citet{2019MNRAS.487..333H} to distinguish between AGN and star formation over a range of redshift ($0 < z \lesssim 6$) based on the implementation of photoionisation models in cosmological simulations.

The diagrams, consisting of the EW of the \CIII doublet and the (dust-corrected) C3He2 and C43 line ratios, are displayed in \cref{fig:UV_line_ratios}. We show the three main LAEs alongside local, metal-poor galaxies \citep{2017MNRAS.472.2608S, 2019ApJ...874...93B}, including Pox186 \citep{2024MNRAS.529..781K}, a dwarf galaxy that is observed to have one of the highest known EWs of \CIII ($34 \, \Angstrom$) as well as the highest far-infrared $\OIII \, 88 \, \mathrm{\upmu m}$ to $\CII \, 158 \, \mathrm{\upmu m}$ line ratio seen among local galaxies ($\OIII / \CII \approx 10$, comparable to what is typically seen in reionisation-era galaxies; \citealt{2022MNRAS.515.1751W}). Following \citet{2021MNRAS.508.1686W}, we also directly compared the C3He2 and C43 ratios with the \citet{2016MNRAS.462.1757G} and \citet{2016MNRAS.456.3354F} photoionisation models computed in \program{Cloudy} \citep{2017RMxAA..53..385F} with star formation and AGN as ionising sources, respectively. The star-formation models are based on an updated version of the \citet{2003MNRAS.344.1000B} SPS models under a \citet{2003PASP..115..763C} IMF with an upper mass limit of $300 \, \mathrm{M_\odot}$, and adopt a constant SFH over $100 \, \mathrm{Myr}$. We considered models with fixed dust-to-metal ratios, $\xi_\text{d} = 0.3$, and hydrogen densities, $n_\text{H} = 10^2 \, \mathrm{cm^{-3}}$ for star-formation models and $10^3 \, \mathrm{cm^{-3}}$ for AGN (noting that varying these parameters does not significantly impact the results). We varied the ionisation parameter ($-4 < \log_{10} U < -1$) and, for the narrow-line region AGN models, the power-law slope of the rest-frame UV (i.e. $S_\nu \propto \nu^\alpha$ at $\lambda_\text{emit} \leq 2500 \, \Angstrom$; see \citealt{2016MNRAS.456.3354F}), $-2 < \alpha < -1.2$. To convert absolute metallicities to solar units, we adopt $Z_\odot = 0.0134$ \citep{2009ARA&A..47..481A}.

We complement the star-formation photoionisation models described above by the (regular, Population II) models from \citet{2022MNRAS.513.5134N}. While these are also based on \program{Cloudy} \citep{2017RMxAA..53..385F} and explore a similar range of metallicity\footnote{Since the exact metallicities in \citet{2022MNRAS.513.5134N} differ slightly from those in \citet{2016MNRAS.462.1757G} and \citet{2016MNRAS.456.3354F} models, for simplicity we use the values that match the rounded values shown in \cref{fig:UV_line_ratios}.} and ionisation parameter as discussed above, some parameters are tweaked to form a representative set of models specifically for galaxies in the very early Universe. A higher density of $n_\text{H} = 10^3 \, \mathrm{cm^{-3}}$ is adopted \citep[see e.g.][]{2016ApJ...816...23S}, and the incident radiation field is instead set by a very young ($1 \, \mathrm{Myr}$ or $10 \, \mathrm{Myr}$) stellar population, as modelled by \program{bpass} v2.2.1 \citep{2017PASA...34...58E} under its default IMF ranging up to $300 \, \mathrm{M_\odot}$ (cf. \cref{ssec:SED_modelling}). To aid visualisation in \cref{fig:UV_line_ratios}, we leave out the \citet{2022MNRAS.513.5134N} AGN models here, noting these occupy broadly the same region in terms of C3He2 and C43 ratios as the \citet{2016MNRAS.456.3354F} AGN models.

From this comparison, we find that \JGNzeightzeroLA shows nebular emission line ratios consistent with what is expected for star formation in the metal-poor regime, both empirically (showing similarity to local metal-poor galaxies) and from the perspective of photoionisation modelling. Interestingly, the high C43 line ratio is only reproduced under extreme circumstances \citep{2024MNRAS.529.3301T}: it requires the highest ionisation parameter ($\log_{10} U = -1$) in the \citet{2016MNRAS.462.1757G} model track that is significantly more metal poor ($\ssim 1\%$ solar) than directly inferred ($\ssim 15\%$ solar; \cref{tab:Source_properties}). This may be a result of the actual SFH rising very steeply over the last $10 \, \mathrm{Myr}$ (while extending further back to account for the existing metal content), whereas the \citet{2016MNRAS.462.1757G} models adopt a constant SFH over $100 \, \mathrm{Myr}$. The \citet{2022MNRAS.513.5134N} models similarly struggle to reproduce such a high C43 ratio, except for the most metal-poor track which however underpredicts the C3He2 ratio.

A lack of detections of the \CIV and \HeII lines in \JGSzeightzeroLA prevents us from conclusively determining its main source of ionisation, and although it is formally located in a region demarcating AGN activity, the upper limits indicate this galaxy could be consistent with star formation. Conversely, \JGSzeightoneLA at $z \simeq 8.72$ shows a reasonably strong \HeII line at $\text{EW}_\text{\HeII} = 4.9_{-2.0}^{+1.9} \, \Angstrom$ (detected consistently between the PRISM and R1000 modes, though only at $\ssim 3\sigma$; see \cref{app:Emission-line_measurements}). Its emission-line ratios, similar to those measured in stacked spectra of $z \sim 2.5$ AGN \citep{2011ApJ...733...31H}, indicate the ionisation source may be a combination of AGN and star formation, if not dominated by an AGN. This finding is corroborated by the high $\OIII \, \lambda \, 4364 \, \Angstrom / \text{\Hgamma} \approx 0.4$ ratio found in \JGSzeightoneLA, typically only observed in AGN \citep{2023MNRAS.525.2087B, 2024MNRAS.531..355U, 2024arXiv240410811M}.
\begin{figure*}
	\centering
	\includegraphics[width=\linewidth]{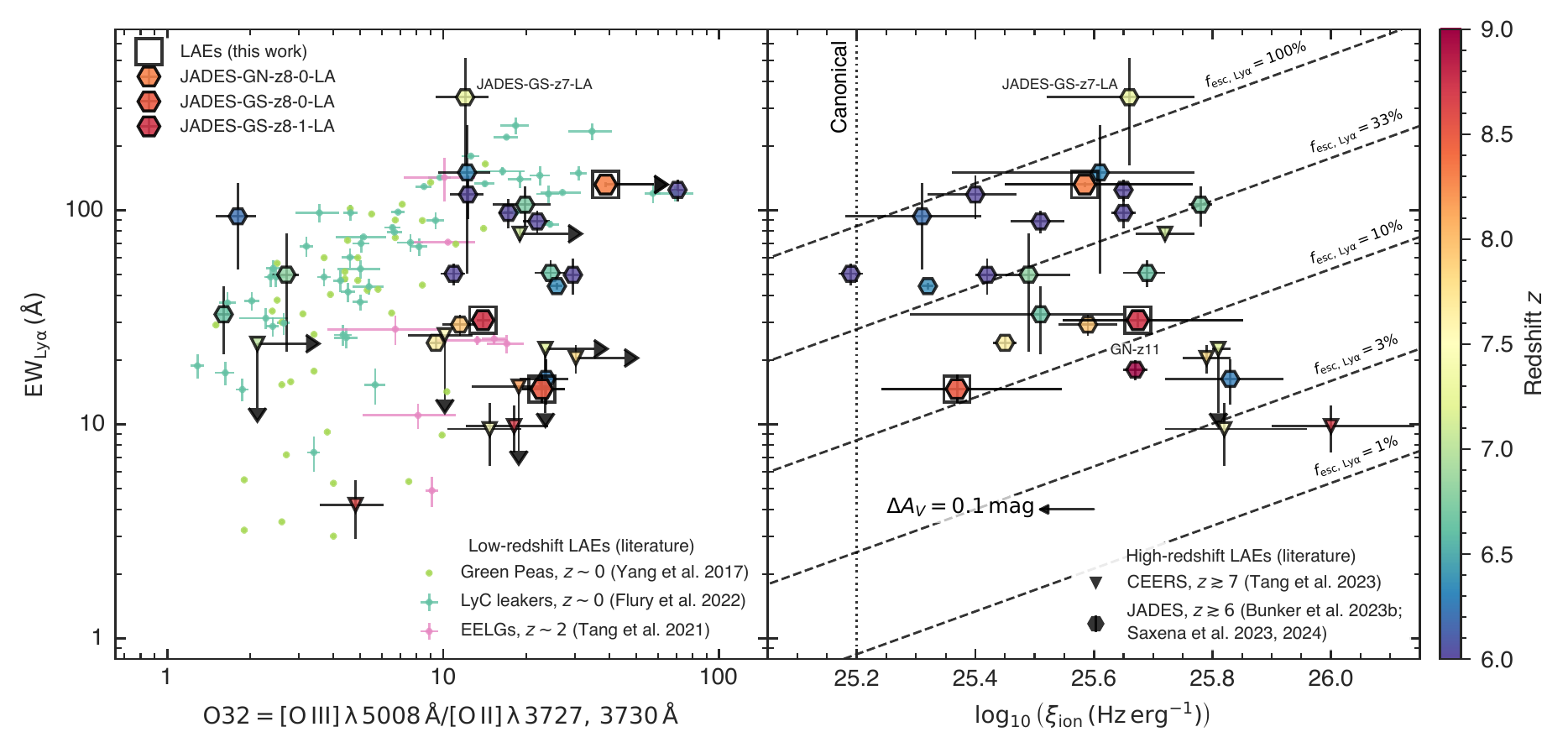}
	\caption{EW of \Lya as a function of the O32 ratio (\textit{left panel}) and ionising-photon production efficiency (\textit{right panel}). The main LAEs considered in this work are shown by hexagons outlined by black squares (representing $\xi_\mathrm{ion, \, 0}$ values; \cref{sssec:Production_and_escape_of_ionising_photons}). Literature measurements include Green Pea galaxies \citep{2017ApJ...844..171Y}, confirmed LyC leakers from the LzLCS \citep{2022ApJS..260....1F}, EELGs \citep{2021MNRAS.503.4105T}, and high-redshift ($z \gtrsim 6$) LAEs \citep{2023MNRAS.526.1657T, 2024A&A...684A..84S}, of which JADES-GS-z7-LA \citep{2023A&A...678A..68S} and GN-z11 \citep{2023A&A...677A..88B} are annotated. All high-redshift LAEs coloured according to the colourbar on the right. In the right panel, the canonical value $\xi_\text{ion} = 10^{25.2} \, \mathrm{Hz \, erg^{-1}}$ \citep{2015ApJ...802L..19R} is shown by a vertical dotted line. An arrow indicates the effect of an additional dust-attenuation correction of $\Delta A_V = 0.1 \, \mathrm{mag}$. Dashed lines illustrate the approximate relation between $\xi_\mathrm{ion, \, 0}$ and $\text{EW}_\text{\Lya}$ given by \cref{eq:xi_ion_Lya_EW_approx} for various $f_\text{esc, \Lya}$ (see \cref{sssec:Comparing_LAE_properties_across_cosmic_time,app:Relating_xi_ion_to_Lya_EW} for details). The two LAEs with modest \Lya EWs (\JGSzeightzeroLA and \JGSzeightoneLA) show relatively weak \Lya given their observed O32 ratios and $\xi_\text{ion}$, suggesting a substantial fraction may be absorbed by the IGM. \JGNzeightzeroLA, on the other hand, exhibits a \Lya EW similar to LAEs at significantly lower redshift, indicating less pronounced IGM attenuation.
	}
	\label{fig:EW_Lya_O32_xi_ion}
\end{figure*}

\subsection{Ionised bubbles: driven by individual sources or in concert?}
\label{ssec:Ionised_bubbles_and_large-scale_environments}

\subsubsection{Comparing LAE properties across cosmic time}
\label{sssec:Comparing_LAE_properties_across_cosmic_time}

To gain insight into the mechanisms regulating the production and escape of \Lya in reionisation-era LAEs ($z \gtrsim 6$), in \cref{fig:EW_Lya_O32_xi_ion}, we compare their \Lya EWs to those of LAEs at lower redshift, where the impact of IGM absorption of \Lya is minimal, while controlling for the effective ionisation parameter by means of the O32 ratio or for the ionising-photon production efficiency through $\xi_\text{ion}$ (\cref{sssec:Production_and_escape_of_ionising_photons}). The low-redshift sample shown here consists of Green Pea galaxies, argued to be analogues of high-redshift galaxies owing to their strong nebular emission \citep{2017ApJ...844..171Y}, galaxies from the low-redshift LyC survey \citep[LzLCS;][]{2022ApJS..260....1F}, of which we only consider those confirmed to be leaking LyC radiation, and extreme emission-line galaxies (EELGs) at $z \sim 2$ \citep{2021MNRAS.503.4105T}.

Even if part of the \Lya flux may also be missed by the NIRSpec micro-shutters due to spatial offsets (\cref{ssec:NIRSpec_data_reduction}), we note some \jwst-observed LAEs \citep[such as JADES-GS-z7-LA;][]{2023A&A...678A..68S} do reach, or even slightly surpass, the most extreme \Lya EWs seen in the sample of lower-redshift galaxies at fixed O32 ratio. Generally, however, we find that high-redshift LAEs predominantly have lower \Lya EWs at fixed O32 than local Green Pea and LyC-leaking galaxies. This indicates that the neutral IGM indeed plays an important role in reducing the observed \Lya EW in the majority of $z > 6$ LAEs \citep[see also][]{2023MNRAS.526.1657T, 2024MNRAS.531.2701T, 2024arXiv240307103R}. This is particularly the case for \JGSzeightzeroLA at $z \simeq 8.49$, whose \Lya strength of $\text{EW}_\text{\Lya} \approx 15 \, \Angstrom$ is unusually small given its high O32 ratio of $\text{O32} \approx 23$. In contrast, we find the location of \JGNzeightzeroLA is consistent with minimal attenuation of \Lya photons by the IGM, in line with the large ionised bubble ($R_\text{ion}^\text{req} \approx 3 \, \mathrm{pMpc}$) required to explain why the peak of the \Lya emission is observed very near the systemic redshift ($\Delta v_\text{\Lya} \approx 140 \, \mathrm{km \, s^{-1}}$). Some residual line flux is even seen close to the systemic redshift ($\Delta v_\text{\Lya} = 0$; \cref{fig:Line_overview_1899}), again indicating very little IGM attenuation, as also confirmed by an independent analysis using a separate data reduction that however yields a similar line profile \citep{2024MNRAS.531.2701T}.

The \Lya escape fraction $f_\text{esc, \Lya}$, which encapsulates the total attenuation of the \Lya photons accumulated along their path through the ISM, circumgalactic medium, and the IGM, can be more directly quantified in the panel where the \Lya EW is shown as a function of the ionising photon production efficiency $\xi_\text{ion}$. Indeed, as outlined in \cref{app:Relating_xi_ion_to_Lya_EW}, we can directly relate $\xi_\mathrm{ion, \, 0}$ to $\text{EW}_\text{\Lya}$ via
\begin{equation}
    \label{eq:xi_ion_Lya_EW}
    \xi_\mathrm{ion, \, 0} = \frac{1}{f_\text{esc, \Lya} \, f_\text{rec, B} \, h \, \lambda_\text{\Lya}} \, \left( \frac{\lambda_\text{\Lya}}{1500 \, \Angstrom} \right)^{2+\beta_\text{UV}} \, \text{EW}_\text{\Lya} \, ,
\end{equation}
\noindent where $h$ is the Planck constant and $f_\text{rec, B}$ represents the fraction of case-B recombination events that result in the emission of a \Lya photon. For $T = \num{15000} \, \mathrm{K}$ where $f_\text{rec, B} = 66\%$ \citep{2014PASA...31...40D}, in approximate form we have
\begin{equation}
    \label{eq:xi_ion_Lya_EW_approx}
    \xi_\mathrm{ion, \, 0} \approx 1.88 \times 10^{23} \, \mathrm{Hz \, erg^{-1} \, \Angstrom^{-1}} \times \left( 0.810 \right)^{2+\beta_\text{UV}} \, \frac{\text{EW}_\text{\Lya}}{f_\text{esc, \Lya}} \, ,
\end{equation}

\noindent which is shown for $\beta_\text{UV} = -2$ under various \Lya escape fractions in \cref{fig:EW_Lya_O32_xi_ion}. We caution that all inferred $\xi_\mathrm{ion, \, 0}$ values shown here, as they are derived using a combination of rest-frame UV and optical measurements (\cref{sssec:Production_and_escape_of_ionising_photons}), are sensitive to the impact of dust attenuation \citep{2023MNRAS.523.5468S}. The magnitude of this effect is visualised by the arrow in \cref{fig:EW_Lya_O32_xi_ion} indicating an increased V-band attenuation of $\Delta A_V = 0.1 \, \mathrm{mag}$. Despite (systematic) uncertainties, however, it again becomes clear that \JGNzeightzeroLA accommodates a remarkably efficient escape of \Lya photons compared to the other highest-redshift LAEs. We will now turn to discuss one of the possible reasons behind the notable \Lya emission strength seen in \JGNzeightzeroLA.

\begin{figure}
	\centering
	\includegraphics[width=\linewidth]{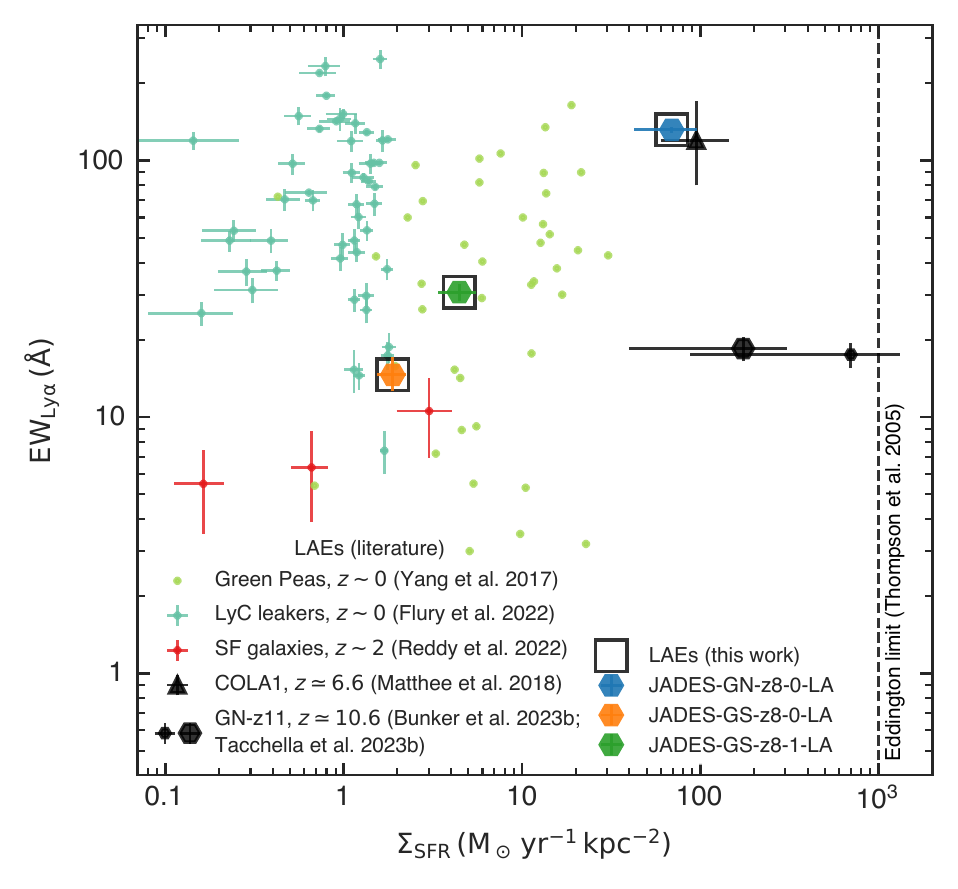}
	\caption{EW of \Lya as a function of $\Sigma_\text{SFR}$, the SFR surface density. The main LAEs considered in this work are shown by coloured hexagons outlined by black squares. Also shown are Green Pea galaxies \citep{2017ApJ...844..171Y}, confirmed LyC leakers from the LzLCS \citep{2022ApJS..260....1F}, composite spectra of star-forming galaxies at $z \simeq 2.3$ \citep{2022ApJ...926...31R}, and the high-redshift LAEs COLA1 \citep{2018A&A...619A.136M} and GN-z11 \citep{2023A&A...677A..88B, 2023ApJ...952...74T}. For GN-z11, two points (slightly offset in \Lya EW for visualisation purposes) are shown to represent the compact (small hexagon) and extended (large hexagon) source components identified by \citet{2023ApJ...952...74T}. A vertical dashed line indicates the Eddington limit for star formation \citep[$\Sigma_\text{SFR} \approx 1000 \, \mathrm{M_\odot \, yr^{-1} \, kpc^{-2}}$;][]{2005ApJ...630..167T}. Strikingly similar to the double-peaked LAE COLA1, \JGNzeightzeroLA is set apart from the local analogues and the other two LAEs by its extremely high $\Sigma_\text{SFR} \approx 80 \, \mathrm{M_\odot \, yr^{-1} \, kpc^{-2}}$.
	}
	\label{fig:EW_Lya_Sigma_SFR}
\end{figure}

Having established the three $z > 8$ LAEs exhibit high SFR surface densities (\cref{sssec:Stellar_properties}), we explore its relation with the observed \Lya EW in \cref{fig:EW_Lya_Sigma_SFR}. Interestingly, we find that \JGNzeightzeroLA has very similar properties to COLA1, a $z \simeq 6.6$ galaxy observed to have double-peaked \Lya emission \citep{2018A&A...619A.136M, 2024A&A...689A..44T}, which indicates the surrounding IGM must be very highly ionised for the blue peak not to be entirely extinguished \citep{2020MNRAS.499.1395M}. Together with GN-z11 \citep{2023A&A...677A..88B, 2023ApJ...952...74T}, their extremely high SFR surface densities clearly set them apart from Green Pea \citep{2017ApJ...844..171Y} and confirmed local LyC-leaking galaxies \citep{2022ApJS..260....1F}, argued to be local analogues of reionisation-era galaxies, and the other two $z > 8$ LAEs considered in this work. In the case of GN-z11, however, as with \JGSzeightoneLA (yet not for \JGNzeightzeroLA; \cref{sssec:Sources_of_ionisation}), we caution its light may be significantly affected by the contribution of an AGN \citep{2024Natur.627...59M}. Even so, their relatively low \Lya EW again suggests a strong suppression by intervening neutral IGM \citep[see also][]{2023A&A...677A..88B, 2023ApJ...954L..14H}. For COLA1 and \JGNzeightzeroLA, meanwhile, a high SFR surface density appears to be an important factor in determining the observed strength of \Lya, likely relating to strong feedback processes \citep{2023A&A...677L...4P} and the efficient escape of ionising photons they may accommodate \citep{2020ApJ...892..109N}.
\begin{figure*}
    \centering
    \includegraphics[width=\linewidth]{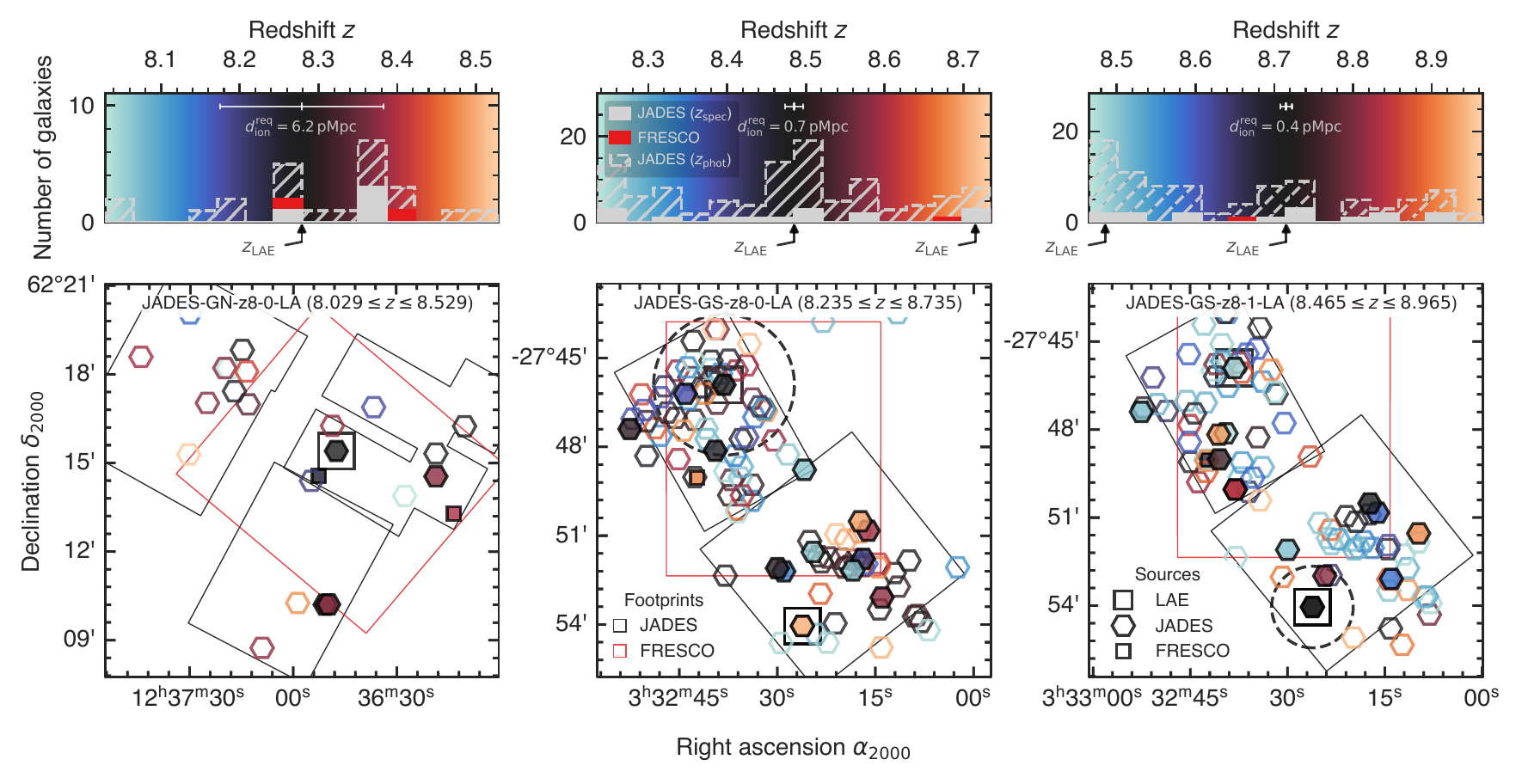}
    \caption{Distribution of distant galaxies ($z > 8$) observed by \jwst. Galaxies spectroscopically confirmed in JADES and FRESCO are shown by filled grey and red histograms (top panels) and filled hexagons and squares (bottom), respectively. Robust photometric candidates in JADES (excluding duplicates from the spectroscopic sample; \cref{sssec:Predicted_ionised_bubble_sizes}) are indicated by the grey-hatched histogram and open hexagons. \textit{Top row}: redshift-space distributions. The redshift of the main LAEs considered in this work is annotated, as is the minimum required size of the ionised bubble around each LAE (inferred based on the observed \Lya properties; see \cref{sssec:Required_ionised_bubble_sizes}), here quoted as a diameter ($d_\text{ion}^\text{req} \equiv 2 R_\text{ion}^\text{req}$). \textit{Bottom row}: on-sky distributions. The LAEs considered in this work are highlighted by black squares. The JADES (FRESCO) footprint is indicated by black (red) outlines. Dashed circles show the projected ionised-bubble sizes around each LAE (note that it exceeds the field of view shown around \JGNzeightzeroLA).
    }
    \label{fig:LAE_on-sky_distribution}
\end{figure*}

\subsubsection{Required ionised bubble sizes}
\label{sssec:Required_ionised_bubble_sizes}

We conservatively estimated the sizes of the ionised bubbles surrounding the LAEs following the methods outlined in \citet{2024A&A...682A..40W}. Specifically, we inferred the minimum radius of a spherical ionised region centred on the LAE (the bubble) that is required to explain the measured \Lya escape fractions $f_\text{esc, \Lya}$ (\cref{sssec:Production_and_escape_of_ionising_photons}) at the observed \Lya velocity offset from the systemic redshift, $\Delta v_\text{\Lya}$. Motivated by growing evidence for an inhomogeneous, late, and rapid reionisation scenario \citep[e.g.][]{2014ApJ...793..113P, 2018ApJ...863...92B, 2018MNRAS.479.2564W, 2019MNRAS.485.1350W, 2019MNRAS.485L..24K, 2022MNRAS.514...55B}, we calculated the IGM transmission in a two-zone model,\footnote{Code available at \url{https://github.com/joriswitstok/lymana_absorption}.} following \citet{2020MNRAS.499.1395M}: the gas within the bubble is highly ionised (residual neutral hydrogen fraction of $x_\text{\HI} = 10^{-8}$) and uniformly at $T = 10^4 \, \mathrm{K}$, while the surrounding IGM ($T = 1 \, \mathrm{K}$) is still fully neutral (all intergalactic gas along the line of sight is taken to be at mean cosmic density). Within this geometry, we find the size of an ionised bubble $R_\text{ion}^\text{req}$ at which the IGM transmission at velocity offset $\Delta v_\text{\Lya}$ equals the \Lya escape fraction,
\begin{equation*}
    \label{eq:Bubble_size_criterion}
    T_\text{IGM} \left( \left. R_\text{ion}^\text{req} \, \right| \Delta v_\text{\Lya} \right) = f_\text{esc, \Lya} \, ,
\end{equation*}

\noindent where we adopt our best-fit value of the skewed Gaussian profile (\cref{ssec:Continuum_and_line_emission}) as the \Lya velocity offset, $\Delta v_\text{\Lya}$. As argued in \citet{2024A&A...682A..40W}, this effectively represents a lower limit on the required size of the ionised region, since different effects (e.g. dust absorption in the ISM, infall velocity of the IGM, higher residual neutral fraction within the bubble, or spatially extended emission) may cause additional \Lya photons to avoid being captured within the NIRSpec micro-shutter. We further refer to \citet{2024A&A...682A..40W} for a detailed assessment of the impact of varying intrinsic line profiles on the estimated IGM transmission properties, and in particular note that given the $\Delta v \approx 150 \, \mathrm{km \, s^{-1}}$ wavelength bin size in the G140M/F070LP spectra, our estimates of these ionised bubble sizes will likely carry a systematic uncertainty (in physical units) of the order of $\Delta l = \Delta v / H(z) \approx 0.14 \, \mathrm{Mpc}$ at $z = 8.5$.
\begingroup
    \setlength{\tabcolsep}{4pt} 
    \begin{table}
        \centering
        \caption{Inferred ionised bubble properties.}
        \label{tab:Bubble_properties}
        \begin{tabular}{llllll}
            \toprule
            Name & $R_\text{ion}^\text{req}$ & $N_\text{spec}$ & $N_\text{phot}$ & $R_\text{ion, LAE}^\text{pred}$ & $R_\text{ion, tot}^\text{pred}$
            \\
            & $(\mathrm{pMpc})$ & & & $(\mathrm{pMpc})$ & $(\mathrm{pMpc})$
            \\
            \midrule
            \begin{filecontents*}{Bubble_sizes_GNGS.csv}
Name,NID,ID,IDfoot,RA,Dec,zspec,MUV,betaUV,EW,dv,fesc,Rion,Nionspec,Nionphot,RiongxHI,RionLAE,RionLAEgxHI,Riontot,RiontotgxHI
JADES-GN-z8-0-LA,1899,JADES-GN+189.19774+62.25696,JADES-GN+189.19774+62.25696,$+189.19774$,$+62.25696$,$8.2789$,$-19.65 \pm 0.10$,$-2.21 \pm 0.25$,$131.8 \pm 3.9$,$140$,$0.71 \pm 0.06$,$3.08$,$2$,$6$,$2.80$,$0.17$,$0.17$,$0.34$,$0.32$
JADES-GS-z8-0-LA,20213084,JADES-GS+53.15891-27.76508,JADES-GS+53.15891-27.76508,$+53.15891$,$-27.76508$,$8.4852$,$-19.19 \pm 0.05$,$-2.43 \pm 0.09$,$14.6 \pm 2.2$,$212$,$0.13 \pm 0.02$,$0.33$,$1$,$2$,$0.29$,$0.12$,$0.12$,$0.15$,$0.14$
JADES-GS-z8-1-LA,20006347,JADES-GS+53.10900-27.90084,JADES-GS+53.10900-27.90084,$+53.10900$,$-27.90084$,$8.7152$,$-20.10 \pm 0.05$,$-1.92 \pm 0.08$,$30.7 \pm 2.4$,$360$,$0.12 \pm 0.01$,$0.19$,$1$,$0$,$0.17$,$0.20$,$0.20$,$0.20$,$0.20$
\end{filecontents*}
\csvreader[late after line=\\, head to column names]{Bubble_sizes_GNGS.csv}{}{\Name & \Rion & \Nionspec & \Nionphot & \RionLAE & \Riontot}
            \bottomrule
        \end{tabular}
        \flushleft
        \textbf{Notes.} Listed properties are the inferred minimum ionised bubble sizes ($R_\text{ion}^\text{req}$), the number of spectroscopically confirmed ($N_\text{spec}$) and photometrically selected ($N_\text{phot}$; excluding those with spectroscopic redshifts) galaxies contained within it, and the bubble sizes that can be produced by the LAEs on their own ($R_\text{ion, LAE}^\text{pred}$) or by all $N_\text{spec} + N_\text{phot}$ galaxies inside the ionised bubble ($R_\text{ion, tot}^\text{pred}$), under the assumption of a fiducial age of $t_* = 50 \, \mathrm{Myr}$ and $f_\text{esc, LyC} = 5\%$ (see \cref{sssec:Predicted_ionised_bubble_sizes} for details).
    \end{table}
\endgroup

\subsubsection{Predicting the creation of ionised bubbles}
\label{sssec:Predicted_ionised_bubble_sizes}

Following \citet{2024A&A...682A..40W}, we further obtain an estimate of the sizes of ionised bubbles that the LAEs would be able to form by themselves, $R_\text{ion, LAE}^\text{pred}$, assuming fiducial ages of $t_* = 50 \, \mathrm{Myr}$ and LyC escape fractions of $f_\text{esc, LyC} = 5\%$ (the latter in agreement with our best estimates for these galaxies; \cref{sssec:Production_and_escape_of_ionising_photons}). We repeated the same calculation including the ionising photon production of all observed galaxies (discussed in more detail below) whose distance to the central LAE $r$ is less than the required bubble radius, $r < R_\text{ion}^\text{req}$. While the assumed galaxy properties are likely not entirely accurate on a source-by-source basis, this gives us a first-order approximation to the size of the ionised bubble that may be in place when assuming ages \citep{2023MNRAS.519..157W, 2023MNRAS.519.5859W} and LyC escape fractions \citep{2019ApJ...879...36F} typically expected at this epoch. We also note that the much younger inferred ages for the three main LAEs ($t_* < 5 \, \mathrm{Myr}$) may be underestimates due to the bursty nature of star formation \citep[e.g.][]{2023ApJ...949L..23S, 2024MNRAS.527.2139D}: the recent episode of strongly rising star formation we infer (\cref{sssec:Stellar_properties}) could suggest the `outshining' of underlying older stellar populations \citep[e.g.][]{2023ApJ...948..126G, 2024A&A...686A..63G}. Furthermore, the scaling between these parameters and predicted ionised bubble sizes \citep{2024A&A...682A..40W}, $R_\text{ion}^\text{pred} \propto (t_* \, f_\text{esc, LyC})^{1/3}$, means that even an order of magnitude change in the product of $t_* \times f_\text{esc, LyC}$ only leads to a factor of $2$ difference in the size of ionised bubble that is created. As will be discussed in \cref{sssec:Have_we_identified_the_sources_inflating_ionised_bubbles}, at least in the case of \JGNzeightzeroLA it will become apparent that these potentially generous assumptions will significantly fall short.

As neighbouring sources, we include spectroscopically confirmed galaxies within the JADES and FRESCO surveys (\cref{ssec:Data_sets}). Additionally, we identified a select number of photometric galaxy candidates, measuring photometric redshifts using \program{eazy} \citep{2008ApJ...686.1503B} as outlined in \citet{2024ApJ...964...71H}. Robust candidates were selected adopting similar criteria as in \citet{2024ApJ...962..124H, 2024ApJ...974...41H}, requiring the estimated redshift to be sharply peaked around the redshifts of interest. More specifically, we selected those sources whose maximum-likelihood redshift falls within $\pm 0.5$ of the redshift of a given LAE, to ensure we capture potential structures on scales of several pMpc given the typical (systematic) uncertainties of photometric redshifts (\cref{sssec:Have_we_identified_the_sources_inflating_ionised_bubbles}). Furthermore, we required the difference between the \nth{16} and \nth{84} (\nth{5} and \nth{95}) percentiles of the posterior redshift distribution to be $\Delta z_1 < 1$ ($\Delta z_2 < 2$). We removed duplicates between the spectroscopic samples from FRESCO and JADES and the photometric sample from JADES to avoid double counting. In total, we identified $41$ galaxy candidates in GOODS-N (around \JGNzeightzeroLA) and $171$ in GOODS-S (around \JGSzeightzeroLA and \JGSzeightoneLA), with UV magnitudes ranging between $-21.4 \, \mathrm{mag} \lesssim M_\text{UV} \lesssim -16.7 \, \mathrm{mag}$.

The ionising photon production rates $\dot{N}_\text{ion}$ of the main LAEs are taken to be the direct, \Hbeta-based measurements (see \cref{sssec:Production_and_escape_of_ionising_photons}). As in \citet{2024A&A...682A..40W}, we modelled the ionising spectrum for other sources as a double power law \citep[equations (7) to (9) in][]{2020MNRAS.499.1395M}, allowing us to convert the UV magnitude $M_\text{UV}$ and UV slope $\beta_\text{UV}$ measured from NIRCam photometry into an estimate of $\dot{N}_\text{ion}$. We assumed a slope of $\alpha = 2$ for the ionising continuum \citep{2023A&A...678A..68S}, which corresponds to an ionising photon production efficiency $\xi_\text{ion} \approx 10^{25.6} \, \mathrm{Hz \, erg^{-1}}$ for $\beta_\text{UV} = -2$ (reflecting the values directly measured for the main LAEs; \cref{tab:Source_properties}). All inferred bubble properties are summarised in \cref{tab:Bubble_properties}.

\subsubsection{Have we identified the sources inflating ionised bubbles?}
\label{sssec:Have_we_identified_the_sources_inflating_ionised_bubbles}

To investigate the potential impact of the broader environment on clearing the path that allows \Lya photons to travel unimpededly through the IGM, we show projections of the three-dimensional distributions of galaxies in the vicinity of the three main LAEs (both spectroscopically confirmed galaxies and photometric candidates; \cref{sssec:Predicted_ionised_bubble_sizes}) in \cref{fig:LAE_on-sky_distribution}. Specifically, we consider a redshift range of $\Delta z = 0.5$ (spanning $13$-$15 \, \mathrm{pMpc}$ along the line of sight) spaced symmetrically around the redshift of each LAE. This choice for a rather sizeable redshift range and the inclusion of sources further along the line of sight that nominally may not contribute to the visibility of \Lya in the more close-by LAEs is made on purpose, motivated by the typical uncertainties on the photometric redshifts, which furthermore tend to be systematically overestimated by up to $\Delta z \sim 0.2$ at this high-redshift regime \citep[e.g.][]{, 2024ApJ...964...71H, 2024arXiv240218543F}.

In both GOODS fields, we are able to identify a considerable number of photometric galaxy candidates within several pMpc, including a subset that have been spectroscopically confirmed in the JADES or FRESCO spectroscopy. We note that all galaxies shown here will carry some bias towards the brighter end based on their selection (photometrically or for spectroscopic follow up). In GOODS-N, the median UV magnitude of the photometric candidates is $M_\text{UV} \approx -18.9 \, \mathrm{mag}$, while in GOODS-S this is $M_\text{UV} \approx -18.2 \, \mathrm{mag}$, reflecting the different imaging depth in the two fields. \JGSzeightzeroLA appears to be located in or near a noticeable overdensity formed by several tens of galaxy candidates within $\ssim 1 \, \mathrm{pMpc}$. Two of these photometric galaxy candidates fall inside of the ionised bubble with inferred minimum radius of $R_\text{ion}^\text{req} \sim 0.3 \, \mathrm{pMpc}$, and it is conceivable some of the actual redshifts could shift with respect to the photometric redshifts such that they do contribute to its formation. At first sight, \JGSzeightoneLA on the other hand does not seem to live in a particularly rich clustering of galaxies itself, but we note this galaxy falls near the edge of the JADES footprint, complicating the identification of surrounding large-scale structures. There is however a dense structure of galaxies located several pMpc in front of it, at a similar redshift as (but spatially offset from) \JGSzeightzeroLA at $z \simeq 8.49$. Moreover, the ionised bubble size it is predicted to form by itself by leaking LyC at $f_\text{esc, LyC} = 5\%$ over $50 \, \mathrm{Myr}$ (or the equivalent thereof), $R_\text{ion, LAE}^\text{pred} \approx 0.2 \, \mathrm{pMpc}$, is already larger than the (minimum) size as inferred from the observed \Lya line, $R_\text{ion}^\text{req} \approx 0.19 \, \mathrm{pMpc}$.

Although the available JADES imaging and spectroscopy is slightly shallower in the GOODS-N field, it does show a number of galaxies ($N \approx 8$; \cref{tab:Bubble_properties}) clustered closely around \JGNzeightzeroLA. Interestingly, the FRESCO data reveal an \OIII detection at $z \approx 8.2713$ in a galaxy separated $\ssim 1\arcmin$ on the sky and $\Delta z \approx 0.0076$ in redshift (translating to $0.2 \, \mathrm{pMpc}$ along the line of sight), which places it only $0.38 \, \mathrm{pMpc}$ from \JGNzeightzeroLA. Several more are confirmed a few pMpc further along the line of sight, hinting at a marginal overdensity of galaxies, especially given the limited ability of the FRESCO spectra to confirm all but the brightest galaxies beyond $z > 8$ \citep{2024ApJ...962..124H}. However, despite our perhaps optimistic assumptions on the LyC escape fraction and stellar ages, the observed sources are not nearly sufficient to create the enormous ionised bubble ($R_\text{ion}^\text{req} \approx 3 \, \mathrm{pMpc}$) that is required to explain the high \Lya escape fraction, $f_\text{esc, \Lya} \approx 72\%$ combined with the small peak velocity offset of $\Delta v_\text{\Lya} \approx 140 \, \mathrm{km \, s^{-1}}$, which we note is a somewhat conservative estimate given the flux appearing at or very close to systemic (\cref{sssec:Comparing_LAE_properties_across_cosmic_time}). Indeed, the mismatch of the required $R_\text{ion}^\text{req}$ with the predicted $R_\text{ion, tot}^\text{pred}$ of an order of magnitude implies a discrepancy in the total ionised volume of more than $10^3$, given the scaling $V \propto R^3$.

Particularly considering the required ionised bubble size we estimate represents a lower limit (\cref{sssec:Required_ionised_bubble_sizes}), it is hard to reconcile with the predicted output of the observed galaxies even under very extreme assumptions, such as all sources leaking significant amounts of ionising photons (e.g. $f_\text{esc, LyC} = 50\%$, which would only increase the ionised volume by a factor of $10$ over our fiducial estimate with $f_\text{esc, LyC} = 5\%$). Moreover, such early large bubbles are in tension with predicted size distributions predicted from simulations, certainly in `late and rapid' reionisation scenarios where the bulk of the IGM is reionised at $z < 8$ \citep{2023MNRAS.524.6124H, 2024MNRAS.528.4872L, 2024MNRAS.531.2943N}. This is similar to the high-EW LAE at $z \sim 7$ where an ionised region of several pMpc is inferred to be in place along the line of sight \citep{2023A&A...678A..68S}, causing a comparable disparity between the expected size of the ionised bubble inferred from the observed neighbouring sources \citep{2024A&A...682A..40W}.

Instead, there could be a number of effects conspiring to explain this large discrepancy. Firstly, the assumed geometry of a spherically symmetric ionised region centred on the LAE itself may not be accurate in these extreme cases. As argued in \citet{2024A&A...682A..40W}, a scenario where the LAE is located towards the far edge of an ionised bubble (though somewhat contrived) could alleviate the required ionised volume, and hence ionising output of the sources responsible for creating it, by a factor of $\ssim 5$. Secondly, the galaxies we do observe are likely undergoing a highly stochastic mode of star formation with episodes of `mini-quenching' \citep[e.g.][]{2024Natur.629...53L, 2024MNRAS.533.1111E}. Such stochasticity would be a natural consequence of strong feedback processes expected in the regime characterised by high SFR surface densities \citep{2023A&A...677L...4P, 2024ApJ...963....9M}, as discussed in \cref{sssec:Comparing_LAE_properties_across_cosmic_time}. This means we might be biased towards seeing the `tip of the iceberg' of galaxies currently in a UV-bright phase characterised by a steeply rising star formation rate \citep[e.g.][]{2023MNRAS.525.3254S}, thereby underestimating the effective ages and potentially missing a significant fraction of galaxies that since contributing have become UV faint, which would imply a mini-quenching timescale of $\ssim 100 \, \mathrm{Myr}$ \citep[e.g.][]{2012ARA&A..50..531K}. Thirdly, even when our limiting UV magnitude approaches $M_\text{UV} = -16.5 \, \mathrm{mag}$, there is likely a significant contribution of even fainter, currently unseen galaxies \citep[e.g.][]{2021A&A...647A.107B, 2018ApJ...865L...1M, 2024A&A...682A..40W}, especially considering their high number density \citep[e.g.][]{2022ApJ...940...55B} and observational evidence that metal-poor, young star clusters are extremely efficient at producing ionising photons \citep{2020MNRAS.493.5120M, 2023A&A...678A.173V}. A detailed census of the entire galaxy population in the EoR, including the very faint end ($M_\text{UV} \gtrsim -17 \, \mathrm{mag}$) where the help of gravitational lensing is required, will be needed to disentangle these possible scenarios.

Still, the existence of considerably large ionised bubbles at early times may not be surprising in more gradual reionisation driven by faint galaxies, where we do occasionally expect to see large regions that have been ionised very early on \citep{2023MNRAS.524.6124H, 2024MNRAS.528.4872L, 2024MNRAS.531.2943N}, and accompanying LAEs to appear well before the midpoint of reionisation.

\section{Summary}
\label{sec:Summary}

We have presented the discovery of three of the most distant known LAEs ($z > 8$), identified by \jwst/NIRSpec as part of the JADES survey and follow-up observations in the JADES Origins Field. We characterise the physical properties and investigate their direct and large-scale environments to explain the production and escape mechanisms of \Lya. We summarise our main findings as follows:
\begin{itemize}
    \item All three $z > 8$ LAEs are similarly UV-bright ($M_\text{UV} \approx -20 \, \mathrm{mag}$) and exhibit small \Lya velocity offsets ($\Delta v_\text{\Lya} \lesssim 200 \, \mathrm{km \, s^{-1}}$), yet span a range of \Lya EWs: \JGSzeightzeroLA at $z \simeq 8.49$ has a relatively small $\text{EW}_\text{\Lya} \approx 15 \, \Angstrom$ \citep[below the classical threshold for strong LAEs;][]{2014ApJ...793..113P}, \JGSzeightoneLA at $z \simeq 8.72$ is in the intermediate regime with $\text{EW}_\text{\Lya} \approx 31 \, \Angstrom$, and finally \JGNzeightzeroLA at $z \simeq 8.28$ is an exceptionally strong LAE, $\text{EW}_\text{\Lya} \approx 132 \, \Angstrom$. We infer moderate \Lya escape fractions for \JGSzeightzeroLA and \JGSzeightoneLA, both $f_\text{esc, \Lya} \approx 10\%$ (assuming case B; fractions are reduced by $1.45 \times$ under case A), while \Lya escapes very efficiently in \JGNzeightzeroLA, $f_\text{esc, \Lya} \approx 72\%$, where \Lya flux near the systemic redshift further points towards little IGM attenuation.
    \item When interpreting their rest-frame UV and optical emission as a combination of stellar and nebular light, we find that they are low-mass galaxies ($M_* < 10^{8} \, \mathrm{M_\odot}$) experiencing a strong and recent upturn in star formation, reflected by high specific star formation rates ($\text{sSFR}_{10} \approx 100 \, \mathrm{Gyr^{-1}}$) that result in remarkably high EWs of several prime emission lines, $\text{EW}_\text{\CIII} \gtrsim 10 \, \Angstrom$ and $\text{EW}_\text{\OIII + \Hbeta} \approx 2000 \, \Angstrom$. The highest redshift among the three, however, \JGSzeightoneLA at $z \simeq 8.72$, shows evidence of photoionisation by an actively accreting black hole, as indicated by various diagnostics based both on UV and optical emission lines.
    \item Given their compact sizes, we find the LAEs are characterised by intensely confined star formation, particularly the high-EW LAE \JGNzeightzeroLA with SFR surface density of $\Sigma_\text{SFR} \approx 70 \, \mathrm{M_\odot \, yr^{-1} \, kpc^{-2}}$. As the LAE with the highest observed \Lya EW and escape fraction out of the three by far, this suggests stellar feedback plays a vital role in regulating the escape of \Lya as well as ionising photons from the galaxy. We further find evidence for nearby ($\ssim 1 \, \mathrm{kpc}$), low-mass ($M_* \approx 10^{7} \, \mathrm{M_\odot}$) companion galaxies, suggesting interactions could play an important role in regulating the production and escape of \Lya.
    \item As reflected by the wide range of \Lya escape fractions, we find the environments of our $z > 8$ LAEs likely represent different stages of the reionisation process. While \JGSzeightzeroLA and \JGSzeightoneLA could reasonably be solely responsible for carving out small ionised bubbles allowing \Lya to become observable, \JGNzeightzeroLA at $z \simeq 8.28$ seemingly occupies a very large ionised bubble ($R_\text{ion}^\text{req} \approx 3 \, \mathrm{pMpc}$) which is difficult to explain as having been produced even considering optimistic contributions from all directly observed neighbours. We interpret this as owing to a favourable geometry, in terms of our ability to observe \Lya, or potentially as evidence for fainter galaxies ($M_\text{UV} \gtrsim -17 \, \mathrm{mag}$) or bursty star formation likely playing an important role in concealing the actors behind reionisation.
\end{itemize}

We conclude that the population of LAEs observed at $z > 8$, which has until \jwst remained largely elusive, provides a unique window into the middle or even early stages of reionisation. Our findings tentatively point towards an important contribution by (temporarily) faint galaxies to cosmic reionisation. The powerful combination of \jwst imaging and spectroscopy proves crucial in characterising the most distant LAEs, and in the coming years will help us understand the larger and, aided by gravitational lensing, yet fainter population of galaxies and their \Lya properties, in order to conclusively reveal the sources behind reionisation.

\section*{Acknowledgements}

We thank Callum Witten, Giovanni Mazzolari, and Naveen Reddy for enlightening conversations. We further thank the anonymous referee for their helpful suggestions. This work is based on observations made with the NASA/ESA/CSA \textit{James Webb Space Telescope} (\jwst). The data were obtained from the Mikulski Archive for Space Telescopes at the Space Telescope Science Institute, which is operated by the Association of Universities for Research in Astronomy, Inc., under NASA contract NAS 5-03127 for \jwst. These observations are associated with programmes 1180, 1181, 1210, 1286, 1287, 1895, 1963, and 3215. The authors acknowledge the FRESCO team led by PI Pascal Oesch for developing their observing program with a zero-exclusive-access period. JW, RM, FDE, and JS acknowledge support from the Science and Technology Facilities Council (STFC), by the ERC through Advanced Grant 695671 ``QUENCH'', by the UKRI Frontier Research grant RISEandFALL. JW also gratefully acknowledges support from the Cosmic Dawn Center through the DAWN Fellowship. The Cosmic Dawn Center (DAWN) is funded by the Danish National Research Foundation under grant No. 140. RM also acknowledges support by the STFC and funding from a research professorship from the Royal Society. RS acknowledges support from a STFC Ernest Rutherford Fellowship (ST/S004831/1). GCJ, AJB, AS, AJC, and JC acknowledge funding from the `FirstGalaxies' Advanced Grant from the ERC under the European Union's Horizon 2020 research and innovation programme (Grant agreement No. 789056). JMH, BDJ, PAC, DJE, MR, BER, and CNAW acknowledge a \jwst/NIRCam contract to the University of Arizona (NAS5-02015). ST acknowledges support by the Royal Society Research Grant G125142. SA acknowledges support from Grant PID2021-127718NB-I00 funded by the Spanish Ministry of Science and Innovation/State Agency of Research (MICIN/AEI/10.13039/501100011033). KB is supported by the Australian Research Council Centre of Excellence for All Sky Astrophysics in 3 Dimensions (ASTRO 3D), through project number CE170100013. SC acknowledges support by European Union's HE ERC Starting Grant 101040227, `WINGS'. ECL acknowledges support of an STFC Webb Fellowship (ST/W001438/1). DJE is supported as a Simons Investigator. Funding for this research was provided by the Johns Hopkins University, Institute for Data Intensive Engineering and Science (IDIES). IL is supported by the National Science Foundation Graduate Research Fellowship under Grant No. 2137424. BER also acknowledges support from \jwst Program 3215. The research of CCW is supported by NOIRLab, which is managed by the Association of Universities for Research in Astronomy (AURA) under a cooperative agreement with the National Science Foundation. This work has also relied on the following \program{python} packages: the \program{SciPy} library \citep{Jones2001}, its packages \program{NumPy} \citep{2011CSE....13b..22V} and \program{Matplotlib} \citep{Hunter2007}, the \program{Astropy} package \citep{2013A&A...558A..33A, 2018AJ....156..123A}, and the \program{pymultinest} package \citep{2009MNRAS.398.1601F, 2014A&A...564A.125B}.

\section*{Data Availability}
 
Reduced data underlying this article will be shared on reasonable request to the corresponding author.



\bibliographystyle{mnras}
\bibliography{Distant_LAEs}

\begin{thebibliography}{}
\makeatletter
\relax
\def\mn@urlcharsother{\let\do\@makeother \do\$\do\&\do\#\do\^\do\_\do\%\do\~}
\def\mn@doi{\begingroup\mn@urlcharsother \@ifnextchar [ {\mn@doi@} {\mn@doi@[]}}
\def\mn@doi@[#1]#2{\def\@tempa{#1}\ifx\@tempa\@empty \href {http://dx.doi.org/#2} {doi:#2}\else \href {http://dx.doi.org/#2} {#1}\fi \endgroup}
\def\mn@eprint#1#2{\mn@eprint@#1:#2::\@nil}
\def\mn@eprint@arXiv#1{\href {http://arxiv.org/abs/#1} {{\tt arXiv:#1}}}
\def\mn@eprint@dblp#1{\href {http://dblp.uni-trier.de/rec/bibtex/#1.xml} {dblp:#1}}
\def\mn@eprint@#1:#2:#3:#4\@nil{\def\@tempa {#1}\def\@tempb {#2}\def\@tempc {#3}\ifx \@tempc \@empty \let \@tempc \@tempb \let \@tempb \@tempa \fi \ifx \@tempb \@empty \def\@tempb {arXiv}\fi \@ifundefined {mn@eprint@\@tempb}{\@tempb:\@tempc}{\expandafter \expandafter \csname mn@eprint@\@tempb\endcsname \expandafter{\@tempc}}}

\bibitem[\protect\citeauthoryear{{Adamo} et~al.,}{{Adamo} et~al.}{2024}]{2024Natur.632..513A}
{Adamo} A.,  et~al., 2024, \mn@doi [\nat] {10.1038/s41586-024-07703-7}, \href {https://ui.adsabs.harvard.edu/abs/2024Natur.632..513A} {632, 513}

\bibitem[\protect\citeauthoryear{{Aggarwal} \& {Keenan}}{{Aggarwal} \& {Keenan}}{1999}]{1999ApJS..123..311A}
{Aggarwal} K.~M.,  {Keenan} F.~P.,  1999, \mn@doi [\apjs] {10.1086/313232}, \href {https://ui.adsabs.harvard.edu/abs/1999ApJS..123..311A} {123, 311}

\bibitem[\protect\citeauthoryear{{Asplund}, {Grevesse}, {Sauval}  \& {Scott}}{{Asplund} et~al.}{2009}]{2009ARA&A..47..481A}
{Asplund} M.,  {Grevesse} N.,  {Sauval} A.~J.,   {Scott} P.,  2009, \mn@doi [\araa] {10.1146/annurev.astro.46.060407.145222}, \href {https://ui.adsabs.harvard.edu/abs/2009ARA&A..47..481A} {47, 481}

\bibitem[\protect\citeauthoryear{{Astropy Collaboration} et~al.,}{{Astropy Collaboration} et~al.}{2013}]{2013A&A...558A..33A}
{Astropy Collaboration} et~al., 2013, \mn@doi [\aap] {10.1051/0004-6361/201322068}, \href {http://adsabs.harvard.edu/abs/2013A%26A...558A..33A} {558, A33}

\bibitem[\protect\citeauthoryear{{Astropy Collaboration} et~al.,}{{Astropy Collaboration} et~al.}{2018}]{2018AJ....156..123A}
{Astropy Collaboration} et~al., 2018, \mn@doi [\aj] {10.3847/1538-3881/aabc4f}, \href {http://adsabs.harvard.edu/abs/2018AJ....156..123A} {156, 123}

\bibitem[\protect\citeauthoryear{{Atek} et~al.,}{{Atek} et~al.}{2024}]{2024Natur.626..975A}
{Atek} H.,  et~al., 2024, \mn@doi [\nat] {10.1038/s41586-024-07043-6}, \href {https://ui.adsabs.harvard.edu/abs/2024Natur.626..975A} {626, 975}

\bibitem[\protect\citeauthoryear{{Azzalini} \& {Capitanio}}{{Azzalini} \& {Capitanio}}{2009}]{2009arXiv0911.2093A}
{Azzalini} A.,  {Capitanio} A.,  2009, preprint, \href {https://ui.adsabs.harvard.edu/abs/2009arXiv0911.2093A} {} (\mn@eprint {arXiv} {0911.2093})

\bibitem[\protect\citeauthoryear{{Bacon} et~al.,}{{Bacon} et~al.}{2021}]{2021A&A...647A.107B}
{Bacon} R.,  et~al., 2021, \mn@doi [\aap] {10.1051/0004-6361/202039887}, \href {https://ui.adsabs.harvard.edu/abs/2021A&A...647A.107B} {647, A107}

\bibitem[\protect\citeauthoryear{{Baker} et~al.,}{{Baker} et~al.}{2024}]{2024NatAs.tmp..246B}
{Baker} W.~M.,  et~al., 2024, \mn@doi [Nature Astronomy] {10.1038/s41550-024-02384-8}, \href {https://ui.adsabs.harvard.edu/abs/2024NatAs.tmp..246B} {}

\bibitem[\protect\citeauthoryear{{Baldwin}, {Phillips}  \& {Terlevich}}{{Baldwin} et~al.}{1981}]{1981PASP...93....5B}
{Baldwin} J.~A.,  {Phillips} M.~M.,   {Terlevich} R.,  1981, \mn@doi [\pasp] {10.1086/130766}, \href {https://ui.adsabs.harvard.edu/abs/1981PASP...93....5B} {93, 5}

\bibitem[\protect\citeauthoryear{{Becker}, {Bolton}, {Madau}, {Pettini}, {Ryan-Weber}  \& {Venemans}}{{Becker} et~al.}{2015}]{2015MNRAS.447.3402B}
{Becker} G.~D.,  {Bolton} J.~S.,  {Madau} P.,  {Pettini} M.,  {Ryan-Weber} E.~V.,   {Venemans} B.~P.,  2015, \mn@doi [\mnras] {10.1093/mnras/stu2646}, \href {https://ui.adsabs.harvard.edu/abs/2015MNRAS.447.3402B} {447, 3402}

\bibitem[\protect\citeauthoryear{{Becker}, {Davies}, {Furlanetto}, {Malkan}, {Boera}  \& {Douglass}}{{Becker} et~al.}{2018}]{2018ApJ...863...92B}
{Becker} G.~D.,  {Davies} F.~B.,  {Furlanetto} S.~R.,  {Malkan} M.~A.,  {Boera} E.,   {Douglass} C.,  2018, \mn@doi [\apj] {10.3847/1538-4357/aacc73}, \href {https://ui.adsabs.harvard.edu/abs/2018ApJ...863...92B} {863, 92}

\bibitem[\protect\citeauthoryear{{Beckwith} et~al.,}{{Beckwith} et~al.}{2006}]{2006AJ....132.1729B}
{Beckwith} S. V.~W.,  et~al., 2006, \mn@doi [\aj] {10.1086/507302}, \href {https://ui.adsabs.harvard.edu/abs/2006AJ....132.1729B} {132, 1729}

\bibitem[\protect\citeauthoryear{{Berg}, {Erb}, {Henry}, {Skillman}  \& {McQuinn}}{{Berg} et~al.}{2019}]{2019ApJ...874...93B}
{Berg} D.~A.,  {Erb} D.~K.,  {Henry} R. B.~C.,  {Skillman} E.~D.,   {McQuinn} K. B.~W.,  2019, \mn@doi [\apj] {10.3847/1538-4357/ab020a}, \href {https://ui.adsabs.harvard.edu/abs/2019ApJ...874...93B} {874, 93}

\bibitem[\protect\citeauthoryear{{B{\"o}ker} et~al.,}{{B{\"o}ker} et~al.}{2023}]{2023PASP..135c8001B}
{B{\"o}ker} T.,  et~al., 2023, \mn@doi [\pasp] {10.1088/1538-3873/acb846}, \href {https://ui.adsabs.harvard.edu/abs/2023PASP..135c8001B} {135, 038001}

\bibitem[\protect\citeauthoryear{{Bosman} et~al.,}{{Bosman} et~al.}{2022}]{2022MNRAS.514...55B}
{Bosman} S. E.~I.,  et~al., 2022, \mn@doi [\mnras] {10.1093/mnras/stac1046}, \href {https://ui.adsabs.harvard.edu/abs/2022MNRAS.514...55B} {514, 55}

\bibitem[\protect\citeauthoryear{{Bouwens} et~al.,}{{Bouwens} et~al.}{2010}]{2010ApJ...709L.133B}
{Bouwens} R.~J.,  et~al., 2010, \mn@doi [\apjl] {10.1088/2041-8205/709/2/L133}, \href {https://ui.adsabs.harvard.edu/abs/2010ApJ...709L.133B} {709, L133}

\bibitem[\protect\citeauthoryear{{Bouwens}, {Illingworth}, {Ellis}, {Oesch}  \& {Stefanon}}{{Bouwens} et~al.}{2022}]{2022ApJ...940...55B}
{Bouwens} R.~J.,  {Illingworth} G.,  {Ellis} R.~S.,  {Oesch} P.,   {Stefanon} M.,  2022, \mn@doi [\apj] {10.3847/1538-4357/ac86d1}, \href {https://ui.adsabs.harvard.edu/abs/2022ApJ...940...55B} {940, 55}

\bibitem[\protect\citeauthoryear{{Boyett} et~al.,}{{Boyett} et~al.}{2024}]{2024arXiv240116934B}
{Boyett} K.,  et~al., 2024, preprint, \href {https://ui.adsabs.harvard.edu/abs/2024arXiv240116934B} {} (\mn@eprint {arXiv} {2401.16934})

\bibitem[\protect\citeauthoryear{{Brammer}, {van Dokkum}  \& {Coppi}}{{Brammer} et~al.}{2008}]{2008ApJ...686.1503B}
{Brammer} G.~B.,  {van Dokkum} P.~G.,   {Coppi} P.,  2008, \mn@doi [\apj] {10.1086/591786}, \href {https://ui.adsabs.harvard.edu/abs/2008ApJ...686.1503B} {686, 1503}

\bibitem[\protect\citeauthoryear{{Brinchmann}}{{Brinchmann}}{2023}]{2023MNRAS.525.2087B}
{Brinchmann} J.,  2023, \mn@doi [\mnras] {10.1093/mnras/stad1704}, \href {https://ui.adsabs.harvard.edu/abs/2023MNRAS.525.2087B} {525, 2087}

\bibitem[\protect\citeauthoryear{{Bruzual} \& {Charlot}}{{Bruzual} \& {Charlot}}{2003}]{2003MNRAS.344.1000B}
{Bruzual} G.,  {Charlot} S.,  2003, \mn@doi [\mnras] {10.1046/j.1365-8711.2003.06897.x}, \href {https://ui.adsabs.harvard.edu/abs/2003MNRAS.344.1000B} {344, 1000}

\bibitem[\protect\citeauthoryear{{Buchner} et~al.,}{{Buchner} et~al.}{2014}]{2014A&A...564A.125B}
{Buchner} J.,  et~al., 2014, \mn@doi [\aap] {10.1051/0004-6361/201322971}, \href {https://ui.adsabs.harvard.edu/abs/2014A&A...564A.125B} {564, A125}

\bibitem[\protect\citeauthoryear{{Bunker} et~al.,}{{Bunker} et~al.}{2010}]{2010MNRAS.409..855B}
{Bunker} A.~J.,  et~al., 2010, \mn@doi [\mnras] {10.1111/j.1365-2966.2010.17350.x}, \href {https://ui.adsabs.harvard.edu/abs/2010MNRAS.409..855B} {409, 855}

\bibitem[\protect\citeauthoryear{{Bunker}, {Caruana}, {Wilkins}, {Stanway}, {Lorenzoni}, {Lacy}, {Jarvis}  \& {Hickey}}{{Bunker} et~al.}{2013}]{2013MNRAS.430.3314B}
{Bunker} A.~J.,  {Caruana} J.,  {Wilkins} S.~M.,  {Stanway} E.~R.,  {Lorenzoni} S.,  {Lacy} M.,  {Jarvis} M.~J.,   {Hickey} S.,  2013, \mn@doi [\mnras] {10.1093/mnras/stt132}, \href {https://ui.adsabs.harvard.edu/abs/2013MNRAS.430.3314B} {430, 3314}

\bibitem[\protect\citeauthoryear{{Bunker} et~al.,}{{Bunker} et~al.}{2023}]{2023A&A...677A..88B}
{Bunker} A.~J.,  et~al., 2023, \mn@doi [\aap] {10.1051/0004-6361/202346159}, \href {https://ui.adsabs.harvard.edu/abs/2023A&A...677A..88B} {677, A88}

\bibitem[\protect\citeauthoryear{{Bunker} et~al.,}{{Bunker} et~al.}{2024}]{2024A&A...690A.288B}
{Bunker} A.~J.,  et~al., 2024, \mn@doi [\aap] {10.1051/0004-6361/202347094}, \href {https://ui.adsabs.harvard.edu/abs/2024A&A...690A.288B} {690, A288}

\bibitem[\protect\citeauthoryear{{Calabro} et~al.,}{{Calabro} et~al.}{2024}]{2024arXiv240312683C}
{Calabro} A.,  et~al., 2024, preprint, \href {https://ui.adsabs.harvard.edu/abs/2024arXiv240312683C} {} (\mn@eprint {arXiv} {2403.12683})

\bibitem[\protect\citeauthoryear{{Calzetti}, {Kinney}  \& {Storchi-Bergmann}}{{Calzetti} et~al.}{1994}]{1994ApJ...429..582C}
{Calzetti} D.,  {Kinney} A.~L.,   {Storchi-Bergmann} T.,  1994, \mn@doi [\apj] {10.1086/174346}, \href {https://ui.adsabs.harvard.edu/abs/1994ApJ...429..582C} {429, 582}

\bibitem[\protect\citeauthoryear{{Cameron}, {Katz}  \& {Rey}}{{Cameron} et~al.}{2023a}]{2023MNRAS.522L..89C}
{Cameron} A.~J.,  {Katz} H.,   {Rey} M.~P.,  2023a, \mn@doi [\mnras] {10.1093/mnrasl/slad046}, \href {https://ui.adsabs.harvard.edu/abs/2023MNRAS.522L..89C} {522, L89}

\bibitem[\protect\citeauthoryear{{Cameron} et~al.,}{{Cameron} et~al.}{2023b}]{2023A&A...677A.115C}
{Cameron} A.~J.,  et~al., 2023b, \mn@doi [\aap] {10.1051/0004-6361/202346107}, \href {https://ui.adsabs.harvard.edu/abs/2023A&A...677A.115C} {677, A115}

\bibitem[\protect\citeauthoryear{{Cameron}, {Katz}, {Witten}, {Saxena}, {Laporte}  \& {Bunker}}{{Cameron} et~al.}{2024}]{2024MNRAS.534..523C}
{Cameron} A.~J.,  {Katz} H.,  {Witten} C.,  {Saxena} A.,  {Laporte} N.,   {Bunker} A.~J.,  2024, \mn@doi [\mnras] {10.1093/mnras/stae1547}, \href {https://ui.adsabs.harvard.edu/abs/2024MNRAS.534..523C} {534, 523}

\bibitem[\protect\citeauthoryear{{Carnall}, {McLure}, {Dunlop}  \& {Dav{\'e}}}{{Carnall} et~al.}{2018}]{2018MNRAS.480.4379C}
{Carnall} A.~C.,  {McLure} R.~J.,  {Dunlop} J.~S.,   {Dav{\'e}} R.,  2018, \mn@doi [\mnras] {10.1093/mnras/sty2169}, \href {https://ui.adsabs.harvard.edu/abs/2018MNRAS.480.4379C} {480, 4379}

\bibitem[\protect\citeauthoryear{{Carnall} et~al.,}{{Carnall} et~al.}{2019a}]{2019MNRAS.490..417C}
{Carnall} A.~C.,  et~al., 2019a, \mn@doi [\mnras] {10.1093/mnras/stz2544}, \href {https://ui.adsabs.harvard.edu/abs/2019MNRAS.490..417C} {490, 417}

\bibitem[\protect\citeauthoryear{{Carnall}, {Leja}, {Johnson}, {McLure}, {Dunlop}  \& {Conroy}}{{Carnall} et~al.}{2019b}]{2019ApJ...873...44C}
{Carnall} A.~C.,  {Leja} J.,  {Johnson} B.~D.,  {McLure} R.~J.,  {Dunlop} J.~S.,   {Conroy} C.,  2019b, \mn@doi [\apj] {10.3847/1538-4357/ab04a2}, \href {https://ui.adsabs.harvard.edu/abs/2019ApJ...873...44C} {873, 44}

\bibitem[\protect\citeauthoryear{{Castellano} et~al.,}{{Castellano} et~al.}{2024}]{2024ApJ...972..143C}
{Castellano} M.,  et~al., 2024, \mn@doi [\apj] {10.3847/1538-4357/ad5f88}, \href {https://ui.adsabs.harvard.edu/abs/2024ApJ...972..143C} {972, 143}

\bibitem[\protect\citeauthoryear{{Chabrier}}{{Chabrier}}{2003}]{2003PASP..115..763C}
{Chabrier} G.,  2003, \mn@doi [\pasp] {10.1086/376392}, \href {https://ui.adsabs.harvard.edu/abs/2003PASP..115..763C} {115, 763}

\bibitem[\protect\citeauthoryear{{Charlot} \& {Fall}}{{Charlot} \& {Fall}}{2000}]{2000ApJ...539..718C}
{Charlot} S.,  {Fall} S.~M.,  2000, \mn@doi [\apj] {10.1086/309250}, \href {https://ui.adsabs.harvard.edu/abs/2000ApJ...539..718C} {539, 718}

\bibitem[\protect\citeauthoryear{{Chatzikos} et~al.,}{{Chatzikos} et~al.}{2023}]{2023RMxAA..59..327C}
{Chatzikos} M.,  et~al., 2023, \mn@doi [\rmxaa] {10.22201/ia.01851101p.2023.59.02.12}, \href {https://ui.adsabs.harvard.edu/abs/2023RMxAA..59..327C} {59, 327}

\bibitem[\protect\citeauthoryear{{Chen}, {Stark}, {Mason}, {Topping}, {Whitler}, {Tang}, {Endsley}  \& {Charlot}}{{Chen} et~al.}{2024}]{2024MNRAS.528.7052C}
{Chen} Z.,  {Stark} D.~P.,  {Mason} C.,  {Topping} M.~W.,  {Whitler} L.,  {Tang} M.,  {Endsley} R.,   {Charlot} S.,  2024, \mn@doi [\mnras] {10.1093/mnras/stae455}, \href {https://ui.adsabs.harvard.edu/abs/2024MNRAS.528.7052C} {528, 7052}

\bibitem[\protect\citeauthoryear{{Chevallard} et~al.,}{{Chevallard} et~al.}{2019}]{2019MNRAS.483.2621C}
{Chevallard} J.,  et~al., 2019, \mn@doi [\mnras] {10.1093/mnras/sty2426}, \href {https://ui.adsabs.harvard.edu/abs/2019MNRAS.483.2621C} {483, 2621}

\bibitem[\protect\citeauthoryear{{Chisholm} et~al.,}{{Chisholm} et~al.}{2022}]{2022MNRAS.517.5104C}
{Chisholm} J.,  et~al., 2022, \mn@doi [\mnras] {10.1093/mnras/stac2874}, \href {https://ui.adsabs.harvard.edu/abs/2022MNRAS.517.5104C} {517, 5104}

\bibitem[\protect\citeauthoryear{{Choustikov} et~al.,}{{Choustikov} et~al.}{2024}]{2024MNRAS.529.3751C}
{Choustikov} N.,  et~al., 2024, \mn@doi [\mnras] {10.1093/mnras/stae776}, \href {https://ui.adsabs.harvard.edu/abs/2024MNRAS.529.3751C} {529, 3751}

\bibitem[\protect\citeauthoryear{{Claeyssens} et~al.,}{{Claeyssens} et~al.}{2022}]{2022A&A...666A..78C}
{Claeyssens} A.,  et~al., 2022, \mn@doi [\aap] {10.1051/0004-6361/202142320}, \href {https://ui.adsabs.harvard.edu/abs/2022A&A...666A..78C} {666, A78}

\bibitem[\protect\citeauthoryear{{Curti}, {Cresci}, {Mannucci}, {Marconi}, {Maiolino}  \& {Esposito}}{{Curti} et~al.}{2017}]{2017MNRAS.465.1384C}
{Curti} M.,  {Cresci} G.,  {Mannucci} F.,  {Marconi} A.,  {Maiolino} R.,   {Esposito} S.,  2017, \mn@doi [\mnras] {10.1093/mnras/stw2766}, \href {https://ui.adsabs.harvard.edu/abs/2017MNRAS.465.1384C} {465, 1384}

\bibitem[\protect\citeauthoryear{{Curti}, {Mannucci}, {Cresci}  \& {Maiolino}}{{Curti} et~al.}{2020}]{2020MNRAS.491..944C}
{Curti} M.,  {Mannucci} F.,  {Cresci} G.,   {Maiolino} R.,  2020, \mn@doi [\mnras] {10.1093/mnras/stz2910}, \href {https://ui.adsabs.harvard.edu/abs/2020MNRAS.491..944C} {491, 944}

\bibitem[\protect\citeauthoryear{{Curti} et~al.,}{{Curti} et~al.}{2023}]{2023MNRAS.518..425C}
{Curti} M.,  et~al., 2023, \mn@doi [\mnras] {10.1093/mnras/stac2737}, \href {https://ui.adsabs.harvard.edu/abs/2023MNRAS.518..425C} {518, 425}

\bibitem[\protect\citeauthoryear{{D'Eugenio} et~al.,}{{D'Eugenio} et~al.}{2024a}]{2024arXiv240406531D}
{D'Eugenio} F.,  et~al., 2024a, preprint, \href {https://ui.adsabs.harvard.edu/abs/2024arXiv240406531D} {} (\mn@eprint {arXiv} {2404.06531})

\bibitem[\protect\citeauthoryear{{D'Eugenio} et~al.,}{{D'Eugenio} et~al.}{2024b}]{2024A&A...689A.152D}
{D'Eugenio} F.,  et~al., 2024b, \mn@doi [\aap] {10.1051/0004-6361/202348636}, \href {https://ui.adsabs.harvard.edu/abs/2024A&A...689A.152D} {689, A152}

\bibitem[\protect\citeauthoryear{{Dayal} \& {Ferrara}}{{Dayal} \& {Ferrara}}{2018}]{2018PhR...780....1D}
{Dayal} P.,  {Ferrara} A.,  2018, \mn@doi [\physrep] {10.1016/j.physrep.2018.10.002}, \href {https://ui.adsabs.harvard.edu/abs/2018PhR...780....1D} {780, 1}

\bibitem[\protect\citeauthoryear{{Dijkstra}}{{Dijkstra}}{2014}]{2014PASA...31...40D}
{Dijkstra} M.,  2014, \mn@doi [\pasa] {10.1017/pasa.2014.33}, \href {https://ui.adsabs.harvard.edu/abs/2014PASA...31...40D} {31, e040}

\bibitem[\protect\citeauthoryear{{Dome}, {Tacchella}, {Fialkov}, {Ceverino}, {Dekel}, {Ginzburg}, {Lapiner}  \& {Looser}}{{Dome} et~al.}{2024}]{2024MNRAS.527.2139D}
{Dome} T.,  {Tacchella} S.,  {Fialkov} A.,  {Ceverino} D.,  {Dekel} A.,  {Ginzburg} O.,  {Lapiner} S.,   {Looser} T.~J.,  2024, \mn@doi [\mnras] {10.1093/mnras/stad3239}, \href {https://ui.adsabs.harvard.edu/abs/2024MNRAS.527.2139D} {527, 2139}

\bibitem[\protect\citeauthoryear{{Eisenhauer} et~al.,}{{Eisenhauer} et~al.}{2003}]{2003SPIE.4841.1548E}
{Eisenhauer} F.,  et~al., 2003, in {Iye} M.,  {Moorwood} A. F.~M.,  eds,  Society of Photo-Optical Instrumentation Engineers (SPIE) Conference Series Vol. 4841, Instrument Design and Performance for Optical/Infrared Ground-based Telescopes. pp 1548--1561 (\mn@eprint {arXiv} {astro-ph/0306191}), \mn@doi{10.1117/12.459468}

\bibitem[\protect\citeauthoryear{{Eisenstein} et~al.,}{{Eisenstein} et~al.}{2023a}]{2023arXiv230602465E}
{Eisenstein} D.~J.,  et~al., 2023a, preprint, \href {https://ui.adsabs.harvard.edu/abs/2023arXiv230602465E} {} (\mn@eprint {arXiv} {2306.02465})

\bibitem[\protect\citeauthoryear{{Eisenstein} et~al.,}{{Eisenstein} et~al.}{2023b}]{2023arXiv231012340E}
{Eisenstein} D.~J.,  et~al., 2023b, preprint, \href {https://ui.adsabs.harvard.edu/abs/2023arXiv231012340E} {} (\mn@eprint {arXiv} {2310.12340})

\bibitem[\protect\citeauthoryear{{Eldridge}, {Stanway}, {Xiao}, {McClelland}, {Taylor}, {Ng}, {Greis}  \& {Bray}}{{Eldridge} et~al.}{2017}]{2017PASA...34...58E}
{Eldridge} J.~J.,  {Stanway} E.~R.,  {Xiao} L.,  {McClelland} L.~A.~S.,  {Taylor} G.,  {Ng} M.,  {Greis} S.~M.~L.,   {Bray} J.~C.,  2017, \mn@doi [\pasa] {10.1017/pasa.2017.51}, \href {https://ui.adsabs.harvard.edu/abs/2017PASA...34...58E} {34, e058}

\bibitem[\protect\citeauthoryear{{Endsley} \& {Stark}}{{Endsley} \& {Stark}}{2022}]{2022MNRAS.511.6042E}
{Endsley} R.,  {Stark} D.~P.,  2022, \mn@doi [\mnras] {10.1093/mnras/stac524}, \href {https://ui.adsabs.harvard.edu/abs/2022MNRAS.511.6042E} {511, 6042}

\bibitem[\protect\citeauthoryear{{Endsley}, {Stark}, {Charlot}, {Chevallard}, {Robertson}, {Bouwens}  \& {Stefanon}}{{Endsley} et~al.}{2021}]{2021MNRAS.502.6044E}
{Endsley} R.,  {Stark} D.~P.,  {Charlot} S.,  {Chevallard} J.,  {Robertson} B.,  {Bouwens} R.~J.,   {Stefanon} M.,  2021, \mn@doi [\mnras] {10.1093/mnras/stab432}, \href {https://ui.adsabs.harvard.edu/abs/2021MNRAS.502.6044E} {502, 6044}

\bibitem[\protect\citeauthoryear{{Endsley} et~al.,}{{Endsley} et~al.}{2022}]{2022MNRAS.517.5642E}
{Endsley} R.,  et~al., 2022, \mn@doi [\mnras] {10.1093/mnras/stac3064}, \href {https://ui.adsabs.harvard.edu/abs/2022MNRAS.517.5642E} {517, 5642}

\bibitem[\protect\citeauthoryear{{Endsley} et~al.,}{{Endsley} et~al.}{2024}]{2024MNRAS.533.1111E}
{Endsley} R.,  et~al., 2024, \mn@doi [\mnras] {10.1093/mnras/stae1857}, \href {https://ui.adsabs.harvard.edu/abs/2024MNRAS.533.1111E} {533, 1111}

\bibitem[\protect\citeauthoryear{{Fan}, {Ba{\~n}ados}  \& {Simcoe}}{{Fan} et~al.}{2023}]{2023ARA&A..61..373F}
{Fan} X.,  {Ba{\~n}ados} E.,   {Simcoe} R.~A.,  2023, \mn@doi [\araa] {10.1146/annurev-astro-052920-102455}, \href {https://ui.adsabs.harvard.edu/abs/2023ARA&A..61..373F} {61, 373}

\bibitem[\protect\citeauthoryear{{Feltre}, {Charlot}  \& {Gutkin}}{{Feltre} et~al.}{2016}]{2016MNRAS.456.3354F}
{Feltre} A.,  {Charlot} S.,   {Gutkin} J.,  2016, \mn@doi [\mnras] {10.1093/mnras/stv2794}, \href {https://ui.adsabs.harvard.edu/abs/2016MNRAS.456.3354F} {456, 3354}

\bibitem[\protect\citeauthoryear{{Ferland} et~al.,}{{Ferland} et~al.}{2017}]{2017RMxAA..53..385F}
{Ferland} G.~J.,  et~al., 2017, \mn@doi [\rmxaa] {10.48550/arXiv.1705.10877}, \href {https://ui.adsabs.harvard.edu/abs/2017RMxAA..53..385F} {53, 385}

\bibitem[\protect\citeauthoryear{{Feroz}, {Hobson}  \& {Bridges}}{{Feroz} et~al.}{2009}]{2009MNRAS.398.1601F}
{Feroz} F.,  {Hobson} M.~P.,   {Bridges} M.,  2009, \mn@doi [\mnras] {10.1111/j.1365-2966.2009.14548.x}, \href {https://ui.adsabs.harvard.edu/abs/2009MNRAS.398.1601F} {398, 1601}

\bibitem[\protect\citeauthoryear{{Ferrara}}{{Ferrara}}{2024}]{2024A&A...684A.207F}
{Ferrara} A.,  2024, \mn@doi [\aap] {10.1051/0004-6361/202348321}, \href {https://ui.adsabs.harvard.edu/abs/2024A&A...684A.207F} {684, A207}

\bibitem[\protect\citeauthoryear{{Ferruit} et~al.,}{{Ferruit} et~al.}{2022}]{2022A&A...661A..81F}
{Ferruit} P.,  et~al., 2022, \mn@doi [\aap] {10.1051/0004-6361/202142673}, \href {https://ui.adsabs.harvard.edu/abs/2022A&A...661A..81F} {661, A81}

\bibitem[\protect\citeauthoryear{{Finkelstein} et~al.,}{{Finkelstein} et~al.}{2013}]{2013Natur.502..524F}
{Finkelstein} S.~L.,  et~al., 2013, \mn@doi [\nat] {10.1038/nature12657}, \href {https://ui.adsabs.harvard.edu/abs/2013Natur.502..524F} {502, 524}

\bibitem[\protect\citeauthoryear{{Finkelstein} et~al.,}{{Finkelstein} et~al.}{2019}]{2019ApJ...879...36F}
{Finkelstein} S.~L.,  et~al., 2019, \mn@doi [\apj] {10.3847/1538-4357/ab1ea8}, \href {https://ui.adsabs.harvard.edu/abs/2019ApJ...879...36F} {879, 36}

\bibitem[\protect\citeauthoryear{{Finkelstein} et~al.,}{{Finkelstein} et~al.}{2023}]{2023ApJ...946L..13F}
{Finkelstein} S.~L.,  et~al., 2023, \mn@doi [\apjl] {10.3847/2041-8213/acade4}, \href {https://ui.adsabs.harvard.edu/abs/2023ApJ...946L..13F} {946, L13}

\bibitem[\protect\citeauthoryear{{Flury} et~al.,}{{Flury} et~al.}{2022}]{2022ApJS..260....1F}
{Flury} S.~R.,  et~al., 2022, \mn@doi [\apjs] {10.3847/1538-4365/ac5331}, \href {https://ui.adsabs.harvard.edu/abs/2022ApJS..260....1F} {260, 1}

\bibitem[\protect\citeauthoryear{{Fujimoto} et~al.,}{{Fujimoto} et~al.}{2023}]{2023arXiv230811609F}
{Fujimoto} S.,  et~al., 2023, preprint, \href {https://ui.adsabs.harvard.edu/abs/2023arXiv230811609F} {} (\mn@eprint {arXiv} {2308.11609})

\bibitem[\protect\citeauthoryear{{Fujimoto} et~al.,}{{Fujimoto} et~al.}{2024}]{2024arXiv240218543F}
{Fujimoto} S.,  et~al., 2024, preprint, \href {https://ui.adsabs.harvard.edu/abs/2024arXiv240218543F} {} (\mn@eprint {arXiv} {2402.18543})

\bibitem[\protect\citeauthoryear{{Furlanetto}, {Hernquist}  \& {Zaldarriaga}}{{Furlanetto} et~al.}{2004}]{2004MNRAS.354..695F}
{Furlanetto} S.~R.,  {Hernquist} L.,   {Zaldarriaga} M.,  2004, \mn@doi [\mnras] {10.1111/j.1365-2966.2004.08225.x}, \href {https://ui.adsabs.harvard.edu/abs/2004MNRAS.354..695F} {354, 695}

\bibitem[\protect\citeauthoryear{{Gardner} et~al.,}{{Gardner} et~al.}{2023}]{2023PASP..135f8001G}
{Gardner} J.~P.,  et~al., 2023, \mn@doi [\pasp] {10.1088/1538-3873/acd1b5}, \href {https://ui.adsabs.harvard.edu/abs/2023PASP..135f8001G} {135, 068001}

\bibitem[\protect\citeauthoryear{{Giavalisco} et~al.,}{{Giavalisco} et~al.}{2004}]{2004ApJ...600L..93G}
{Giavalisco} M.,  et~al., 2004, \mn@doi [\apjl] {10.1086/379232}, \href {https://ui.adsabs.harvard.edu/abs/2004ApJ...600L..93G} {600, L93}

\bibitem[\protect\citeauthoryear{{Gim{\'e}nez-Arteaga} et~al.,}{{Gim{\'e}nez-Arteaga} et~al.}{2023}]{2023ApJ...948..126G}
{Gim{\'e}nez-Arteaga} C.,  et~al., 2023, \mn@doi [\apj] {10.3847/1538-4357/acc5ea}, \href {https://ui.adsabs.harvard.edu/abs/2023ApJ...948..126G} {948, 126}

\bibitem[\protect\citeauthoryear{{Gim{\'e}nez-Arteaga} et~al.,}{{Gim{\'e}nez-Arteaga} et~al.}{2024}]{2024A&A...686A..63G}
{Gim{\'e}nez-Arteaga} C.,  et~al., 2024, \mn@doi [\aap] {10.1051/0004-6361/202349135}, \href {https://ui.adsabs.harvard.edu/abs/2024A&A...686A..63G} {686, A63}

\bibitem[\protect\citeauthoryear{{Gordon}, {Clayton}, {Misselt}, {Landolt}  \& {Wolff}}{{Gordon} et~al.}{2003}]{2003ApJ...594..279G}
{Gordon} K.~D.,  {Clayton} G.~C.,  {Misselt} K.~A.,  {Landolt} A.~U.,   {Wolff} M.~J.,  2003, \mn@doi [\apj] {10.1086/376774}, \href {https://ui.adsabs.harvard.edu/abs/2003ApJ...594..279G} {594, 279}

\bibitem[\protect\citeauthoryear{{\VAN{Graaff}{De}{de} Graaff} et~al.,}{{\VAN{Graaff}{De}{de} Graaff} et~al.}{2024}]{2024A&A...684A..87D}
{\VAN{Graaff}{De}{de} Graaff} A.,  et~al., 2024, \mn@doi [\aap] {10.1051/0004-6361/202347755}, \href {https://ui.adsabs.harvard.edu/abs/2024A&A...684A..87D} {684, A87}

\bibitem[\protect\citeauthoryear{{Gunn} \& {Peterson}}{{Gunn} \& {Peterson}}{1965}]{1965ApJ...142.1633G}
{Gunn} J.~E.,  {Peterson} B.~A.,  1965, \mn@doi [\apj] {10.1086/148444}, \href {https://ui.adsabs.harvard.edu/abs/1965ApJ...142.1633G} {142, 1633}

\bibitem[\protect\citeauthoryear{{Guo} et~al.,}{{Guo} et~al.}{2024}]{2024A&A...688A..37G}
{Guo} Y.,  et~al., 2024, \mn@doi [\aap] {10.1051/0004-6361/202347658}, \href {https://ui.adsabs.harvard.edu/abs/2024A&A...688A..37G} {688, A37}

\bibitem[\protect\citeauthoryear{{Gutkin}, {Charlot}  \& {Bruzual}}{{Gutkin} et~al.}{2016}]{2016MNRAS.462.1757G}
{Gutkin} J.,  {Charlot} S.,   {Bruzual} G.,  2016, \mn@doi [\mnras] {10.1093/mnras/stw1716}, \href {https://ui.adsabs.harvard.edu/abs/2016MNRAS.462.1757G} {462, 1757}

\bibitem[\protect\citeauthoryear{{Hainline}, {Shapley}, {Greene}  \& {Steidel}}{{Hainline} et~al.}{2011}]{2011ApJ...733...31H}
{Hainline} K.~N.,  {Shapley} A.~E.,  {Greene} J.~E.,   {Steidel} C.~C.,  2011, \mn@doi [\apj] {10.1088/0004-637X/733/1/31}, \href {https://ui.adsabs.harvard.edu/abs/2011ApJ...733...31H} {733, 31}

\bibitem[\protect\citeauthoryear{{Hainline} et~al.,}{{Hainline} et~al.}{2024a}]{2024arXiv240404325H}
{Hainline} K.~N.,  et~al., 2024a, preprint, \href {https://ui.adsabs.harvard.edu/abs/2024arXiv240404325H} {} (\mn@eprint {arXiv} {2404.04325})

\bibitem[\protect\citeauthoryear{{Hainline} et~al.,}{{Hainline} et~al.}{2024b}]{2024ApJ...964...71H}
{Hainline} K.~N.,  et~al., 2024b, \mn@doi [\apj] {10.3847/1538-4357/ad1ee4}, \href {https://ui.adsabs.harvard.edu/abs/2024ApJ...964...71H} {964, 71}

\bibitem[\protect\citeauthoryear{{Harikane} et~al.,}{{Harikane} et~al.}{2023}]{2023ApJ...959...39H}
{Harikane} Y.,  et~al., 2023, \mn@doi [\apj] {10.3847/1538-4357/ad029e}, \href {https://ui.adsabs.harvard.edu/abs/2023ApJ...959...39H} {959, 39}

\bibitem[\protect\citeauthoryear{{Hayes} \& {Scarlata}}{{Hayes} \& {Scarlata}}{2023}]{2023ApJ...954L..14H}
{Hayes} M.~J.,  {Scarlata} C.,  2023, \mn@doi [\apjl] {10.3847/2041-8213/acee6a}, \href {https://ui.adsabs.harvard.edu/abs/2023ApJ...954L..14H} {954, L14}

\bibitem[\protect\citeauthoryear{{Hayes}, {Schaerer}, {{\"O}stlin}, {Mas-Hesse}, {Atek}  \& {Kunth}}{{Hayes} et~al.}{2011}]{2011ApJ...730....8H}
{Hayes} M.,  {Schaerer} D.,  {{\"O}stlin} G.,  {Mas-Hesse} J.~M.,  {Atek} H.,   {Kunth} D.,  2011, \mn@doi [\apj] {10.1088/0004-637X/730/1/8}, \href {https://ui.adsabs.harvard.edu/abs/2011ApJ...730....8H} {730, 8}

\bibitem[\protect\citeauthoryear{{Heintz} et~al.,}{{Heintz} et~al.}{2024a}]{2024arXiv240402211H}
{Heintz} K.~E.,  et~al., 2024a, preprint, \href {https://ui.adsabs.harvard.edu/abs/2024arXiv240402211H} {} (\mn@eprint {arXiv} {2404.02211})

\bibitem[\protect\citeauthoryear{{Heintz} et~al.,}{{Heintz} et~al.}{2024b}]{2024Sci...384..890H}
{Heintz} K.~E.,  et~al., 2024b, \mn@doi [Science] {10.1126/science.adj0343}, \href {https://ui.adsabs.harvard.edu/abs/2024Sci...384..890H} {384, 890}

\bibitem[\protect\citeauthoryear{{Helton} et~al.,}{{Helton} et~al.}{2024a}]{2024ApJ...962..124H}
{Helton} J.~M.,  et~al., 2024a, \mn@doi [\apj] {10.3847/1538-4357/ad0da7}, \href {https://ui.adsabs.harvard.edu/abs/2024ApJ...962..124H} {962, 124}

\bibitem[\protect\citeauthoryear{{Helton} et~al.,}{{Helton} et~al.}{2024b}]{2024ApJ...974...41H}
{Helton} J.~M.,  et~al., 2024b, \mn@doi [\apj] {10.3847/1538-4357/ad6867}, \href {https://ui.adsabs.harvard.edu/abs/2024ApJ...974...41H} {974, 41}

\bibitem[\protect\citeauthoryear{{Hirschmann}, {Charlot}, {Feltre}, {Naab}, {Somerville}  \& {Choi}}{{Hirschmann} et~al.}{2019}]{2019MNRAS.487..333H}
{Hirschmann} M.,  {Charlot} S.,  {Feltre} A.,  {Naab} T.,  {Somerville} R.~S.,   {Choi} E.,  2019, \mn@doi [\mnras] {10.1093/mnras/stz1256}, \href {https://ui.adsabs.harvard.edu/abs/2019MNRAS.487..333H} {487, 333}

\bibitem[\protect\citeauthoryear{{Hirschmann} et~al.,}{{Hirschmann} et~al.}{2023}]{2023MNRAS.526.3610H}
{Hirschmann} M.,  et~al., 2023, \mn@doi [\mnras] {10.1093/mnras/stad2955}, \href {https://ui.adsabs.harvard.edu/abs/2023MNRAS.526.3610H} {526, 3610}

\bibitem[\protect\citeauthoryear{{Hu} et~al.,}{{Hu} et~al.}{2021}]{2021NatAs...5..485H}
{Hu} W.,  et~al., 2021, \mn@doi [Nature Astronomy] {10.1038/s41550-020-01291-y}, \href {https://ui.adsabs.harvard.edu/abs/2021NatAs...5..485H} {5, 485}

\bibitem[\protect\citeauthoryear{Hunter}{Hunter}{2007}]{Hunter2007}
Hunter J.~D.,  2007, \mn@doi [Computing in Science {\&} Engineering] {10.1109/MCSE.2007.55}, 9, 90

\bibitem[\protect\citeauthoryear{{Hutter}, {Trebitsch}, {Dayal}, {Gottl{\"o}ber}, {Yepes}  \& {Legrand}}{{Hutter} et~al.}{2023}]{2023MNRAS.524.6124H}
{Hutter} A.,  {Trebitsch} M.,  {Dayal} P.,  {Gottl{\"o}ber} S.,  {Yepes} G.,   {Legrand} L.,  2023, \mn@doi [\mnras] {10.1093/mnras/stad2230}, \href {https://ui.adsabs.harvard.edu/abs/2023MNRAS.524.6124H} {524, 6124}

\bibitem[\protect\citeauthoryear{{Inoue}, {Shimizu}, {Iwata}  \& {Tanaka}}{{Inoue} et~al.}{2014}]{2014MNRAS.442.1805I}
{Inoue} A.~K.,  {Shimizu} I.,  {Iwata} I.,   {Tanaka} M.,  2014, \mn@doi [\mnras] {10.1093/mnras/stu936}, \href {https://ui.adsabs.harvard.edu/abs/2014MNRAS.442.1805I} {442, 1805}

\bibitem[\protect\citeauthoryear{{Jakobsen} et~al.,}{{Jakobsen} et~al.}{2022}]{2022A&A...661A..80J}
{Jakobsen} P.,  et~al., 2022, \mn@doi [\aap] {10.1051/0004-6361/202142663}, \href {https://ui.adsabs.harvard.edu/abs/2022A&A...661A..80J} {661, A80}

\bibitem[\protect\citeauthoryear{Jones, Oliphant, Peterson  et~al.}{Jones et~al.}{2001}]{Jones2001}
Jones E.,  Oliphant T.,  Peterson P.,   et~al., 2001, {SciPy}: Open source scientific tools for {Python}, \url {http://www.scipy.org/}

\bibitem[\protect\citeauthoryear{{Jones} et~al.,}{{Jones} et~al.}{2024a}]{2024arXiv240906405J}
{Jones} G.~C.,  et~al., 2024a, preprint, \href {https://ui.adsabs.harvard.edu/abs/2024arXiv240906405J} {} (\mn@eprint {arXiv} {2409.06405})

\bibitem[\protect\citeauthoryear{{Jones} et~al.,}{{Jones} et~al.}{2024b}]{2024A&A...683A.238J}
{Jones} G.~C.,  et~al., 2024b, \mn@doi [\aap] {10.1051/0004-6361/202347099}, \href {https://ui.adsabs.harvard.edu/abs/2024A&A...683A.238J} {683, A238}

\bibitem[\protect\citeauthoryear{{Jung} et~al.,}{{Jung} et~al.}{2022a}]{2022arXiv221209850J}
{Jung} I.,  et~al., 2022a, preprint, \href {https://ui.adsabs.harvard.edu/abs/2022arXiv221209850J} {} (\mn@eprint {arXiv} {2212.09850})

\bibitem[\protect\citeauthoryear{{Jung} et~al.,}{{Jung} et~al.}{2022b}]{2022ApJ...933...87J}
{Jung} I.,  et~al., 2022b, \mn@doi [\apj] {10.3847/1538-4357/ac6fe7}, \href {https://ui.adsabs.harvard.edu/abs/2022ApJ...933...87J} {933, 87}

\bibitem[\protect\citeauthoryear{{Jung} et~al.,}{{Jung} et~al.}{2024}]{2024ApJ...967...73J}
{Jung} I.,  et~al., 2024, \mn@doi [\apj] {10.3847/1538-4357/ad3913}, \href {https://ui.adsabs.harvard.edu/abs/2024ApJ...967...73J} {967, 73}

\bibitem[\protect\citeauthoryear{{Kauffmann} et~al.,}{{Kauffmann} et~al.}{2003}]{2003MNRAS.346.1055K}
{Kauffmann} G.,  et~al., 2003, \mn@doi [\mnras] {10.1111/j.1365-2966.2003.07154.x}, \href {https://ui.adsabs.harvard.edu/abs/2003MNRAS.346.1055K} {346, 1055}

\bibitem[\protect\citeauthoryear{{Keating}, {Weinberger}, {Kulkarni}, {Haehnelt}, {Chardin}  \& {Aubert}}{{Keating} et~al.}{2020}]{2020MNRAS.491.1736K}
{Keating} L.~C.,  {Weinberger} L.~H.,  {Kulkarni} G.,  {Haehnelt} M.~G.,  {Chardin} J.,   {Aubert} D.,  2020, \mn@doi [\mnras] {10.1093/mnras/stz3083}, \href {https://ui.adsabs.harvard.edu/abs/2020MNRAS.491.1736K} {491, 1736}

\bibitem[\protect\citeauthoryear{{Kennicutt} \& {Evans}}{{Kennicutt} \& {Evans}}{2012}]{2012ARA&A..50..531K}
{Kennicutt} R.~C.,  {Evans} N.~J.,  2012, \mn@doi [\araa] {10.1146/annurev-astro-081811-125610}, \href {https://ui.adsabs.harvard.edu/abs/2012ARA&A..50..531K} {50, 531}

\bibitem[\protect\citeauthoryear{{Kewley}, {Dopita}, {Sutherland}, {Heisler}  \& {Trevena}}{{Kewley} et~al.}{2001}]{2001ApJ...556..121K}
{Kewley} L.~J.,  {Dopita} M.~A.,  {Sutherland} R.~S.,  {Heisler} C.~A.,   {Trevena} J.,  2001, \mn@doi [\apj] {10.1086/321545}, \href {https://ui.adsabs.harvard.edu/abs/2001ApJ...556..121K} {556, 121}

\bibitem[\protect\citeauthoryear{{Kewley}, {Nicholls}, {Sutherland}, {Rigby}, {Acharya}, {Dopita}  \& {Bayliss}}{{Kewley} et~al.}{2019}]{2019ApJ...880...16K}
{Kewley} L.~J.,  {Nicholls} D.~C.,  {Sutherland} R.,  {Rigby} J.~R.,  {Acharya} A.,  {Dopita} M.~A.,   {Bayliss} M.~B.,  2019, \mn@doi [\apj] {10.3847/1538-4357/ab16ed}, \href {https://ui.adsabs.harvard.edu/abs/2019ApJ...880...16K} {880, 16}

\bibitem[\protect\citeauthoryear{{Kulkarni}, {Keating}, {Haehnelt}, {Bosman}, {Puchwein}, {Chardin}  \& {Aubert}}{{Kulkarni} et~al.}{2019}]{2019MNRAS.485L..24K}
{Kulkarni} G.,  {Keating} L.~C.,  {Haehnelt} M.~G.,  {Bosman} S. E.~I.,  {Puchwein} E.,  {Chardin} J.,   {Aubert} D.,  2019, \mn@doi [\mnras] {10.1093/mnrasl/slz025}, \href {https://ui.adsabs.harvard.edu/abs/2019MNRAS.485L..24K} {485, L24}

\bibitem[\protect\citeauthoryear{{Kumari}, {Smit}, {Leitherer}, {Witstok}, {Irwin}, {Sirianni}  \& {Aloisi}}{{Kumari} et~al.}{2024}]{2024MNRAS.529..781K}
{Kumari} N.,  {Smit} R.,  {Leitherer} C.,  {Witstok} J.,  {Irwin} M.~J.,  {Sirianni} M.,   {Aloisi} A.,  2024, \mn@doi [\mnras] {10.1093/mnras/stae252}, \href {https://ui.adsabs.harvard.edu/abs/2024MNRAS.529..781K} {529, 781}

\bibitem[\protect\citeauthoryear{{Kusakabe} et~al.,}{{Kusakabe} et~al.}{2022}]{2022A&A...660A..44K}
{Kusakabe} H.,  et~al., 2022, \mn@doi [\aap] {10.1051/0004-6361/202142302}, \href {https://ui.adsabs.harvard.edu/abs/2022A&A...660A..44K} {660, A44}

\bibitem[\protect\citeauthoryear{{Larson} et~al.,}{{Larson} et~al.}{2022}]{2022ApJ...930..104L}
{Larson} R.~L.,  et~al., 2022, \mn@doi [\apj] {10.3847/1538-4357/ac5dbd}, \href {https://ui.adsabs.harvard.edu/abs/2022ApJ...930..104L} {930, 104}

\bibitem[\protect\citeauthoryear{{Laseter} et~al.,}{{Laseter} et~al.}{2024}]{2024A&A...681A..70L}
{Laseter} I.~H.,  et~al., 2024, \mn@doi [\aap] {10.1051/0004-6361/202347133}, \href {https://ui.adsabs.harvard.edu/abs/2024A&A...681A..70L} {681, A70}

\bibitem[\protect\citeauthoryear{{Laursen}, {Sommer-Larsen}, {Milvang-Jensen}, {Fynbo}  \& {Razoumov}}{{Laursen} et~al.}{2019}]{2019A&A...627A..84L}
{Laursen} P.,  {Sommer-Larsen} J.,  {Milvang-Jensen} B.,  {Fynbo} J. P.~U.,   {Razoumov} A.~O.,  2019, \mn@doi [\aap] {10.1051/0004-6361/201833645}, \href {https://ui.adsabs.harvard.edu/abs/2019A&A...627A..84L} {627, A84}

\bibitem[\protect\citeauthoryear{{Leclercq} et~al.,}{{Leclercq} et~al.}{2017}]{2017A&A...608A...8L}
{Leclercq} F.,  et~al., 2017, \mn@doi [\aap] {10.1051/0004-6361/201731480}, \href {https://ui.adsabs.harvard.edu/abs/2017A&A...608A...8L} {608, A8}

\bibitem[\protect\citeauthoryear{{Lehnert} et~al.,}{{Lehnert} et~al.}{2010}]{2010Natur.467..940L}
{Lehnert} M.~D.,  et~al., 2010, \mn@doi [\nat] {10.1038/nature09462}, \href {https://ui.adsabs.harvard.edu/abs/2010Natur.467..940L} {467, 940}

\bibitem[\protect\citeauthoryear{{Leja}, {Carnall}, {Johnson}, {Conroy}  \& {Speagle}}{{Leja} et~al.}{2019}]{2019ApJ...876....3L}
{Leja} J.,  {Carnall} A.~C.,  {Johnson} B.~D.,  {Conroy} C.,   {Speagle} J.~S.,  2019, \mn@doi [\apj] {10.3847/1538-4357/ab133c}, \href {https://ui.adsabs.harvard.edu/abs/2019ApJ...876....3L} {876, 3}

\bibitem[\protect\citeauthoryear{{Leonova} et~al.,}{{Leonova} et~al.}{2022}]{2022MNRAS.515.5790L}
{Leonova} E.,  et~al., 2022, \mn@doi [\mnras] {10.1093/mnras/stac1908}, \href {https://ui.adsabs.harvard.edu/abs/2022MNRAS.515.5790L} {515, 5790}

\bibitem[\protect\citeauthoryear{{Looser} et~al.,}{{Looser} et~al.}{2024}]{2024Natur.629...53L}
{Looser} T.~J.,  et~al., 2024, \mn@doi [\nat] {10.1038/s41586-024-07227-0}, \href {https://ui.adsabs.harvard.edu/abs/2024Natur.629...53L} {629, 53}

\bibitem[\protect\citeauthoryear{{Lu}, {Mason}, {Hutter}, {Mesinger}, {Qin}, {Stark}  \& {Endsley}}{{Lu} et~al.}{2024}]{2024MNRAS.528.4872L}
{Lu} T.-Y.,  {Mason} C.~A.,  {Hutter} A.,  {Mesinger} A.,  {Qin} Y.,  {Stark} D.~P.,   {Endsley} R.,  2024, \mn@doi [\mnras] {10.1093/mnras/stae266}, \href {https://ui.adsabs.harvard.edu/abs/2024MNRAS.528.4872L} {528, 4872}

\bibitem[\protect\citeauthoryear{{Luridiana}, {Morisset}  \& {Shaw}}{{Luridiana} et~al.}{2015}]{2015A&A...573A..42L}
{Luridiana} V.,  {Morisset} C.,   {Shaw} R.~A.,  2015, \mn@doi [\aap] {10.1051/0004-6361/201323152}, \href {https://ui.adsabs.harvard.edu/abs/2015A&A...573A..42L} {573, A42}

\bibitem[\protect\citeauthoryear{{Maiolino} et~al.,}{{Maiolino} et~al.}{2024}]{2024Natur.627...59M}
{Maiolino} R.,  et~al., 2024, \mn@doi [\nat] {10.1038/s41586-024-07052-5}, \href {https://ui.adsabs.harvard.edu/abs/2024Natur.627...59M} {627, 59}

\bibitem[\protect\citeauthoryear{{Marconi} et~al.,}{{Marconi} et~al.}{2024}]{2024A&A...689A..78M}
{Marconi} A.,  et~al., 2024, \mn@doi [\aap] {10.1051/0004-6361/202449240}, \href {https://ui.adsabs.harvard.edu/abs/2024A&A...689A..78M} {689, A78}

\bibitem[\protect\citeauthoryear{{Maseda} et~al.,}{{Maseda} et~al.}{2018}]{2018ApJ...865L...1M}
{Maseda} M.~V.,  et~al., 2018, \mn@doi [\apjl] {10.3847/2041-8213/aade4b}, \href {https://ui.adsabs.harvard.edu/abs/2018ApJ...865L...1M} {865, L1}

\bibitem[\protect\citeauthoryear{{Maseda} et~al.,}{{Maseda} et~al.}{2020}]{2020MNRAS.493.5120M}
{Maseda} M.~V.,  et~al., 2020, \mn@doi [\mnras] {10.1093/mnras/staa622}, \href {https://ui.adsabs.harvard.edu/abs/2020MNRAS.493.5120M} {493, 5120}

\bibitem[\protect\citeauthoryear{{Mason} \& {Gronke}}{{Mason} \& {Gronke}}{2020}]{2020MNRAS.499.1395M}
{Mason} C.~A.,  {Gronke} M.,  2020, \mn@doi [\mnras] {10.1093/mnras/staa2910}, \href {https://ui.adsabs.harvard.edu/abs/2020MNRAS.499.1395M} {499, 1395}

\bibitem[\protect\citeauthoryear{{Mason}, {Treu}, {Dijkstra}, {Mesinger}, {Trenti}, {Pentericci}, {de Barros}  \& {Vanzella}}{{Mason} et~al.}{2018a}]{2018ApJ...856....2M}
{Mason} C.~A.,  {Treu} T.,  {Dijkstra} M.,  {Mesinger} A.,  {Trenti} M.,  {Pentericci} L.,  {de Barros} S.,   {Vanzella} E.,  2018a, \mn@doi [\apj] {10.3847/1538-4357/aab0a7}, \href {https://ui.adsabs.harvard.edu/abs/2018ApJ...856....2M} {856, 2}

\bibitem[\protect\citeauthoryear{{Mason} et~al.,}{{Mason} et~al.}{2018b}]{2018ApJ...857L..11M}
{Mason} C.~A.,  et~al., 2018b, \mn@doi [\apjl] {10.3847/2041-8213/aabbab}, \href {https://ui.adsabs.harvard.edu/abs/2018ApJ...857L..11M} {857, L11}

\bibitem[\protect\citeauthoryear{{Mason} et~al.,}{{Mason} et~al.}{2019}]{2019MNRAS.485.3947M}
{Mason} C.~A.,  et~al., 2019, \mn@doi [\mnras] {10.1093/mnras/stz632}, \href {https://ui.adsabs.harvard.edu/abs/2019MNRAS.485.3947M} {485, 3947}

\bibitem[\protect\citeauthoryear{{Matthee}, {Sobral}, {Gronke}, {Paulino-Afonso}, {Stefanon}  \& {R{\"o}ttgering}}{{Matthee} et~al.}{2018}]{2018A&A...619A.136M}
{Matthee} J.,  {Sobral} D.,  {Gronke} M.,  {Paulino-Afonso} A.,  {Stefanon} M.,   {R{\"o}ttgering} H.,  2018, \mn@doi [\aap] {10.1051/0004-6361/201833528}, \href {https://ui.adsabs.harvard.edu/abs/2018A&A...619A.136M} {619, A136}

\bibitem[\protect\citeauthoryear{{Mazzolari} et~al.,}{{Mazzolari} et~al.}{2024}]{2024arXiv240410811M}
{Mazzolari} G.,  et~al., 2024, preprint, \href {https://ui.adsabs.harvard.edu/abs/2024arXiv240410811M} {} (\mn@eprint {arXiv} {2404.10811})

\bibitem[\protect\citeauthoryear{{McClymont} et~al.,}{{McClymont} et~al.}{2024}]{2024arXiv240515859M}
{McClymont} W.,  et~al., 2024, preprint, \href {https://ui.adsabs.harvard.edu/abs/2024arXiv240515859M} {} (\mn@eprint {arXiv} {2405.15859})

\bibitem[\protect\citeauthoryear{{McElwain} et~al.,}{{McElwain} et~al.}{2023}]{2023PASP..135e8001M}
{McElwain} M.~W.,  et~al., 2023, \mn@doi [\pasp] {10.1088/1538-3873/acada0}, \href {https://ui.adsabs.harvard.edu/abs/2023PASP..135e8001M} {135, 058001}

\bibitem[\protect\citeauthoryear{{McLure}, {Dunlop}, {Cirasuolo}, {Koekemoer}, {Sabbi}, {Stark}, {Targett}  \& {Ellis}}{{McLure} et~al.}{2010}]{2010MNRAS.403..960M}
{McLure} R.~J.,  {Dunlop} J.~S.,  {Cirasuolo} M.,  {Koekemoer} A.~M.,  {Sabbi} E.,  {Stark} D.~P.,  {Targett} T.~A.,   {Ellis} R.~S.,  2010, \mn@doi [\mnras] {10.1111/j.1365-2966.2009.16176.x}, \href {https://ui.adsabs.harvard.edu/abs/2010MNRAS.403..960M} {403, 960}

\bibitem[\protect\citeauthoryear{{Morishita} et~al.,}{{Morishita} et~al.}{2024}]{2024ApJ...963....9M}
{Morishita} T.,  et~al., 2024, \mn@doi [\apj] {10.3847/1538-4357/ad1404}, \href {https://ui.adsabs.harvard.edu/abs/2024ApJ...963....9M} {963, 9}

\bibitem[\protect\citeauthoryear{{Naidu}, {Tacchella}, {Mason}, {Bose}, {Oesch}  \& {Conroy}}{{Naidu} et~al.}{2020}]{2020ApJ...892..109N}
{Naidu} R.~P.,  {Tacchella} S.,  {Mason} C.~A.,  {Bose} S.,  {Oesch} P.~A.,   {Conroy} C.,  2020, \mn@doi [\apj] {10.3847/1538-4357/ab7cc9}, \href {https://ui.adsabs.harvard.edu/abs/2020ApJ...892..109N} {892, 109}

\bibitem[\protect\citeauthoryear{{Nakajima} \& {Maiolino}}{{Nakajima} \& {Maiolino}}{2022}]{2022MNRAS.513.5134N}
{Nakajima} K.,  {Maiolino} R.,  2022, \mn@doi [\mnras] {10.1093/mnras/stac1242}, \href {https://ui.adsabs.harvard.edu/abs/2022MNRAS.513.5134N} {513, 5134}

\bibitem[\protect\citeauthoryear{{Nakane} et~al.,}{{Nakane} et~al.}{2024}]{2024ApJ...967...28N}
{Nakane} M.,  et~al., 2024, \mn@doi [\apj] {10.3847/1538-4357/ad38c2}, \href {https://ui.adsabs.harvard.edu/abs/2024ApJ...967...28N} {967, 28}

\bibitem[\protect\citeauthoryear{{Napolitano} et~al.,}{{Napolitano} et~al.}{2024}]{2024A&A...688A.106N}
{Napolitano} L.,  et~al., 2024, \mn@doi [\aap] {10.1051/0004-6361/202449644}, \href {https://ui.adsabs.harvard.edu/abs/2024A&A...688A.106N} {688, A106}

\bibitem[\protect\citeauthoryear{{Neyer} et~al.,}{{Neyer} et~al.}{2024}]{2024MNRAS.531.2943N}
{Neyer} M.,  et~al., 2024, \mn@doi [\mnras] {10.1093/mnras/stae1325}, \href {https://ui.adsabs.harvard.edu/abs/2024MNRAS.531.2943N} {531, 2943}

\bibitem[\protect\citeauthoryear{{Oesch} et~al.,}{{Oesch} et~al.}{2015}]{2015ApJ...804L..30O}
{Oesch} P.~A.,  et~al., 2015, \mn@doi [\apjl] {10.1088/2041-8205/804/2/L30}, \href {https://ui.adsabs.harvard.edu/abs/2015ApJ...804L..30O} {804, L30}

\bibitem[\protect\citeauthoryear{{Oesch} et~al.,}{{Oesch} et~al.}{2016}]{2016ApJ...819..129O}
{Oesch} P.~A.,  et~al., 2016, \mn@doi [\apj] {10.3847/0004-637X/819/2/129}, \href {https://ui.adsabs.harvard.edu/abs/2016ApJ...819..129O} {819, 129}

\bibitem[\protect\citeauthoryear{{Oesch} et~al.,}{{Oesch} et~al.}{2023}]{2023MNRAS.525.2864O}
{Oesch} P.~A.,  et~al., 2023, \mn@doi [\mnras] {10.1093/mnras/stad2411}, \href {https://ui.adsabs.harvard.edu/abs/2023MNRAS.525.2864O} {525, 2864}

\bibitem[\protect\citeauthoryear{{Oke} \& {Gunn}}{{Oke} \& {Gunn}}{1983}]{1983ApJ...266..713O}
{Oke} J.~B.,  {Gunn} J.~E.,  1983, \mn@doi [\apj] {10.1086/160817}, \href {https://ui.adsabs.harvard.edu/abs/1983ApJ...266..713O} {266, 713}

\bibitem[\protect\citeauthoryear{{Ono} et~al.,}{{Ono} et~al.}{2012}]{2012ApJ...744...83O}
{Ono} Y.,  et~al., 2012, \mn@doi [\apj] {10.1088/0004-637X/744/2/83}, \href {https://ui.adsabs.harvard.edu/abs/2012ApJ...744...83O} {744, 83}

\bibitem[\protect\citeauthoryear{{Osterbrock} \& {Ferland}}{{Osterbrock} \& {Ferland}}{2006}]{2006agna.book.....O}
{Osterbrock} D.~E.,  {Ferland} G.~J.,  2006, {Astrophysics of gaseous nebulae and active galactic nuclei}.
{University Science Books (Sausalito, Calif.)}

\bibitem[\protect\citeauthoryear{{Ouchi}, {Ono}  \& {Shibuya}}{{Ouchi} et~al.}{2020}]{2020ARA&A..58..617O}
{Ouchi} M.,  {Ono} Y.,   {Shibuya} T.,  2020, \mn@doi [\araa] {10.1146/annurev-astro-032620-021859}, \href {https://ui.adsabs.harvard.edu/abs/2020ARA&A..58..617O} {58, 617}

\bibitem[\protect\citeauthoryear{{Pahl} et~al.,}{{Pahl} et~al.}{2024}]{2024arXiv240703399P}
{Pahl} A.~J.,  et~al., 2024, preprint, \href {https://ui.adsabs.harvard.edu/abs/2024arXiv240703399P} {} (\mn@eprint {arXiv} {2407.03399})

\bibitem[\protect\citeauthoryear{{Palay}, {Nahar}, {Pradhan}  \& {Eissner}}{{Palay} et~al.}{2012}]{2012MNRAS.423L..35P}
{Palay} E.,  {Nahar} S.~N.,  {Pradhan} A.~K.,   {Eissner} W.,  2012, \mn@doi [\mnras] {10.1111/j.1745-3933.2012.01252.x}, \href {https://ui.adsabs.harvard.edu/abs/2012MNRAS.423L..35P} {423, L35}

\bibitem[\protect\citeauthoryear{{Pallottini} \& {Ferrara}}{{Pallottini} \& {Ferrara}}{2023}]{2023A&A...677L...4P}
{Pallottini} A.,  {Ferrara} A.,  2023, \mn@doi [\aap] {10.1051/0004-6361/202347384}, \href {https://ui.adsabs.harvard.edu/abs/2023A&A...677L...4P} {677, L4}

\bibitem[\protect\citeauthoryear{{Partridge} \& {Peebles}}{{Partridge} \& {Peebles}}{1967}]{1967ApJ...147..868P}
{Partridge} R.~B.,  {Peebles} P.~J.~E.,  1967, \mn@doi [\apj] {10.1086/149079}, \href {https://ui.adsabs.harvard.edu/abs/1967ApJ...147..868P} {147, 868}

\bibitem[\protect\citeauthoryear{{Pasha} \& {Miller}}{{Pasha} \& {Miller}}{2023}]{2023JOSS....8.5703P}
{Pasha} I.,  {Miller} T.~B.,  2023, \mn@doi [The Journal of Open Source Software] {10.21105/joss.05703}, \href {https://ui.adsabs.harvard.edu/abs/2023JOSS....8.5703P} {8, 5703}

\bibitem[\protect\citeauthoryear{{Pentericci} et~al.,}{{Pentericci} et~al.}{2011}]{2011ApJ...743..132P}
{Pentericci} L.,  et~al., 2011, \mn@doi [\apj] {10.1088/0004-637X/743/2/132}, \href {https://ui.adsabs.harvard.edu/abs/2011ApJ...743..132P} {743, 132}

\bibitem[\protect\citeauthoryear{{Pentericci} et~al.,}{{Pentericci} et~al.}{2014}]{2014ApJ...793..113P}
{Pentericci} L.,  et~al., 2014, \mn@doi [\apj] {10.1088/0004-637X/793/2/113}, \href {https://ui.adsabs.harvard.edu/abs/2014ApJ...793..113P} {793, 113}

\bibitem[\protect\citeauthoryear{{Pilyugin}, {Mattsson}, {V{\'\i}lchez}  \& {Cedr{\'e}s}}{{Pilyugin} et~al.}{2009}]{2009MNRAS.398..485P}
{Pilyugin} L.~S.,  {Mattsson} L.,  {V{\'\i}lchez} J.~M.,   {Cedr{\'e}s} B.,  2009, \mn@doi [\mnras] {10.1111/j.1365-2966.2009.15182.x}, \href {https://ui.adsabs.harvard.edu/abs/2009MNRAS.398..485P} {398, 485}

\bibitem[\protect\citeauthoryear{{Planck Collaboration} et~al.,}{{Planck Collaboration} et~al.}{2020}]{2020A&A...641A...6P}
{Planck Collaboration} et~al., 2020, \mn@doi [\aap] {10.1051/0004-6361/201833910}, \href {https://ui.adsabs.harvard.edu/abs/2020A&A...641A...6P} {641, A6}

\bibitem[\protect\citeauthoryear{{Pradhan}, {Montenegro}, {Nahar}  \& {Eissner}}{{Pradhan} et~al.}{2006}]{2006MNRAS.366L...6P}
{Pradhan} A.~K.,  {Montenegro} M.,  {Nahar} S.~N.,   {Eissner} W.,  2006, \mn@doi [\mnras] {10.1111/j.1745-3933.2005.00119.x}, \href {https://ui.adsabs.harvard.edu/abs/2006MNRAS.366L...6P} {366, L6}

\bibitem[\protect\citeauthoryear{{Prieto-Lyon} et~al.,}{{Prieto-Lyon} et~al.}{2023}]{2023A&A...672A.186P}
{Prieto-Lyon} G.,  et~al., 2023, \mn@doi [\aap] {10.1051/0004-6361/202245532}, \href {https://ui.adsabs.harvard.edu/abs/2023A&A...672A.186P} {672, A186}

\bibitem[\protect\citeauthoryear{{Qin}, {Wyithe}, {Oesch}, {Illingworth}, {Leonova}, {Mutch}  \& {Naidu}}{{Qin} et~al.}{2022}]{2022MNRAS.510.3858Q}
{Qin} Y.,  {Wyithe} J. S.~B.,  {Oesch} P.~A.,  {Illingworth} G.~D.,  {Leonova} E.,  {Mutch} S.~J.,   {Naidu} R.~P.,  2022, \mn@doi [\mnras] {10.1093/mnras/stab3733}, \href {https://ui.adsabs.harvard.edu/abs/2022MNRAS.510.3858Q} {510, 3858}

\bibitem[\protect\citeauthoryear{{Reddy} et~al.,}{{Reddy} et~al.}{2015}]{2015ApJ...806..259R}
{Reddy} N.~A.,  et~al., 2015, \mn@doi [\apj] {10.1088/0004-637X/806/2/259}, \href {https://ui.adsabs.harvard.edu/abs/2015ApJ...806..259R} {806, 259}

\bibitem[\protect\citeauthoryear{{Reddy} et~al.,}{{Reddy} et~al.}{2022}]{2022ApJ...926...31R}
{Reddy} N.~A.,  et~al., 2022, \mn@doi [\apj] {10.3847/1538-4357/ac3b4c}, \href {https://ui.adsabs.harvard.edu/abs/2022ApJ...926...31R} {926, 31}

\bibitem[\protect\citeauthoryear{{Rieke} et~al.,}{{Rieke} et~al.}{2023a}]{2023PASP..135b8001R}
{Rieke} M.~J.,  et~al., 2023a, \mn@doi [\pasp] {10.1088/1538-3873/acac53}, \href {https://ui.adsabs.harvard.edu/abs/2023PASP..135b8001R} {135, 028001}

\bibitem[\protect\citeauthoryear{{Rieke} et~al.,}{{Rieke} et~al.}{2023b}]{2023ApJS..269...16R}
{Rieke} M.~J.,  et~al., 2023b, \mn@doi [\apjs] {10.3847/1538-4365/acf44d}, \href {https://ui.adsabs.harvard.edu/abs/2023ApJS..269...16R} {269, 16}

\bibitem[\protect\citeauthoryear{{Rigby} et~al.,}{{Rigby} et~al.}{2023}]{2023PASP..135d8001R}
{Rigby} J.,  et~al., 2023, \mn@doi [\pasp] {10.1088/1538-3873/acb293}, \href {https://ui.adsabs.harvard.edu/abs/2023PASP..135d8001R} {135, 048001}

\bibitem[\protect\citeauthoryear{{Roberts-Borsani} et~al.,}{{Roberts-Borsani} et~al.}{2016}]{2016ApJ...823..143R}
{Roberts-Borsani} G.~W.,  et~al., 2016, \mn@doi [\apj] {10.3847/0004-637X/823/2/143}, \href {https://ui.adsabs.harvard.edu/abs/2016ApJ...823..143R} {823, 143}

\bibitem[\protect\citeauthoryear{{Roberts-Borsani} et~al.,}{{Roberts-Borsani} et~al.}{2024}]{2024arXiv240307103R}
{Roberts-Borsani} G.,  et~al., 2024, preprint, \href {https://ui.adsabs.harvard.edu/abs/2024arXiv240307103R} {} (\mn@eprint {arXiv} {2403.07103})

\bibitem[\protect\citeauthoryear{{Robertson}}{{Robertson}}{2022}]{2022ARA&A..60..121R}
{Robertson} B.~E.,  2022, \mn@doi [\araa] {10.1146/annurev-astro-120221-044656}, \href {https://ui.adsabs.harvard.edu/abs/2022ARA&A..60..121R} {60, 121}

\bibitem[\protect\citeauthoryear{{Robertson}, {Ellis}, {Furlanetto}  \& {Dunlop}}{{Robertson} et~al.}{2015}]{2015ApJ...802L..19R}
{Robertson} B.~E.,  {Ellis} R.~S.,  {Furlanetto} S.~R.,   {Dunlop} J.~S.,  2015, \mn@doi [\apjl] {10.1088/2041-8205/802/2/L19}, \href {https://ui.adsabs.harvard.edu/abs/2015ApJ...802L..19R} {802, L19}

\bibitem[\protect\citeauthoryear{{Robertson} et~al.,}{{Robertson} et~al.}{2023}]{2023NatAs...7..611R}
{Robertson} B.~E.,  et~al., 2023, \mn@doi [Nature Astronomy] {10.1038/s41550-023-01921-1}, \href {https://ui.adsabs.harvard.edu/abs/2023NatAs...7..611R} {7, 611}

\bibitem[\protect\citeauthoryear{{Sanders} et~al.,}{{Sanders} et~al.}{2016}]{2016ApJ...816...23S}
{Sanders} R.~L.,  et~al., 2016, \mn@doi [\apj] {10.3847/0004-637X/816/1/23}, \href {https://ui.adsabs.harvard.edu/abs/2016ApJ...816...23S} {816, 23}

\bibitem[\protect\citeauthoryear{{Sanders} et~al.,}{{Sanders} et~al.}{2023}]{2023ApJ...943...75S}
{Sanders} R.~L.,  et~al., 2023, \mn@doi [\apj] {10.3847/1538-4357/aca9cc}, \href {https://ui.adsabs.harvard.edu/abs/2023ApJ...943...75S} {943, 75}

\bibitem[\protect\citeauthoryear{{Sanders}, {Shapley}, {Topping}, {Reddy}  \& {Brammer}}{{Sanders} et~al.}{2024}]{2024ApJ...962...24S}
{Sanders} R.~L.,  {Shapley} A.~E.,  {Topping} M.~W.,  {Reddy} N.~A.,   {Brammer} G.~B.,  2024, \mn@doi [\apj] {10.3847/1538-4357/ad15fc}, \href {https://ui.adsabs.harvard.edu/abs/2024ApJ...962...24S} {962, 24}

\bibitem[\protect\citeauthoryear{{Saxena} et~al.,}{{Saxena} et~al.}{2023}]{2023A&A...678A..68S}
{Saxena} A.,  et~al., 2023, \mn@doi [\aap] {10.1051/0004-6361/202346245}, \href {https://ui.adsabs.harvard.edu/abs/2023A&A...678A..68S} {678, A68}

\bibitem[\protect\citeauthoryear{{Saxena} et~al.,}{{Saxena} et~al.}{2024}]{2024A&A...684A..84S}
{Saxena} A.,  et~al., 2024, \mn@doi [\aap] {10.1051/0004-6361/202347132}, \href {https://ui.adsabs.harvard.edu/abs/2024A&A...684A..84S} {684, A84}

\bibitem[\protect\citeauthoryear{{Schenker}, {Ellis}, {Konidaris}  \& {Stark}}{{Schenker} et~al.}{2014}]{2014ApJ...795...20S}
{Schenker} M.~A.,  {Ellis} R.~S.,  {Konidaris} N.~P.,   {Stark} D.~P.,  2014, \mn@doi [\apj] {10.1088/0004-637X/795/1/20}, \href {https://ui.adsabs.harvard.edu/abs/2014ApJ...795...20S} {795, 20}

\bibitem[\protect\citeauthoryear{{Scholtz} et~al.,}{{Scholtz} et~al.}{2023}]{2023arXiv231118731S}
{Scholtz} J.,  et~al., 2023, preprint, \href {https://ui.adsabs.harvard.edu/abs/2023arXiv231118731S} {} (\mn@eprint {arXiv} {2311.18731})

\bibitem[\protect\citeauthoryear{{Scholtz} et~al.,}{{Scholtz} et~al.}{2024}]{2024A&A...687A.283S}
{Scholtz} J.,  et~al., 2024, \mn@doi [\aap] {10.1051/0004-6361/202347187}, \href {https://ui.adsabs.harvard.edu/abs/2024A&A...687A.283S} {687, A283}

\bibitem[\protect\citeauthoryear{{Senchyna} et~al.,}{{Senchyna} et~al.}{2017}]{2017MNRAS.472.2608S}
{Senchyna} P.,  et~al., 2017, \mn@doi [\mnras] {10.1093/mnras/stx2059}, \href {https://ui.adsabs.harvard.edu/abs/2017MNRAS.472.2608S} {472, 2608}

\bibitem[\protect\citeauthoryear{{S{\'e}rsic}}{{S{\'e}rsic}}{1963}]{1963BAAA....6...41S}
{S{\'e}rsic} J.~L.,  1963, Boletin de la Asociacion Argentina de Astronomia La Plata Argentina, \href {https://ui.adsabs.harvard.edu/abs/1963BAAA....6...41S} {6, 41}

\bibitem[\protect\citeauthoryear{{Shen}, {Vogelsberger}, {Boylan-Kolchin}, {Tacchella}  \& {Kannan}}{{Shen} et~al.}{2023}]{2023MNRAS.525.3254S}
{Shen} X.,  {Vogelsberger} M.,  {Boylan-Kolchin} M.,  {Tacchella} S.,   {Kannan} R.,  2023, \mn@doi [\mnras] {10.1093/mnras/stad2508}, \href {https://ui.adsabs.harvard.edu/abs/2023MNRAS.525.3254S} {525, 3254}

\bibitem[\protect\citeauthoryear{{Simmonds} et~al.,}{{Simmonds} et~al.}{2023}]{2023MNRAS.523.5468S}
{Simmonds} C.,  et~al., 2023, \mn@doi [\mnras] {10.1093/mnras/stad1749}, \href {https://ui.adsabs.harvard.edu/abs/2023MNRAS.523.5468S} {523, 5468}

\bibitem[\protect\citeauthoryear{{Simmonds} et~al.,}{{Simmonds} et~al.}{2024}]{2024arXiv240901286S}
{Simmonds} C.,  et~al., 2024, preprint, \href {https://ui.adsabs.harvard.edu/abs/2024arXiv240901286S} {} (\mn@eprint {arXiv} {2409.01286})

\bibitem[\protect\citeauthoryear{{Smit} et~al.,}{{Smit} et~al.}{2014}]{2014ApJ...784...58S}
{Smit} R.,  et~al., 2014, \mn@doi [\apj] {10.1088/0004-637X/784/1/58}, \href {https://ui.adsabs.harvard.edu/abs/2014ApJ...784...58S} {784, 58}

\bibitem[\protect\citeauthoryear{{Smit} et~al.,}{{Smit} et~al.}{2015}]{2015ApJ...801..122S}
{Smit} R.,  et~al., 2015, \mn@doi [\apj] {10.1088/0004-637X/801/2/122}, \href {https://ui.adsabs.harvard.edu/abs/2015ApJ...801..122S} {801, 122}

\bibitem[\protect\citeauthoryear{{Stark}, {Ellis}, {Chiu}, {Ouchi}  \& {Bunker}}{{Stark} et~al.}{2010}]{2010MNRAS.408.1628S}
{Stark} D.~P.,  {Ellis} R.~S.,  {Chiu} K.,  {Ouchi} M.,   {Bunker} A.,  2010, \mn@doi [\mnras] {10.1111/j.1365-2966.2010.17227.x}, \href {https://ui.adsabs.harvard.edu/abs/2010MNRAS.408.1628S} {408, 1628}

\bibitem[\protect\citeauthoryear{{Stark}, {Schenker}, {Ellis}, {Robertson}, {McLure}  \& {Dunlop}}{{Stark} et~al.}{2013}]{2013ApJ...763..129S}
{Stark} D.~P.,  {Schenker} M.~A.,  {Ellis} R.,  {Robertson} B.,  {McLure} R.,   {Dunlop} J.,  2013, \mn@doi [\apj] {10.1088/0004-637X/763/2/129}, \href {https://ui.adsabs.harvard.edu/abs/2013ApJ...763..129S} {763, 129}

\bibitem[\protect\citeauthoryear{{Stark} et~al.,}{{Stark} et~al.}{2017}]{2017MNRAS.464..469S}
{Stark} D.~P.,  et~al., 2017, \mn@doi [\mnras] {10.1093/mnras/stw2233}, \href {https://ui.adsabs.harvard.edu/abs/2017MNRAS.464..469S} {464, 469}

\bibitem[\protect\citeauthoryear{{Stefanon}, {Bouwens}, {Illingworth}, {Labb{\'e}}, {Oesch}  \& {Gonzalez}}{{Stefanon} et~al.}{2022}]{2022ApJ...935...94S}
{Stefanon} M.,  {Bouwens} R.~J.,  {Illingworth} G.~D.,  {Labb{\'e}} I.,  {Oesch} P.~A.,   {Gonzalez} V.,  2022, \mn@doi [\apj] {10.3847/1538-4357/ac7e44}, \href {https://ui.adsabs.harvard.edu/abs/2022ApJ...935...94S} {935, 94}

\bibitem[\protect\citeauthoryear{{Steidel} et~al.,}{{Steidel} et~al.}{2014}]{2014ApJ...795..165S}
{Steidel} C.~C.,  et~al., 2014, \mn@doi [\apj] {10.1088/0004-637X/795/2/165}, \href {https://ui.adsabs.harvard.edu/abs/2014ApJ...795..165S} {795, 165}

\bibitem[\protect\citeauthoryear{{Strait} et~al.,}{{Strait} et~al.}{2023}]{2023ApJ...949L..23S}
{Strait} V.,  et~al., 2023, \mn@doi [\apjl] {10.3847/2041-8213/acd457}, \href {https://ui.adsabs.harvard.edu/abs/2023ApJ...949L..23S} {949, L23}

\bibitem[\protect\citeauthoryear{{Sun} et~al.,}{{Sun} et~al.}{2023}]{2023ApJ...953...53S}
{Sun} F.,  et~al., 2023, \mn@doi [\apj] {10.3847/1538-4357/acd53c}, \href {https://ui.adsabs.harvard.edu/abs/2023ApJ...953...53S} {953, 53}

\bibitem[\protect\citeauthoryear{{Tacchella} et~al.,}{{Tacchella} et~al.}{2022}]{2022ApJ...927..170T}
{Tacchella} S.,  et~al., 2022, \mn@doi [\apj] {10.3847/1538-4357/ac4cad}, \href {https://ui.adsabs.harvard.edu/abs/2022ApJ...927..170T} {927, 170}

\bibitem[\protect\citeauthoryear{{Tacchella} et~al.,}{{Tacchella} et~al.}{2023a}]{2023MNRAS.522.6236T}
{Tacchella} S.,  et~al., 2023a, \mn@doi [\mnras] {10.1093/mnras/stad1408}, \href {https://ui.adsabs.harvard.edu/abs/2023MNRAS.522.6236T} {522, 6236}

\bibitem[\protect\citeauthoryear{{Tacchella} et~al.,}{{Tacchella} et~al.}{2023b}]{2023ApJ...952...74T}
{Tacchella} S.,  et~al., 2023b, \mn@doi [\apj] {10.3847/1538-4357/acdbc6}, \href {https://ui.adsabs.harvard.edu/abs/2023ApJ...952...74T} {952, 74}

\bibitem[\protect\citeauthoryear{{Tang}, {Stark}, {Chevallard}, {Charlot}, {Endsley}  \& {Congiu}}{{Tang} et~al.}{2021}]{2021MNRAS.503.4105T}
{Tang} M.,  {Stark} D.~P.,  {Chevallard} J.,  {Charlot} S.,  {Endsley} R.,   {Congiu} E.,  2021, \mn@doi [\mnras] {10.1093/mnras/stab705}, \href {https://ui.adsabs.harvard.edu/abs/2021MNRAS.503.4105T} {503, 4105}

\bibitem[\protect\citeauthoryear{{Tang} et~al.,}{{Tang} et~al.}{2023}]{2023MNRAS.526.1657T}
{Tang} M.,  et~al., 2023, \mn@doi [\mnras] {10.1093/mnras/stad2763}, \href {https://ui.adsabs.harvard.edu/abs/2023MNRAS.526.1657T} {526, 1657}

\bibitem[\protect\citeauthoryear{{Tang}, {Stark}, {Topping}, {Mason}  \& {Ellis}}{{Tang} et~al.}{2024a}]{2024arXiv240801507T}
{Tang} M.,  {Stark} D.~P.,  {Topping} M.~W.,  {Mason} C.,   {Ellis} R.~S.,  2024a, preprint, \href {https://ui.adsabs.harvard.edu/abs/2024arXiv240801507T} {} (\mn@eprint {arXiv} {2408.01507})

\bibitem[\protect\citeauthoryear{{Tang} et~al.,}{{Tang} et~al.}{2024b}]{2024MNRAS.531.2701T}
{Tang} M.,  et~al., 2024b, \mn@doi [\mnras] {10.1093/mnras/stae1338}, \href {https://ui.adsabs.harvard.edu/abs/2024MNRAS.531.2701T} {531, 2701}

\bibitem[\protect\citeauthoryear{{Tang}, {Stark}, {Ellis}, {Topping}, {Mason}, {Li}  \& {Plat}}{{Tang} et~al.}{2024c}]{2024ApJ...972...56T}
{Tang} M.,  {Stark} D.~P.,  {Ellis} R.~S.,  {Topping} M.~W.,  {Mason} C.,  {Li} Z.,   {Plat} A.,  2024c, \mn@doi [\apj] {10.3847/1538-4357/ad5ae0}, \href {https://ui.adsabs.harvard.edu/abs/2024ApJ...972...56T} {972, 56}

\bibitem[\protect\citeauthoryear{{Tayal}}{{Tayal}}{2007}]{2007ApJS..171..331T}
{Tayal} S.~S.,  2007, \mn@doi [\apjs] {10.1086/513107}, \href {https://ui.adsabs.harvard.edu/abs/2007ApJS..171..331T} {171, 331}

\bibitem[\protect\citeauthoryear{{Thompson}, {Quataert}  \& {Murray}}{{Thompson} et~al.}{2005}]{2005ApJ...630..167T}
{Thompson} T.~A.,  {Quataert} E.,   {Murray} N.,  2005, \mn@doi [\apj] {10.1086/431923}, \href {https://ui.adsabs.harvard.edu/abs/2005ApJ...630..167T} {630, 167}

\bibitem[\protect\citeauthoryear{{Tilvi} et~al.,}{{Tilvi} et~al.}{2014}]{2014ApJ...794....5T}
{Tilvi} V.,  et~al., 2014, \mn@doi [\apj] {10.1088/0004-637X/794/1/5}, \href {https://ui.adsabs.harvard.edu/abs/2014ApJ...794....5T} {794, 5}

\bibitem[\protect\citeauthoryear{{Tilvi} et~al.,}{{Tilvi} et~al.}{2020}]{2020ApJ...891L..10T}
{Tilvi} V.,  et~al., 2020, \mn@doi [\apjl] {10.3847/2041-8213/ab75ec}, \href {https://ui.adsabs.harvard.edu/abs/2020ApJ...891L..10T} {891, L10}

\bibitem[\protect\citeauthoryear{{Topping} et~al.,}{{Topping} et~al.}{2024}]{2024MNRAS.529.3301T}
{Topping} M.~W.,  et~al., 2024, \mn@doi [\mnras] {10.1093/mnras/stae682}, \href {https://ui.adsabs.harvard.edu/abs/2024MNRAS.529.3301T} {529, 3301}

\bibitem[\protect\citeauthoryear{{Torralba-Torregrosa} et~al.,}{{Torralba-Torregrosa} et~al.}{2024}]{2024A&A...689A..44T}
{Torralba-Torregrosa} A.,  et~al., 2024, \mn@doi [\aap] {10.1051/0004-6361/202450318}, \href {https://ui.adsabs.harvard.edu/abs/2024A&A...689A..44T} {689, A44}

\bibitem[\protect\citeauthoryear{{Trapp}, {Furlanetto}  \& {Davies}}{{Trapp} et~al.}{2023}]{2023MNRAS.524.5891T}
{Trapp} A.~C.,  {Furlanetto} S.~R.,   {Davies} F.~B.,  2023, \mn@doi [\mnras] {10.1093/mnras/stad2228}, \href {https://ui.adsabs.harvard.edu/abs/2023MNRAS.524.5891T} {524, 5891}

\bibitem[\protect\citeauthoryear{{{\"U}bler} et~al.,}{{{\"U}bler} et~al.}{2023}]{2023A&A...677A.145U}
{{\"U}bler} H.,  et~al., 2023, \mn@doi [\aap] {10.1051/0004-6361/202346137}, \href {https://ui.adsabs.harvard.edu/abs/2023A&A...677A.145U} {677, A145}

\bibitem[\protect\citeauthoryear{{{\"U}bler} et~al.,}{{{\"U}bler} et~al.}{2024}]{2024MNRAS.531..355U}
{{\"U}bler} H.,  et~al., 2024, \mn@doi [\mnras] {10.1093/mnras/stae943}, \href {https://ui.adsabs.harvard.edu/abs/2024MNRAS.531..355U} {531, 355}

\bibitem[\protect\citeauthoryear{{Umeda}, {Ouchi}, {Nakajima}, {Harikane}, {Ono}, {Xu}, {Isobe}  \& {Zhang}}{{Umeda} et~al.}{2024}]{2024ApJ...971..124U}
{Umeda} H.,  {Ouchi} M.,  {Nakajima} K.,  {Harikane} Y.,  {Ono} Y.,  {Xu} Y.,  {Isobe} Y.,   {Zhang} Y.,  2024, \mn@doi [\apj] {10.3847/1538-4357/ad554e}, \href {https://ui.adsabs.harvard.edu/abs/2024ApJ...971..124U} {971, 124}

\bibitem[\protect\citeauthoryear{{Vanzella} et~al.,}{{Vanzella} et~al.}{2023}]{2023A&A...678A.173V}
{Vanzella} E.,  et~al., 2023, \mn@doi [\aap] {10.1051/0004-6361/202346981}, \href {https://ui.adsabs.harvard.edu/abs/2023A&A...678A.173V} {678, A173}

\bibitem[\protect\citeauthoryear{{Veilleux} \& {Osterbrock}}{{Veilleux} \& {Osterbrock}}{1987}]{1987ApJS...63..295V}
{Veilleux} S.,  {Osterbrock} D.~E.,  1987, \mn@doi [\apjs] {10.1086/191166}, \href {https://ui.adsabs.harvard.edu/abs/1987ApJS...63..295V} {63, 295}

\bibitem[\protect\citeauthoryear{{Verhamme}, {Schaerer}  \& {Maselli}}{{Verhamme} et~al.}{2006}]{2006A&A...460..397V}
{Verhamme} A.,  {Schaerer} D.,   {Maselli} A.,  2006, \mn@doi [\aap] {10.1051/0004-6361:20065554}, \href {https://ui.adsabs.harvard.edu/abs/2006A&A...460..397V} {460, 397}

\bibitem[\protect\citeauthoryear{{\VAN{Walt}{Van der}{van der} Walt}, {Colbert}  \& {Varoquaux}}{{\VAN{Walt}{Van der}{van der} Walt} et~al.}{2011}]{2011CSE....13b..22V}
{\VAN{Walt}{Van der}{van der} Walt} S.,  {Colbert} S.~C.,   {Varoquaux} G.,  2011, \mn@doi [Computing in Science and Engineering] {10.1109/MCSE.2011.37}, \href {https://ui.adsabs.harvard.edu/abs/2011CSE....13b..22V} {13, 22}

\bibitem[\protect\citeauthoryear{{Weinberger}, {Kulkarni}, {Haehnelt}, {Choudhury}  \& {Puchwein}}{{Weinberger} et~al.}{2018}]{2018MNRAS.479.2564W}
{Weinberger} L.~H.,  {Kulkarni} G.,  {Haehnelt} M.~G.,  {Choudhury} T.~R.,   {Puchwein} E.,  2018, \mn@doi [\mnras] {10.1093/mnras/sty1563}, \href {https://ui.adsabs.harvard.edu/abs/2018MNRAS.479.2564W} {479, 2564}

\bibitem[\protect\citeauthoryear{{Weinberger}, {Haehnelt}  \& {Kulkarni}}{{Weinberger} et~al.}{2019}]{2019MNRAS.485.1350W}
{Weinberger} L.~H.,  {Haehnelt} M.~G.,   {Kulkarni} G.,  2019, \mn@doi [\mnras] {10.1093/mnras/stz481}, \href {https://ui.adsabs.harvard.edu/abs/2019MNRAS.485.1350W} {485, 1350}

\bibitem[\protect\citeauthoryear{{Whitler}, {Mason}, {Ren}, {Dijkstra}, {Mesinger}, {Pentericci}, {Trenti}  \& {Treu}}{{Whitler} et~al.}{2020}]{2020MNRAS.495.3602W}
{Whitler} L.~R.,  {Mason} C.~A.,  {Ren} K.,  {Dijkstra} M.,  {Mesinger} A.,  {Pentericci} L.,  {Trenti} M.,   {Treu} T.,  2020, \mn@doi [\mnras] {10.1093/mnras/staa1178}, \href {https://ui.adsabs.harvard.edu/abs/2020MNRAS.495.3602W} {495, 3602}

\bibitem[\protect\citeauthoryear{{Whitler}, {Endsley}, {Stark}, {Topping}, {Chen}  \& {Charlot}}{{Whitler} et~al.}{2023a}]{2023MNRAS.519..157W}
{Whitler} L.,  {Endsley} R.,  {Stark} D.~P.,  {Topping} M.,  {Chen} Z.,   {Charlot} S.,  2023a, \mn@doi [\mnras] {10.1093/mnras/stac3535}, \href {https://ui.adsabs.harvard.edu/abs/2023MNRAS.519..157W} {519, 157}

\bibitem[\protect\citeauthoryear{{Whitler}, {Stark}, {Endsley}, {Leja}, {Charlot}  \& {Chevallard}}{{Whitler} et~al.}{2023b}]{2023MNRAS.519.5859W}
{Whitler} L.,  {Stark} D.~P.,  {Endsley} R.,  {Leja} J.,  {Charlot} S.,   {Chevallard} J.,  2023b, \mn@doi [\mnras] {10.1093/mnras/stad004}, \href {https://ui.adsabs.harvard.edu/abs/2023MNRAS.519.5859W} {519, 5859}

\bibitem[\protect\citeauthoryear{{Whitler}, {Stark}, {Endsley}, {Chen}, {Mason}, {Topping}  \& {Charlot}}{{Whitler} et~al.}{2024}]{2024MNRAS.529..855W}
{Whitler} L.,  {Stark} D.~P.,  {Endsley} R.,  {Chen} Z.,  {Mason} C.,  {Topping} M.~W.,   {Charlot} S.,  2024, \mn@doi [\mnras] {10.1093/mnras/stae516}, \href {https://ui.adsabs.harvard.edu/abs/2024MNRAS.529..855W} {529, 855}

\bibitem[\protect\citeauthoryear{{Williams} et~al.,}{{Williams} et~al.}{2023}]{2023ApJS..268...64W}
{Williams} C.~C.,  et~al., 2023, \mn@doi [\apjs] {10.3847/1538-4365/acf130}, \href {https://ui.adsabs.harvard.edu/abs/2023ApJS..268...64W} {268, 64}

\bibitem[\protect\citeauthoryear{{Wisotzki} et~al.,}{{Wisotzki} et~al.}{2018}]{2018Natur.562..229W}
{Wisotzki} L.,  et~al., 2018, \mn@doi [\nat] {10.1038/s41586-018-0564-6}, \href {https://ui.adsabs.harvard.edu/abs/2018Natur.562..229W} {562, 229}

\bibitem[\protect\citeauthoryear{{Witstok}, {Smit}, {Maiolino}, {Curti}, {Laporte}, {Massey}, {Richard}  \& {Swinbank}}{{Witstok} et~al.}{2021a}]{2021MNRAS.508.1686W}
{Witstok} J.,  {Smit} R.,  {Maiolino} R.,  {Curti} M.,  {Laporte} N.,  {Massey} R.,  {Richard} J.,   {Swinbank} M.,  2021a, \mn@doi [\mnras] {10.1093/mnras/stab2591}, \href {https://ui.adsabs.harvard.edu/abs/2021MNRAS.508.1686W} {508, 1686}

\bibitem[\protect\citeauthoryear{{Witstok}, {Puchwein}, {Kulkarni}, {Smit}  \& {Haehnelt}}{{Witstok} et~al.}{2021b}]{2021A&A...650A..98W}
{Witstok} J.,  {Puchwein} E.,  {Kulkarni} G.,  {Smit} R.,   {Haehnelt} M.~G.,  2021b, \mn@doi [\aap] {10.1051/0004-6361/202040187}, \href {https://ui.adsabs.harvard.edu/abs/2021A&A...650A..98W} {650, A98}

\bibitem[\protect\citeauthoryear{{Witstok} et~al.,}{{Witstok} et~al.}{2022}]{2022MNRAS.515.1751W}
{Witstok} J.,  et~al., 2022, \mn@doi [\mnras] {10.1093/mnras/stac1905}, \href {https://ui.adsabs.harvard.edu/abs/2022MNRAS.515.1751W} {515, 1751}

\bibitem[\protect\citeauthoryear{{Witstok} et~al.,}{{Witstok} et~al.}{2023}]{2023Natur.621..267W}
{Witstok} J.,  et~al., 2023, \mn@doi [\nat] {10.1038/s41586-023-06413-w}, \href {https://ui.adsabs.harvard.edu/abs/2023Natur.621..267W} {621, 267}

\bibitem[\protect\citeauthoryear{{Witstok} et~al.,}{{Witstok} et~al.}{2024a}]{2024arXiv240816608W}
{Witstok} J.,  et~al., 2024a, preprint, \href {https://ui.adsabs.harvard.edu/abs/2024arXiv240816608W} {} (\mn@eprint {arXiv} {2408.16608})

\bibitem[\protect\citeauthoryear{{Witstok} et~al.,}{{Witstok} et~al.}{2024b}]{2024A&A...682A..40W}
{Witstok} J.,  et~al., 2024b, \mn@doi [\aap] {10.1051/0004-6361/202347176}, \href {https://ui.adsabs.harvard.edu/abs/2024A&A...682A..40W} {682, A40}

\bibitem[\protect\citeauthoryear{{Witten} et~al.,}{{Witten} et~al.}{2024}]{2024NatAs...8..384W}
{Witten} C.,  et~al., 2024, \mn@doi [Nature Astronomy] {10.1038/s41550-023-02179-3}, \href {https://ui.adsabs.harvard.edu/abs/2024NatAs...8..384W} {8, 384}

\bibitem[\protect\citeauthoryear{{Yang} et~al.,}{{Yang} et~al.}{2017}]{2017ApJ...844..171Y}
{Yang} H.,  et~al., 2017, \mn@doi [\apj] {10.3847/1538-4357/aa7d4d}, \href {https://ui.adsabs.harvard.edu/abs/2017ApJ...844..171Y} {844, 171}

\bibitem[\protect\citeauthoryear{{Zitrin} et~al.,}{{Zitrin} et~al.}{2015}]{2015ApJ...810L..12Z}
{Zitrin} A.,  et~al., 2015, \mn@doi [\apjl] {10.1088/2041-8205/810/1/L12}, \href {https://ui.adsabs.harvard.edu/abs/2015ApJ...810L..12Z} {810, L12}

\makeatother
\end{thebibliography}






\appendix

\section*{Affiliations}
\noindent
\textit{\hypertarget{inst:Kavli}$^{1}$Kavli Institute for Cosmology, University of Cambridge, Madingley Road, Cambridge CB3 0HA, UK
    \\
    \hypertarget{inst:Cav}$^{2}$Cavendish Laboratory, University of Cambridge, 19 JJ Thomson Avenue, Cambridge CB3 0HE, UK
    \\
    \hypertarget{inst:DAWN}$^{3}$Cosmic Dawn Center (DAWN), Copenhagen, Denmark
    \\
    \hypertarget{inst:NBI}$^{4}$Niels Bohr Institute, University of Copenhagen, Jagtvej 128, DK-2200, Copenhagen, Denmark
    \\
    \hypertarget{inst:UCL}$^{5}$Department of Physics and Astronomy, University College London, Gower Street, London WC1E 6BT, UK
    \\
    \hypertarget{inst:LJMU}$^{6}$Astrophysics Research Institute, Liverpool John Moores University, 146 Brownlow Hill, Liverpool L3 5RF, UK
    \\
    \hypertarget{inst:Oxford}$^{7}$Department of Physics, University of Oxford, Denys Wilkinson Building, Keble Road, Oxford OX1 3RH, UK
    \\
    \hypertarget{inst:Steward}$^{8}$Steward Observatory, University of Arizona, 933 N. Cherry Avenue, Tucson AZ 85721, USA
    \\
    \hypertarget{inst:CfA}$^{9}$Center for Astrophysics $|$ Harvard \& Smithsonian, 60 Garden St., Cambridge MA 02138, USA
    \\
    \hypertarget{inst:CAB}$^{10}$Centro de Astrobiolog\'ia (CAB), CSIC–INTA, Cra. de Ajalvir Km.~4, 28850- Torrej\'on de Ardoz, Madrid, Spain
    \\
    \hypertarget{inst:ESAC}$^{11}$European Space Agency (ESA), European Space Astronomy Centre (ESAC), Camino Bajo del Castillo s/n, 28692 Villanueva de la Cañada, Madrid, Spain
    \\
    \hypertarget{inst:Melbourne}$^{12}$School of Physics, University of Melbourne, Parkville 3010, VIC, Australia
    \\
    \hypertarget{inst:ARC3D}$^{13}$ARC Centre of Excellence for All Sky Astrophysics in 3 Dimensions (ASTRO 3D), Australia
    \\
    \hypertarget{inst:SNS}$^{14}$Scuola Normale Superiore, Piazza dei Cavalieri 7, I-56126 Pisa, Italy
    \\
    \hypertarget{inst:IAP}$^{15}$Sorbonne Universit\'e, CNRS, UMR 7095, Institut d'Astrophysique de Paris, 98 bis bd Arago, 75014 Paris, France
    \\
    \hypertarget{inst:ESO}$^{16}$European Southern Observatory, Karl-Schwarzschild-Strasse 2, 85748 Garching, Germany
    \\
    \hypertarget{inst:Herts}$^{17}$Centre for Astrophysics Research, Department of Physics, Astronomy and Mathematics, University of Hertfordshire, Hatfield, AL10 9AB UK
    \\
    \hypertarget{inst:JHU}$^{18}$Department of Physics and Astronomy, The Johns Hopkins University, 3400 N. Charles St., Baltimore MD 21218, USA
    \\
    \hypertarget{inst:AURA}$^{19}$AURA for European Space Agency, Space Telescope Science Institute, 3700 San Martin Drive. Baltimore, MD 21210, USA
    \\
    \hypertarget{inst:Wisconsin}$^{20}$Department of Astronomy, University of Wisconsin-Madison, 475 N. Charter St., Madison WI 53706, USA
    \\
    \hypertarget{inst:UCSC}$^{21}$Department of Astronomy and Astrophysics University of California, Santa Cruz, 1156 High Street, Santa Cruz CA 96054, USA
    \\
    \hypertarget{inst:NSF}$^{22}$NSF's National Optical-Infrared Astronomy Research Laboratory, 950 North Cherry Avenue, Tucson AZ 85719, USA
    \\
    \hypertarget{inst:NRC}$^{23}$NRC Herzberg, 5071 West Saanich Rd, Victoria BC V9E 2E7, Canada
}

\section{NIRCam observations of LAEs and candidate companion galaxies}
\label{app:Neighbour_SED_modelling}
\begin{figure*}
	\centering
	\includegraphics[width=0.95\linewidth]{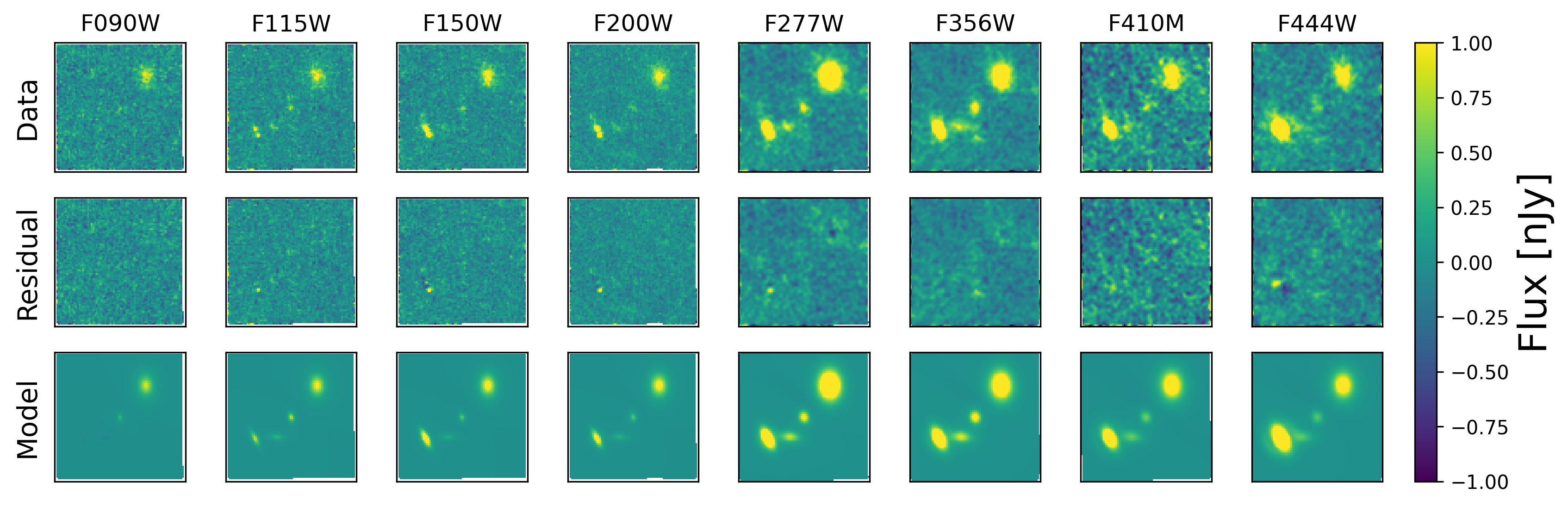}
	\caption{NIRCam imaging across $8$ filters (top row) in the neighbourhood of \JGSzeightoneLA at $z \simeq 8.72$ (source on the bottom left, resolved into two components in the SW filters). The \program{ForcePho} model of the scene (\cref{app:Neighbour_SED_modelling}) is shown in the bottom row, with residuals shown in the middle row. Note these cutouts are in detector coordinates, with different orientation from \cref{fig:LAE_SED_models}.
	}
	\label{fig:ForcePho_scene}
\end{figure*}

To obtain accurate photometry in the crowded neighbourhood of \JGSzeightoneLA at $z \simeq 8.72$ (see \cref{fig:LAE_SED_models}), we perform full-scene modelling with \program{ForcePho} \citetext{Johnson et al. in prep.}. \program{ForcePho} models the light profiles of all sources in a scene as PSF-convolved \citeauthor{1963BAAA....6...41S} profiles, fitting to individual NIRCam exposures to avoid inter-pixel correlations introduced in drizzled mosaic images. The results are shown in \cref{fig:ForcePho_scene}, revealing that two sources (including \JGSzeightoneLA) drop out of the F090W filter, while two others are detected in the F090W filter and thus at lower redshift.

From SED modelling of the nearby sources identified in the NIRCam imaging (\cref{ssec:SED_modelling}), we find that these may be low-mass ($M_* \approx 10^7 \, \mathrm{M_\odot}$) satellite galaxies. Given the close separation between the LAEs and these photometric companions ($\ssim 1 \, \mathrm{kpc}$), it is conceivable that these systems are undergoing merger events, potentially further suggested by low-ionisation line emission (discussed below), which could play an important role in supplying and compressing gas, thereby triggering an episode of star formation and/or AGN activity (\cref{sssec:Sources_of_ionisation}). As suggested by \citet{2024NatAs...8..384W}, by supplying an ample amount of ionising photons as well as feedback, such events may be crucial for the observability of \Lya, facilitating both its production via recombinations in \HII regions and/or collisional excitation in dense gas \citep[e.g.][]{2021A&A...650A..98W} and its escape, which requires neutral gas to be cleared from ISM scales to the surrounding IGM (discussed further in \cref{sssec:Have_we_identified_the_sources_inflating_ionised_bubbles}) and may be facilitated by outflowing gas \citep{2006A&A...460..397V, 2024A&A...684A.207F}.

None of these secondary sources show a characteristic F444W excess that would be expected from high-EW $\OIII \, \lambda \, 4960, 5008 \, \Angstrom$ and \Hbeta lines at $z \sim 8$ (as indeed seen in the photometry of the three LAEs). For the neighbouring source of \JGSzeightoneLA, we do however note an F356W excess compared to other bands, with an additional source seeming to appear only in the F356W filter without a clear counterpart in the other filters (cf. \cref{fig:ForcePho_scene} in \cref{app:Neighbour_SED_modelling}). The excess is relative to both bluer (F277W) and redder bands (F444W), ruling out a strong Balmer break and instead suggesting high-EW line emission in $\OII \, \lambda \, 3727, 3730 \, \Angstrom$ (\OII) and $\NeIII \, \lambda \, 3870 \, \Angstrom$. However, the weak $\OIII \, \lambda \, 4960, 5008 \, \Angstrom$ emission inferred from the lack of F444W excess rules out a strong contribution of $\NeIII \, \lambda \, 3870 \, \Angstrom$. Such a low \OIII/\OII line ratio indicates low ionisation, which is not likely due to high-metallicity stars, perhaps indicating that the nebular emission in this galaxy is powered by shocks induced by interactions with \JGSzeightoneLA. Intriguingly, we further note the neighbour of \JGNzeightzeroLA is almost undetected even in the F115W filter, suggesting it has an even higher redshift ($z \sim 9$), or, perhaps more likely, it is impacted by strong damped \Lya absorption \citep{2024arXiv240402211H, 2024Sci...384..890H, 2024A&A...689A.152D, 2024arXiv240404325H, 2024ApJ...971..124U}, which has been argued by \citet{2024MNRAS.528.7052C} to be prevalent in the dense gaseous environments of very early galaxies. Follow-up spectroscopy will be required to confirm the nature of these neighbouring sources.

\section{Relating \texorpdfstring{$\xi_\mathrm{ion, \, 0}$}{ξion,0} to \texorpdfstring{\Lya}{\Lyatext} EW}
\label{app:Relating_xi_ion_to_Lya_EW}

Multiplying the rate of non-escaping ionising photon $\dot{N}_\mathrm{ion, \, 0}$ by the fraction of case-B recombination events that result in the emission of a \Lya photon \citep[$f_\text{rec, B} = 66\%$ at $T = \num{15000} \, \mathrm{K}$; see e.g.][]{2014PASA...31...40D}, we arrive at the emission rate of \Lya photons, which in turn when multiplied by the energy of a \Lya photon ($E_\text{\Lya} = h \nu_\Lya$) yields the \Lya luminosity, $L_\text{\Lya}$ \citep[see also][]{2024arXiv240816608W}. From the definition of $\xi_\text{ion}$ given in \cref{sssec:Production_and_escape_of_ionising_photons}, we can therefore write
\begin{align*}
    \xi_\mathrm{ion, \, 0} & = \frac{\dot{N}_\mathrm{ion, \, 0}}{L_\mathrm{\nu, \, UV}} = \frac{L_\text{\Lya}}{f_\text{rec, B} \, E_\text{\Lya} \, L_\mathrm{\nu, \, UV}}
    \\
    & = \frac{F_\text{\Lya} \, 4 \pi d_L^2(z) / f_\text{esc, \Lya}}{f_\text{rec, B} \, E_\text{\Lya} \, F_\mathrm{\nu, \, UV} \, 4 \pi d_L^2(z) / (1+z)}
    \\
    & = \frac{1 + z}{f_\text{esc, \Lya} \, f_\text{rec, B} \, E_\text{\Lya}} \frac{F_\text{\Lya}}{F_\mathrm{\nu, \, UV}} \, ,
\end{align*}
\noindent where, in converting luminosity (density) to flux (density) using the luminosity distance $d_L(z)$, we can exploit the cancellation of the $4\pi d_L^2(z)$ terms, although two factors remain: the observed \Lya flux is also modulated by the \Lya escape fraction, $f_\text{esc, \Lya}$, and a factor $(1+z)$ appears to account for (red)shifting the flux density $F_\mathrm{\nu, \, UV}$ into the observed frame. We further note that we assume knowledge of the intrinsic $\xi_\mathrm{ion, \, 0}$ here, and hence observed flux (density) refers to the intrinsic values in the absence of, or corrected for, dust absorption. Typically, however, LAEs are expected to have a low dust content \citep[e.g.][]{2011ApJ...730....8H}. Meanwhile, empirical measurements of $\dot{N}_\text{ion}$ typically rely on optical Balmer lines (\cref{sssec:Production_and_escape_of_ionising_photons}), thereby introducing a source of uncertainty on the inferred $\xi_\mathrm{ion, \, 0}$ via the (non-negligible) dust-attenuation correction between the UV and optical \citep{2023MNRAS.523.5468S}. For the continuum flux density $F_\mathrm{\lambda, \, cont}$ taking the form of a power law\footnote{Specifically, the power-law continuum is normalised at $1500 \, \Angstrom$ to $F_\mathrm{\lambda, \, UV}$, the observed flux-density equivalent of the UV luminosity density $L_\mathrm{\nu, \, UV}$.} with a UV slope $\beta_\text{UV}$, it follows that the continuum flux density at the wavelength of \Lya is related to $F_\mathrm{\nu, \, UV}$ as
\begin{align*}
    F_\mathrm{\lambda, \, cont} (\lambda_\text{\Lya}) & = F_\mathrm{\lambda, \, UV} \left( \frac{\lambda_\text{\Lya}}{1500 \, \Angstrom} \right)^{\beta_\text{UV}}
    \\
    & = \frac{c}{\left(1+z\right)^2} \frac{(\lambda_\text{\Lya})^{\beta_\text{UV}}}{(1500 \, \Angstrom)^{2+\beta_\text{UV}}} \, F_\mathrm{\nu, \, UV} \, ,
\end{align*}
\noindent where an additional factor $1/(1+z)^2$ is introduced through the conversion between the observed flux densities $F_\mathrm{\lambda, \, UV}$ to $F_\mathrm{\nu, \, UV}$, which occurs at the redshifted location of $1500 \, \Angstrom$ in the rest frame. Finally, we can directly relate $\xi_\mathrm{ion, \, 0}$ to $\text{EW}_\text{\Lya}$ via
\begin{align}
    \label{eq:xi_ion_Lya_EW2}
    \xi_\mathrm{ion, \, 0} & = \frac{c \, (\lambda_\text{\Lya})^{\beta_\text{UV}}}{f_\text{esc, \Lya} \, f_\text{rec, B} \, E_\text{\Lya} \, (1500 \, \Angstrom)^{2+\beta_\text{UV}}} \frac{F_\text{\Lya}}{F_\mathrm{\lambda, \, cont} (\lambda_\text{\Lya}) \, \left( 1 + z \right)} \nonumber
    \\
    & = \frac{1}{f_\text{esc, \Lya} \, f_\text{rec, B} \, h \, \lambda_\text{\Lya}} \, \left( \frac{\lambda_\text{\Lya}}{1500 \, \Angstrom} \right)^{2+\beta_\text{UV}} \, \text{EW}_\text{\Lya} \, ,
\end{align}

\noindent to arrive at the expression given in \cref{eq:xi_ion_Lya_EW}.

\section{Emission-line measurements}
\label{app:Emission-line_measurements}

\begin{figure*}
	\centering
	\includegraphics[width=0.95\linewidth]{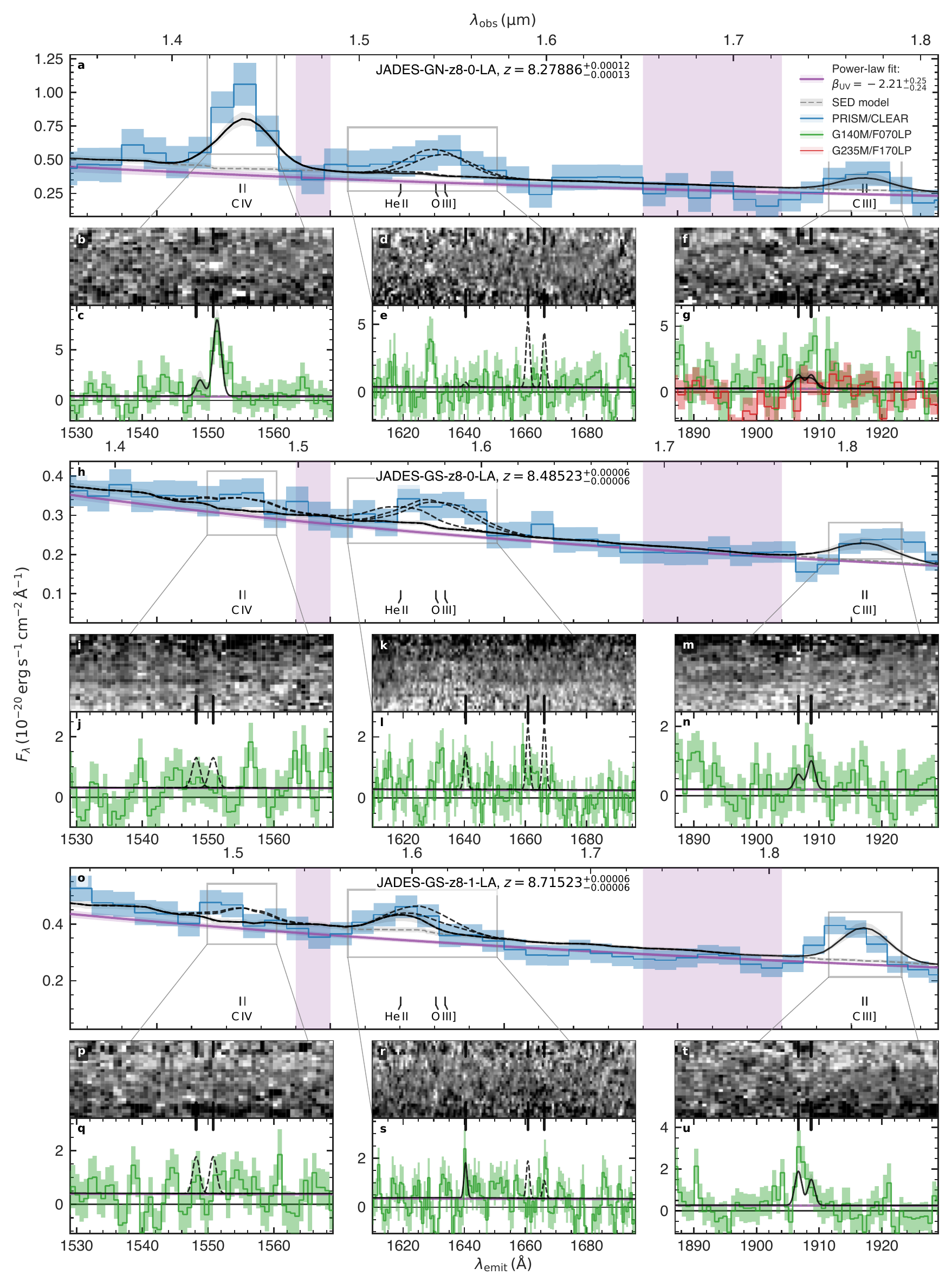}
	\caption{Overview of the rest-frame UV spectra. Differently coloured lines are as in \cref{fig:Line_overview_1899}. The empirical emission-line fit at the appropriate spectral resolution (\cref{ssec:Continuum_and_line_emission}) is shown by the solid black line, while dashed black lines indicate upper limits.
	}
	\label{fig:UV_lines_overview}
\end{figure*}

In \cref{fig:UV_lines_overview}, we present an overview of the PRISM and R1000 spectra covering the rest-frame UV of the three main LAEs considered in this work (\cref{ssec:LAE_sample}). We note that in the R1000 spectrum of \JGNzeightzeroLA, a tentative ($\ssim 3 \sigma$) feature at $\lambda_\text{obs} \approx 1.512 \, \mathrm{\upmu m}$ appears. This may be due to contamination from a low-redshift interloper, or, at rest-frame wavelength of $\lambda_\text{emit} \approx 1629 \, \Angstrom$, a blueshifted ($\ssim 2000 \, \mathrm{km \, s^{-1}}$) \HeII line. However, given the short exposure time we deem this feature sufficiently insignificant such that it can reasonably be attributed to noise (see, for instance, the noise features with similar amplitude at $\lambda_\text{emit} \approx 1693 \, \Angstrom$). Moreover, we do not find evidence for associated emission lines as would be expected when interpreted as a low-redshift contaminant or a nearby satellite galaxy with strong emission lines along the line of sight.

All measurements of observed emission lines via the methods discussed in \cref{ssec:Continuum_and_line_emission} are presented in \cref{tab:Line_measurements1,tab:Line_measurements2}.

\begingroup
    \setlength{\tabcolsep}{6pt} 
    \renewcommand{\arraystretch}{1.5} 
    \begin{table*}
        \centering
        \caption
        {Emission-line measurements of the sources studied in this work.}
        \label{tab:Line_measurements1}
        \begin{tabular}{cclllll}
\toprule
Line & $\lambda_\text{emit} \, (\Angstrom)$ & Quantity & Unit & JADES-GN-z8-0-LA & JADES-GS-z8-0-LA & JADES-GS-z8-1-LA
\\
\midrule
\multicolumn{2}{c}{All lines} & $\sigma_v$ & $\mathrm{km \, s^{-1}}$ & $70.4_{-6.2}^{+6.0}$ & $25.2_{-5.3}^{+5.1}$ & $50.6_{-3.6}^{+3.8}$
\vspace{2.0ex}
\\
\multirow{5}{*}{\Lya} & \multirow{5}{*}{$1215.67$} & $F$ & $10^{-20} \, \mathrm{erg \, s^{-1} \, cm^{-2}}$ & $772_{-23}^{+23}$ & $67.4_{-9.3}^{+11.4}$ & $166_{-13}^{+13}$
\\
& & $\text{EW}$ & $\Angstrom$ & $131.8_{-3.9}^{+4.0}$ & $14.6_{-2.0}^{+2.5}$ & $30.7_{-2.3}^{+2.4}$
\\
& & $a_\text{asym}$ &  & $3.5_{-0.8}^{+0.7}$ & $2.2_{-5.9}^{+4.9}$ & $-6.1_{-2.6}^{+2.7}$
\\
& & $\Delta v$ & $\mathrm{km \, s^{-1}}$ & $140_{-23}^{+23}$ & $212_{-41}^{+33}$ & $360_{-37}^{+30}$
\\
& & $\sigma_v$ & $\mathrm{km \, s^{-1}}$ & $269_{-16}^{+15}$ & $37_{-18}^{+79}$ & $187_{-27}^{+30}$
\vspace{2.0ex}
\\
\multirow{3}{*}{\CIV} & \multirow{3}{*}{$1548.19$} & $F$ & $10^{-20} \, \mathrm{erg \, s^{-1} \, cm^{-2}}$ & $23_{-13}^{+14}$ & $<14$ & $<17$
\\
& & $\text{EW}$ & $\Angstrom$ & $5.9_{-3.3}^{+3.0}$ & $<4.7$ & $<4.3$
\\
& & $\Delta v$ & $\mathrm{km \, s^{-1}}$ & $116_{-28}^{+29}$ & \dots & \dots
\vspace{2.0ex}
\\
\multirow{3}{*}{\CIV} & \multirow{3}{*}{$1550.77$} & $F$ & $10^{-20} \, \mathrm{erg \, s^{-1} \, cm^{-2}}$ & $110_{-15}^{+14}$ & $<14$ & $<17$
\\
& & $\text{EW}$ & $\Angstrom$ & $26.3_{-3.6}^{+3.2}$ & $<4.7$ & $<4.3$
\\
& & $\Delta v$ & $\mathrm{km \, s^{-1}}$ & $116_{-28}^{+29}$ & \dots & \dots
\vspace{2.0ex}
\\
\multirow{2}{*}{\HeII} & \multirow{2}{*}{$1640.42$} & $F$ & $10^{-20} \, \mathrm{erg \, s^{-1} \, cm^{-2}}$ & $<5.0$ & $<18$ & $18.4_{-7.7}^{+7.1}$
\\
& & $\text{EW}$ & $\Angstrom$ & $<1.3$ & $<6.5$ & $4.9_{-2.0}^{+1.9}$
\vspace{2.0ex}
\\
\multirow{2}{*}{\OIIIs} & \multirow{2}{*}{$1660.81$} & $F$ & $10^{-20} \, \mathrm{erg \, s^{-1} \, cm^{-2}}$ & $<71$ & $<29$ & $<20$
\\
& & $\text{EW}$ & $\Angstrom$ & $<19$ & $<11$ & $<5.4$
\vspace{2.0ex}
\\
\multirow{2}{*}{\OIIIs} & \multirow{2}{*}{$1666.15$} & $F$ & $10^{-20} \, \mathrm{erg \, s^{-1} \, cm^{-2}}$ & $<59$ & $<29$ & $<9.5$
\\
& & $\text{EW}$ & $\Angstrom$ & $<16$ & $<11$ & $<2.6$
\vspace{2.0ex}
\\
\multirow{2}{*}{\CIIIf} & \multirow{2}{*}{$1906.68$} & $F$ & $10^{-20} \, \mathrm{erg \, s^{-1} \, cm^{-2}}$ & $16.0_{-8.3}^{+7.7}$ & $6.2_{-4.0}^{+4.4}$ & $21.9_{-4.3}^{+4.6}$
\\
& & $\text{EW}$ & $\Angstrom$ & $6.2_{-3.1}^{+2.9}$ & $3.4_{-2.0}^{+2.7}$ & $8.3_{-1.7}^{+1.7}$
\vspace{2.0ex}
\\
\multirow{2}{*}{\CIIIs} & \multirow{2}{*}{$1908.73$} & $F$ & $10^{-20} \, \mathrm{erg \, s^{-1} \, cm^{-2}}$ & $16.0_{-8.3}^{+7.7}$ & $11.7_{-4.7}^{+5.9}$ & $16.6_{-3.2}^{+3.6}$
\\
& & $\text{EW}$ & $\Angstrom$ & $6.2_{-3.1}^{+2.9}$ & $6.6_{-2.8}^{+3.2}$ & $6.2_{-1.1}^{+1.4}$
\vspace{2.0ex}
\\
\multirow{2}{*}{\MgII} & \multirow{2}{*}{$2796.35$} & $F$ & $10^{-20} \, \mathrm{erg \, s^{-1} \, cm^{-2}}$ & $<2.3$ & $<2.1$ & $<9.4$
\\
& & $\text{EW}$ & $\Angstrom$ & $<1.9$ & $<3.2$ & $<7.4$
\vspace{2.0ex}
\\
\multirow{2}{*}{\MgII} & \multirow{2}{*}{$2803.53$} & $F$ & $10^{-20} \, \mathrm{erg \, s^{-1} \, cm^{-2}}$ & $<2.1$ & $<3.6$ & $<7.2$
\\
& & $\text{EW}$ & $\Angstrom$ & $<1.8$ & $<5.7$ & $<5.7$
\vspace{2.0ex}
\\
\bottomrule
\end{tabular}
        \flushleft
        \textbf{Notes.} Listed emission-line properties are the integrated flux ($F$) and equivalent width ($\text{EW}$). Quoted upper limits are $3\sigma$ (see \cref{ssec:Continuum_and_line_emission}). If applicable, the asymmetry parameter ($a_\text{asym}$), velocity offset ($\Delta v$), and velocity dispersion ($\sigma_v$) are also shown.
    \end{table*}
    \begin{table*}
        \flushleft
        \textbf{\cref{tab:Line_measurements1}} (continued).
        \\ \phantom{x}
        \\
        \centering
        \label{tab:Line_measurements2}
        \centering
        \begin{tabular}{cclllll}
\toprule
Line & $\lambda_\text{emit} \, (\Angstrom)$ & Quantity & Unit & JADES-GN-z8-0-LA & JADES-GS-z8-0-LA & JADES-GS-z8-1-LA
\\
\midrule
\multirow{2}{*}{\OII} & \multirow{2}{*}{$3727.09$} & $F$ & $10^{-20} \, \mathrm{erg \, s^{-1} \, cm^{-2}}$ & $<11$ & $4.5_{-1.0}^{+1.1}$ & $22.3_{-3.2}^{+3.6}$
\\
& & $\text{EW}$ & $\Angstrom$ & $<19$ & $17.8_{-3.9}^{+3.6}$ & $35.7_{-4.6}^{+6.2}$
\vspace{2.0ex}
\\
\multirow{2}{*}{\OII} & \multirow{2}{*}{$3729.88$} & $F$ & $10^{-20} \, \mathrm{erg \, s^{-1} \, cm^{-2}}$ & $<4.7$ & $2.82_{-0.79}^{+1.10}$ & $19.3_{-3.5}^{+3.5}$
\\
& & $\text{EW}$ & $\Angstrom$ & $<8.3$ & $10.8_{-3.1}^{+4.4}$ & $31.4_{-5.6}^{+5.5}$
\vspace{2.0ex}
\\
\multirow{2}{*}{\NeIII} & \multirow{2}{*}{$3869.85$} & $F$ & $10^{-20} \, \mathrm{erg \, s^{-1} \, cm^{-2}}$ & $45.8_{-4.9}^{+4.6}$ & $10.6_{-1.3}^{+1.2}$ & $45.8_{-2.4}^{+2.3}$
\\
& & $\text{EW}$ & $\Angstrom$ & $82.1_{-9.2}^{+8.5}$ & $43.5_{-4.9}^{+4.9}$ & $76.1_{-3.7}^{+3.8}$
\vspace{2.0ex}
\\
\multirow{3}{*}{\HeI} & \multirow{3}{*}{$3889.75$} & $F$ & $10^{-20} \, \mathrm{erg \, s^{-1} \, cm^{-2}}$ & $16.6_{-4.5}^{+4.8}$ & $12.1_{-1.2}^{+1.2}$ & $16.4_{-2.2}^{+2.3}$
\\
& & $\text{EW}$ & $\Angstrom$ & $29.6_{-7.9}^{+8.6}$ & $50.2_{-4.3}^{+5.1}$ & $27.5_{-3.5}^{+3.9}$
\\
& & $\Delta v$ & $\mathrm{km \, s^{-1}}$ & $90_{-58}^{+54}$ & $59_{-22}^{+22}$ & $56_{-42}^{+43}$
\vspace{2.0ex}
\\
\multirow{2}{*}{\NeIII} & \multirow{2}{*}{$3968.59$} & $F$ & $10^{-20} \, \mathrm{erg \, s^{-1} \, cm^{-2}}$ & $13.8_{-1.5}^{+1.4}$ & $3.19_{-0.38}^{+0.36}$ & $13.80_{-0.72}^{+0.68}$
\\
& & $\text{EW}$ & $\Angstrom$ & $25.8_{-2.9}^{+2.7}$ & $14.1_{-1.6}^{+1.6}$ & $24.1_{-1.2}^{+1.2}$
\vspace{2.0ex}
\\
\multirow{2}{*}{\Hepsilon} & \multirow{2}{*}{$3971.20$} & $F$ & $10^{-20} \, \mathrm{erg \, s^{-1} \, cm^{-2}}$ & $12.6_{-4.8}^{+4.7}$ & $5.8_{-1.2}^{+1.2}$ & $10.7_{-2.1}^{+2.3}$
\\
& & $\text{EW}$ & $\Angstrom$ & $23.3_{-8.4}^{+8.6}$ & $25.6_{-5.3}^{+5.3}$ & $18.9_{-3.8}^{+4.2}$
\vspace{2.0ex}
\\
\multirow{2}{*}{\Hdelta} & \multirow{2}{*}{$4102.89$} & $F$ & $10^{-20} \, \mathrm{erg \, s^{-1} \, cm^{-2}}$ & $18.6_{-4.1}^{+4.1}$ & $7.6_{-1.1}^{+1.1}$ & $21.2_{-2.2}^{+2.0}$
\\
& & $\text{EW}$ & $\Angstrom$ & $36.3_{-8.4}^{+8.2}$ & $37.1_{-5.9}^{+5.6}$ & $39.4_{-4.4}^{+4.0}$
\vspace{2.0ex}
\\
\multirow{2}{*}{\Hgamma} & \multirow{2}{*}{$4341.69$} & $F$ & $10^{-20} \, \mathrm{erg \, s^{-1} \, cm^{-2}}$ & $36.8_{-5.6}^{+5.5}$ & $11.8_{-1.2}^{+1.2}$ & $42.5_{-2.3}^{+2.3}$
\\
& & $\text{EW}$ & $\Angstrom$ & $80_{-13}^{+13}$ & $66.7_{-6.4}^{+6.8}$ & $87.7_{-4.5}^{+4.9}$
\vspace{2.0ex}
\\
\multirow{2}{*}{\OIII} & \multirow{2}{*}{$4364.44$} & $F$ & $10^{-20} \, \mathrm{erg \, s^{-1} \, cm^{-2}}$ & $15.2_{-4.6}^{+4.4}$ & $5.7_{-1.2}^{+1.2}$ & $16.8_{-2.3}^{+2.2}$
\\
& & $\text{EW}$ & $\Angstrom$ & $33.1_{-9.5}^{+10.9}$ & $32.9_{-7.2}^{+7.1}$ & $35.3_{-5.0}^{+4.8}$
\vspace{2.0ex}
\\
\multirow{2}{*}{\Hbeta} & \multirow{2}{*}{$4862.71$} & $F$ & $10^{-20} \, \mathrm{erg \, s^{-1} \, cm^{-2}}$ & $66.1_{-5.5}^{+5.2}$ & $32.1_{-1.4}^{+1.3}$ & $83.6_{-2.6}^{+2.7}$
\\
& & $\text{EW}$ & $\Angstrom$ & $168_{-13}^{+13}$ & $253.1_{-10.3}^{+9.5}$ & $204.3_{-6.1}^{+7.5}$
\vspace{2.0ex}
\\
\multirow{2}{*}{\OIII} & \multirow{2}{*}{$4960.30$} & $F$ & $10^{-20} \, \mathrm{erg \, s^{-1} \, cm^{-2}}$ & $204.7_{-3.1}^{+3.2}$ & $58.94_{-0.58}^{+0.58}$ & $200.7_{-1.4}^{+1.3}$
\\
& & $\text{EW}$ & $\Angstrom$ & $525.5_{-8.1}^{+7.7}$ & $479.8_{-4.2}^{+4.9}$ & $505.5_{-3.5}^{+3.5}$
\vspace{2.0ex}
\\
\multirow{2}{*}{\OIII} & \multirow{2}{*}{$5008.24$} & $F$ & $10^{-20} \, \mathrm{erg \, s^{-1} \, cm^{-2}}$ & $610.0_{-9.1}^{+9.7}$ & $175.6_{-1.7}^{+1.7}$ & $598.0_{-4.1}^{+3.9}$
\\
& & $\text{EW}$ & $\Angstrom$ & $1600_{-25}^{+24}$ & $1479_{-13}^{+15}$ & $1534_{-10}^{+10}$
\vspace{2.0ex}
\\
\bottomrule
\end{tabular}
    \end{table*}
\endgroup

\bsp	
\label{lastpage}
\end{document}